\documentclass[12pt]{article}
\usepackage{latexsym,amsfonts}
\def\a{\alpha}
\def\b{\beta}
\def\g{\gamma}
\def\d{\delta}
\def\e{\eta}

\def\l{\lambda}

\def\m{\mu}
\def\n{\nu}
\def\r{\rho}
\def\o{\omega}
\def\s{\sigma}
\def\S{\Sigma}
\def\t{\tau}
\def\p{\pi}
\def\e{\varepsilon}

\def\rg{{\rm g}}
\def\be{\begin{equation}}
\def\ee{\end{equation}}
\def\beq{\begin{eqnarray}}
\def\eeq{\end{eqnarray}}
\def\nn{\nonumber}
\def\ca{{\cal A}}

\def\cd{{\cal D}}
\def\ce{{\cal E}}
\def\cf{{\cal F}}

\def\ch{{\cal H}}

\def\ck{{\cal K}}
\def\cl{{\cal L}}
\def\cm{{\cal M}}
\def\cn{{\cal N}}

\def\cv{{\cal V}}

\def\RR{{\mathbb{R}}}

\def\ZZ{{\mathbb{Z}}}

\newcommand{\ft}[2]{{\textstyle {\frac{#1}{#2}} }}
\newcommand{\tr}{{\rm tr \,}}

\newcommand{\frakg}{{\mathfrak{g}}}
\newcommand{\frakh}{{\mathfrak{h}}}
\newcommand{\frakn}{{\mathfrak{n}}}
\newcommand{\frakk}{{\mathfrak{k}}}

\newcommand{\Rn}{{\mathbb{R}}}
\newcommand{\E}{E_{10}}
\newcommand{\SD}{\sum_{\a\in\Delta_+}}
\newcommand{\Smult}{\sum_{s=1}^{{\rm mult}(\a)}}
\newcommand{\mult}{{{\rm mult}\,}}
\newcommand{\Ref}[1]{(\ref{#1})}
\newcommand{\non}{\nonumber\\}

\newcommand{\bqn}{\begin{eqnarray}}\newcommand{\eqn}{\end{eqnarray}}

\begin{document}

\begin{titlepage}
\begin{flushright}
IHES/P/02/80 \\
AEI-2002-092 \\
ULB-TH/02-33 
\end{flushright}
\vskip 1.0cm

\begin{centering}

{\huge {\bf Cosmological Billiards}}

\vspace{.5cm}

{\Large T.~Damour$^{a}$, M.~Henneaux$^{b,c}$ and H.~Nicolai$^{d}$} \\
\vspace{.7cm} $^a$ Institut des Hautes Etudes Scientifiques,  35,
Route de
Chartres,  F-91440 Bures-sur-Yvette, France \\
\vspace{.2cm} $^b$ Physique Th\'eorique et Math\'ematique,
Universit\'e Libre
de Bruxelles,  C.P. 231, B-1050, Bruxelles, Belgium      \\
\vspace{.2cm}
$^c$ Centro de Estudios Cient\'{\i}ficos, Casilla 1469, Valdivia, Chile \\
\vspace{.2cm} $^d $Max-Planck-Institut f\"ur Gravitationsphysik,
Albert-Einstein-Institut, Am M\"uhlenberg 1, D-14476 Golm, Germany

\vspace{1cm}

\end{centering}

\begin{abstract}
It is shown in detail that the dynamics of the Einstein-dilaton-$p$-form
system in the vicinity of a spacelike singularity can be asymptotically
described, at a generic spatial point, as a billiard motion in a region
of  Lobachevskii space (realized as an hyperboloid in the space of
logarithmic scale factors). This is done within the Hamiltonian
formalism, and for an arbitrary number of spacetime dimensions $D \geq 4$.
A key r\^ole in the derivation is played by the Iwasawa decomposition
of the spatial metric, and by the fact that the off-diagonal degrees
of freedom, as well as the $p$-form degrees of freedom, get
``asymptotically frozen'' in this description. For those models
admitting a Kac-Moody theoretic interpretation of the billiard
dynamics we outline how to set up an asymptotically equivalent
description in terms of a one-dimensional non-linear $\sigma$-model
formally invariant under the corresponding Kac-Moody group.
\end{abstract}

\vfill
\end{titlepage}

\section{Introduction}
\setcounter{equation}{0}
\setcounter{theorem}{0}
\setcounter{lemma}{0}

\subsection{BKL analysis in spacetime dimension $D = 4$}
The non-linearities of the Einstein equations are notably known to
prevent the construction of an exact general solution.  Only
peculiar solutions, corresponding to idealized situations, have
been explicitly derived. The  singularity theorems of
\cite{singularity} predict the generic appearance of spacetime
singularities under certain conditions, but do not provide
a detailed description of how the spacetime becomes singular.
{}From this perspective, the work of Belinskii, Khalatnikov and
Lifshitz \cite{BKL,BKL2}, also known as ``BKL", is quite remarkable
as it gives a description of the generic asymptotic behaviour
of the gravitational field in {\it four spacetime dimensions} in the vicinity
of a spacelike singularity. As argued by these authors, near the
singularity the spatial points essentially decouple, in the sense
that the dynamical evolution of the spatial metric at each
spatial point is asymptotically governed by a set of second-order,
nonlinear ordinary differential equations with respect to time.

These differential equations are the same, at each spatial point,
as those that arise in some spatially homogeneous cosmological models,
which therefore provide valuable insight into the qualitative features
of the general solution. In the vacuum case, the spatially
homogeneous models that capture the behaviour of the general
solution are of Bianchi type IX or VIII (with homogeneity groups
$SU(2)$ or $SL(2,\RR)$). The asymptotic evolution of the metric
can then be pictured as an infinite sequence of ``oscillations''
of the scale factors along independent spatial
directions \cite{BKL,BKL2}.  This regime is
called ``oscillatory", or ``of mixmaster type" \cite{Misner0}, or also ``chaotic'' because it
exhibits strong chaotic features \cite{KLL,Bar}. The coupling to
matter fields does not change the picture, except if one includes
a massless scalar field (equivalent to a perfect fluid with
``stiff" equation of state $p= \rho$), in which case
the chaotic evolution is replaced by a monotonic power law
evolution of the scale factors \cite{BK1,AR}, which mimicks the
Kasner solution at each spatial point and is therefore called
``Kasner-like". In this case the spatially homogeneous
model that captures the behaviour of the general solution is the
Bianchi I model (with the abelian group of translations in
$\RR^3$ as homogeneity group).

\subsection{BKL analysis in spacetime dimensions $D > 4$}

The extension of the BKL analysis to higher
dimensions was addressed within the context of pure
gravity (with no symmetry assumption) in \cite{DHS,DHHST}, where
it was shown that the general BKL approach remains valid: spatial
points decouple as one approaches a spacelike singularity, i.e.,
the dynamical evolution at each spatial point of the scale factors
is again governed by ordinary differential equations. The main
result of \cite{DHS,DHHST}, was that, while the general behaviour
of solutions of the vacuum Einstein equations remains oscillatory
for spacetime dimensions $D \leq 10$, it ceases to be so for
spacetime dimensions $D \geq 11$, where it becomes Kasner-like.
Let us also note that, just as in four spacetime dimensions, the
coupling to a massless scalar field suppresses the chaotic
behaviour in any number of spacetime dimensions and makes the
solution monotonic (see, e.g., \cite{DHRW}).

The authors of \cite{DHS} did not consider the inclusion of
massless $p$-forms, which are part of the low energy bosonic
sector of superstring/M-theory models.  This task was undertaken
in \cite{DH1,DH2}, with the finding that these $p$-forms play a
crucial role and can reinstate chaos when it is otherwise
suppressed.  In particular, even though pure gravity is
non-chaotic in eleven spacetime dimensions, the $3$-form of $D=11$
supergravity renders the system chaotic. Similarly, the bosonic
sectors of all $D=10$ supergravities related to string models
define chaotic dynamical systems, thanks again to the $p$-forms,
and in spite of the presence of a massless scalar dilaton.
It is remarkable and significant that the (maximally supersymmetric)
candidate models for a unified description of the fundamental forces
not only have difficulties accomodating de Sitter-type spacetimes
in any straightforward fashion \cite{nogo1,nogo2}, but furthermore,
and without exception, exhibit BKL chaos as one approaches the
initial singularity.

\subsection{Billiard description of BKL behaviour}

An efficient way to grasp the asymptotic behaviour of the fields as
one approaches a spacelike singularity is based on the qualitative
Hamiltonian methods initiated by Misner \cite{Misner1} in the context
of the Bianchi IX models (in four spacetime dimensions). The Hamiltonian
approach naturally leads to a billiard description of the asymptotic
evolution, in which the logarithms of the spatial scale factors
define (after projecting out the dynamics of the overall volume
factor) a geodesic motion in a region of the Lobachevskii plane
$H_2$, interrupted by geometric reflections against the walls
bounding this region \cite{Chitre,Misnerb}.  Chaos follows from
the fact that the Bianchi IX billiard has finite volume\footnote{Throughout
this paper, the word {\em billiard} used as a noun in the singular
will denote the dynamical system consisting of a ball moving
freely on a ``table" (region in some Riemannian space), with
elastic bounces against the edges.  {\em Billiard} will also
sometimes mean the table itself.}.

As pointed out in  \cite{Kirillov1993,KiMe,IvKiMe94,IvMe,DH3} this
useful billiard description is quite general and can be extended
to higher spacetime dimensions, with $p$-forms and dilaton. If $d
\equiv D-1$ is the number of spatial dimensions, and if there are
$n$ dilatons, the billiard is a region of hyperbolic space
$H_{d+n-1}$, each dilaton being equivalent, in the Hamiltonian, to
the logarithm of a new scale factor. Besides the dimension of the
hyperbolic billiard, the other ingredients that enter its
definition are the walls that bound it. These walls can be of
different types \cite{DH1,DH3}: symmetry walls related to the
off-diagonal components of the spatial metric, gravitational walls
related to the spatial curvature, and $p$-form walls (electric and
magnetic) arising from the $p$-form energy-density. All these
walls are  hyperplanar. The billiard is a convex polyhedron with
finitely many vertices, some of which are at infinity. In many
important cases, the billiard can be identified with the Weyl
chamber of a Kac-Moody algebra, and the reflections against the
billiard walls with the fundamental Weyl reflections
\cite{DH3,DHJN,DdBHS}. This suggests deep connections with
infinite symmetries.

The main purpose of this paper is to provide a self-contained
derivation of the billiard picture, in the general context of
inhomogeneous solutions in $D$ dimensions, with dilaton and
$p$-form gauge fields. In particular, we shall present a detailed
derivation of the general results, announced and used in
Ref. \cite{DH1,DH3}, on the form of the various possible walls.
For that purpose, we shall rely on the
Iwasawa decomposition (see e.g. \cite{Helgason}) of the spatial
metric. This provides an efficient derivation of the symmetry
walls in any number of spacetime dimensions, which we obtain by
working out explicitly the Hamiltonian that governs the dynamics
in the ``BKL limit'' or ``small volume limit''.

The present work is exclusively concerned with properties of
solutions of Einstein's equations and their generalizations
in the vicinity of a {\em spacelike singularity}. Accordingly,
our treatment does not apply to timelike singularities, for which
no analog of causal decoupling exists (the situation may be
more subtle for the borderline case of a null singularity). Furthermore,
we are not making the claim here that all spacelike singularities
are necessarily and uniformly of the BKL type. Rather, the results
presented here are intended to refine and to generalize the work
of \cite{BKL,BKL72} by showing that, in higher dimensions and
for many  theoretically relevant matter sources, one can
{\it self-consistently} describe the behaviour of all fields, in the
vicinity of a spacelike singularity, and at a generic spatial point  $x$, in
terms of $(i)$ a simple hyperbolic billiard description of the
``angular'' dynamics ($\g , \p_{\g}$) of the logarithmic scale factors
(after projection of the ``radial'' motion ($\rho , \p_{\rho}$)),
and $(ii)$ an asymptotic ``freezing'' of the other phase space
variables. This self-consistent asymptotic solution is general in
the sense that it involves as many arbitrary functions of space
as the most general solution.

\subsection{Organization of the paper}

After fixing our conventions and notations, and defining the class of
Lagrangians we shall consider in section 2, we discuss in section 3
the homogeneous and diagonal Kasner solution in $D$ spacetime
dimensions, with a dilaton (as usual, the term ``homogeneous'' implies
invariance under spatial translations). This solution plays a
crucial  r\^ole in the BKL approach because it describes the
``free motion between collisions'' and allows us to develop
an important tool of our approach: the (Minkowskian) geometry
of the scale factors. We then introduce (in section 4) the other
main tool of our investigation: the Iwasawa decomposition of the
spatial metric. To gain some familiarity with it, we discuss
the asymptotic behaviour of non-diagonal Kasner metrics.

In section \ref{general}, we explain the appearance of sharp
potential walls in full generality, without imposing any
homogeneity conditions on the metric and the matter fields. We
then discuss in great detail the various walls that appear in
physical models: symmetry walls, gravitational walls and $p$-form
walls. The resulting geometry of the ``billiard'' made from all
these walls is analyzed in section 7. Section \ref{KM0} is devoted
to the case when the billiard can be identified with the Weyl
chamber of a Kac-Moody algebra. To deal with this case we set up a
Kac-Moody theoretic formulation of the billiard in terms of a
non-linear $\sigma$-model based on the relevant Kac-Moody group in
the last section. As we will show there, the asymptotic limit of
these $\sigma$-models coincides with the asymptotic limit of the
models discussed in the main part of this paper. An appendix
illustrating by a toy model the asymptotic freezing of the
off-diagonal degrees of freedom
 concludes this paper.

We should stress that our analysis is purely classical and accordingly,
as it stands, is valid only up to the Planck (or string) scale. We shall also
ignore fermionic fields throughout. Nevertheless, it is reasonable
to expect that some of the ideas discussed here will remain relevant
in a more general quantum mechanical context, at least qualitatively
(see e.g. \cite{P} for some recent ideas in this direction).
The subject of Hamiltonian cosmology has a long history in the
context of four-dimensional, spatially homogeneous spacetimes and
provides useful insight on the general discussion presented here.
For reviews on this subject, with an extensive bibliography,
see \cite{Ryan,RyanSh,Jantzen}; see also the topical review
on multidimensional gravity \cite{IM}.

\section{Models and Conventions}

\subsection{The models}

We consider models of the general form
\beq
&&S[G_{MN}, \phi, A^{(p)}] = \int d^D x \, \sqrt{- G} \;
\Bigg[R - \partial_M \phi \partial^M \phi \nonumber \\
&& \hspace{1.5cm} - \frac{1}{2} \sum_p \frac{1}{(p+1)!}
e^{\l_p \phi} F^{(p)}_{M_1 \cdots M_{p+1}} F^{(p)  \, M_1
\cdots M_{p+1}} \Bigg] + \dots \label{keyaction}
\eeq
where units are chosen such that $16 \pi G_N = 1$ (where $G_N$
is Newton's constant) and the spacetime dimension $D \equiv d+1$
is left unspecified. Besides the standard
Einstein-Hilbert term the above Lagrangian contains a dilaton field
$\phi$ and a number of $p$-form fields $A^{(p)}_{M_1 \cdots M_p}$
(for $p\geq 0$). As a convenient common
formulation we adopt the Einstein conformal frame and normalize
the kinetic term of the dilaton $\phi$ with weight one w.r.t. to
the Ricci scalar. The Einstein metric $G_{MN}$ has Lorentz
signature $(- + \cdots +)$ and is used to lower or raise the
indices; its determinant is denoted by $G$. The $p$-form field
strengths $F^{(p)} = dA^{(p)}$ are normalized as
\be
F^{(p)}_{M_1 \cdots M_{p+1}} =
(p+1)  \partial_{[M_1} A^{(p)}_{M_2 \cdots M_{p+1}]} \equiv
\partial_{M_1} A^{(p)}_{M_2 \cdots M_{p+1}} \pm p \hbox{
permutations }.
\ee
The dots in the action (\ref{keyaction}) indicate possible
modifications of the field strength by additional Yang-Mills
or Chapline-Manton-type couplings \cite{pvnetal,CM}, such as
$F_C = dC^{(2)} - C^{(0)} dB^{(2)}$ for two $2$-forms $C^{(2)}$
and $B^{(2)}$ and a $0$-form $C^{(0)}$, as they occur in type
IIB supergravity. Further modifications include Chern-Simons
terms, as in the action for $D=11$ supergravity \cite{CJS}.
The real parameter $\l_p$ measures the strength of the coupling
of $A^{(p)}$ to the dilaton. When $p=0$, we assume that $\l_0\neq 0$
so that there is only one dilaton. This is done mostly for
notational convenience. If there were other dilatons among the
$0$-forms, these should be separated off from the $p$-forms
because they play a distinct r\^ole. They would define additional spacelike
directions in the space of the
(logarithmic) scale factors and  would correspondingly increase the dimension of the
relevant hyperbolic billiard.

The metric $G_{MN}$, the dilaton field(s)  $\phi$ and the $p$-form fields
$A^{(p)}_{M_1 \cdots M_p}$ are {\em a priori} arbitrary functions
of both space and time, on which {\it no symmetry conditions are imposed}.
Nevertheless  it will turn out that the evolution equations near the
singularity will be asymptotically the same as those of certain
homogeneous cosmological models. It is important to keep in mind that
this simplification does not follow from imposing extra dimensional reduction
conditions but emerges as a direct consequence of the general dynamics.

\subsection{Gauge conditions}

Our analysis applies both to past and future singularities, and in
particular to Schwarzschild-type singularities inside black holes.
To follow historical usage, we shall assume for definiteness that
the spacelike singularity lies in the past, at finite distance in
proper time. More specifically, we shall adopt a space-time
slicing such that the singularity ``occurs'' on a constant time
slice ($t= 0$ in proper time). The slicing is built by use of
pseudo-Gaussian coordinates defined by vanishing lapse $N^i = 0$,
with metric \be\label{metric1} ds^2 = - \big(N(x^0,x^i)
dx^0\big)^2 + g_{ij} (x^0,x^i) dx^i dx^j \ee In order to simplify
various formulas later, we shall find it useful to introduce a
rescaled lapse function \be\label{metric2} \tilde{N}  \equiv
N/\sqrt{g} \ee where $g \equiv {\rm det}\,g_{ij}$. We shall see
that a useful gauge, within the Hamiltonian approach, is that
defined by requiring

\be\label{tildeN=rho2}
\tilde{N}= \rho^2,
\ee
where $\rho^2$ is a quadratic combination of the logarithms of the
scale factors and the dilaton(s), which we will define below in
terms of the Iwasawa decomposition. After fixing the time zero
hypersurface the only coordinate freedom left in the pseudo-Gaussian
gauge (\ref{tildeN=rho2}) is that of making time-independent changes
of spatial coordinates $x^i \rightarrow x'^i = f^i(x^j)$. Since the
gauge condition Eq.(\ref{tildeN=rho2}) is not invariant under spatial
coordinate transformations, such changes of coordinates have the unusual
feature of also changing the slicing.

Throughout this paper, we will reserve the label $t$ for the {\em proper time}
\be\label{propertime}
dt = - N dx^0 =  - \tilde{N}\sqrt{g} dx^0,
\ee
whereas the time coordinate associated with the special gauge
Eq.(\ref{tildeN=rho2}) will be designated by $T$, viz.
\be\label{timeT}
dT = - \frac{dt}{\rho^2 \sqrt{g}}.
\ee
Sometimes, it will also be useful to introduce the ``intermediate''
time coordinate $\tau$ that would correspond to the gauge condition
$\tilde{N} =1$.  It is explicitly defined by:
\be\label{timetau}
d\tau = - \frac{dt}{\sqrt{g}} = \rho^2  dT
\ee
At the singularity the proper time $t$ is assumed to remain finite
and to decrease toward $0^+$. By contrast, the coordinates $T$ and
$\tau$ both increase toward $+\infty$, as ensured by the minus
sign in (\ref{propertime}). Irrespective of the choice of coordinates,
the spatial volume density $g$ is assumed to collapse to zero at each spatial
point in this limit.

As for the $p$-form fields, we shall assume, throughout this paper,
a generalized temporal gauge, viz.
\be
\label{temporal}
A^{(p)}_{0 i_2 \cdots i_p} = 0
\ee
where small Latin letter $i,j,...$ denote spatial indices from now on.
This choice leaves the freedom of performing time-independent gauge
transformations, and therefore fixes the gauge only partially.

\section{Geometry of the space of the scale factors}
\label{geomspace}

\subsection{Supermetric and Hamiltonian}

To set the stage for our general Hamiltonian approach, it is useful to
study first in detail the dynamics defined by considering {\it only
the kinetic terms of the metric and of the dilaton(s)}.
The corresponding reduced action is obtained from (\ref{keyaction}) by
setting $A^{(p)} = 0$ and assuming that all the other fields depend only on
time. In terms of a general time coordinate $x^0$ this reduced action reads
\begin{equation}
S[g_{ij}, \phi, \tilde{N}] = \int dx^0 {\tilde{N}}^{-1} \left[
\frac{1}{4} \left(\tr (\rg^{-1} \dot{\rg})^2 - (\tr \rg^{-1}
\dot{\rg})^2\right) + \dot{\phi}^2 \right].
\label{KasnerAction}
\end{equation}
where we have suppressed an integral $\int d^dx$ over the spatial volume
for notational simplicity. Furthermore, we make use of the notations
introduced in (\ref{metric1}) and (\ref{metric2}) with
$\dot{F} \equiv dF/dx^0$, and adopt a matrix notation where
$\rg(t)\in GL(d,\Rn)$ stands for the matrix $(g_{ij})$ representing
the spatial components of the metric at each spatial point.

The action (\ref{KasnerAction}) is the (quadratic-in-velocities)
action for a massless free particle with coordinates $(g_{ij},
\phi)$ moving in a curved target space with metric
\begin{equation}
d\s^2 = \ft14 \left[ \tr (\rg^{-1} d\rg)^2 - (\tr \rg^{-1} d\rg)^2
\right]  + d\phi^2
\label{DeWitt}
\end{equation}
We designate by $d\s^2$ the line element in this target ``superspace'' to
distinguish it from the line element in physical space time, which we denote
by $ds^2$. The first two terms in the r.h.s. of (\ref{DeWitt}) define
the so-called DeWitt supermetric in the space of the metric coefficients
$g_{ij}$ \cite{DeWitt0}.  If several dilatons $\phi^i$ (for $i= 1,\ldots,n$)
were present the term $d\phi^2$ in Eq.(\ref{DeWitt}) would be replaced by
$ \Sigma_i (d\phi^i)^2$. That is, each dilaton
adds a (flat) direction in the target superspace.

The rescaled lapse  $\tilde{N}$ plays the role of an ``einbein'' in
the geodesic action  (\ref{KasnerAction}). As usual, extremizing over
$\tilde{N}$ yields the ``zero-mass constraint''
\begin{equation}\label{EoM1}
\ft14 \left(\tr (\rg^{-1} \dot{\rg})^2 - (\tr \rg^{-1}
\dot{\rg})^2\right) + \dot{\phi}^2 = 0.
\end{equation}
Thus, the motion is given by a null geodesic of the metric (\ref{DeWitt}).
An affine parameter along those geodesics is
$d\tau = + \tilde{N} d x^0= - dt/ \sqrt{g}$, cf. Eq.(\ref{timetau})
above. In terms of the parameter $\tau$ the equations of motion read:
\be\label{EoM2}
\frac{d}{d\tau} \left( \rg^{-1} \frac{d \rg}{d\tau} \right) = 0
\quad , \quad \frac{d^2}{d\tau^2} \phi = 0.
\ee
For {\it diagonal metrics }
\be\label{diagonalKasner}
\rg_K = \exp \big[{\rm diag}(-2\beta)\big] \quad \Longleftrightarrow
\quad  g^K_{ij} = \exp(-2 \b^i) \d_{ij}
\ee
the supermetric (\ref{DeWitt}) reduces to
\beq
d\s^2 &=& \tr d \beta^2 - (\tr d\beta)^2 + d \phi^2 = \nonumber\\
&=&\sum_{i=1}^d  (d \b^i)^2 -  (\sum_{i=1}^d  d\b^i)^2 + d\phi^2
\equiv G _{\m \n} d\b^{\m} d\b^{\n} . \label{DeWittbis} \eeq Here
we have introduced a $(d+1)$-dimensional space with coordinates
$\b^\m$, with indices running over $\m = 1, \cdots, d+1$, such
that the first $d$ coordinates $\b^i$ correspond to the logarithms
of the scale factors of the spatial metric [cf. Eq.
(\ref{diagonalKasner})], and the $(d+1)$-th coordinate
$\beta^{d+1} \equiv \phi$  represents the
dilaton\footnote{Obviously, the range of indices would be extended
to $\mu , \nu = 1, \cdots, d+n$ in the presence of $n$ dilatons.}.
The explicit form of the (flat) target space metric $G_{\mu\nu}$
can be read off directly from (\ref{DeWittbis}). The action for
diagonal metrics is
\begin{equation}
S[\beta^\mu, \tilde{N}] = \int dx^0 \, {\tilde{N}}^{-1} G_{\mu
\nu} \dot{\beta}^\mu \dot{\beta}^\nu
\label{betaaction}
\end{equation}
In the sequel, we shall refer to this space as the ``extended space
of (logarithmic) scale factors'' or just ``the $\b$-space'' for short.

Combining the scale factors and the dilaton(s) in a single space is
natural because we know from Kaluza-Klein theory that the dilaton can be
viewed as the logarithm of a scale factor in one extra spatial dimension.
Independently of whether the original metric (\ref{metric1}) has non-vanishing
curvature or not, the metric (\ref{DeWittbis}) induced in the space of
the scale factors (possibly including the dilaton) is flat. More precisely,
the metric $G_{\mu \nu} $ of the (extended) space of scale factors is a
Minkowski metric in ${\RR}^{d+1}$ with signature $(- + + \cdots+)$.
For instance, the direction in which only the dilaton varies
[i.e. $ d\b^{\m} \propto (0,\cdots,0,1)]$ is spacelike, while the
direction in which only one scale factor varies
[e.g., $d\b^{\m} \propto (1,0,\cdots,0)$] is null. A timelike
direction in this space is the direction $d\b^{\m} \propto (1,1,\cdots,1,0)$.
This reflects the familiar fact that the gravitational
action is not bounded from below (even with Euclidean signature):
conformal transformations of the metric, in which the scale factors are
all scaled in the same fashion, make $d\s^2$ negative.  It is this
characteristic feature of gravity which is responsible for the
Lorentzian nature of the Kac-Moody algebras which emerge in the
analysis of the billiard symmetries \cite{DH3}. The Lorentzian
signature of the metric in the space of the scale factors enables
one to define the light cone through any point. We define the
time-orientation to be such that future-pointing vectors $v^\mu$
have $\sum_i v^i>0$. Geometrically, small volumes (small $g$) are
associated with large positive values of $\sum_i \beta^i$. Large
volumes (large $g$), on the other hand, mean large negative values
of $\sum_i \beta^i$. We are interested in the small volume limit,
i.e. $\sum_i \beta^i \to + \infty$.

The Hamiltonian form of the action for the diagonal metric degrees
of freedom and the dilaton is
\begin{equation}
S[\beta^\mu,\pi_\mu, \tilde{N}] = \int dx^0\left[ \pi_\mu
\dot{\beta}^\mu - \ft14 {\tilde{N}} G^{\mu \nu} \pi_\mu \pi_\nu
\right]
\end{equation}
where $G^{\mu \nu}$ is the inverse of  $G_{\mu \nu}$. Explicitly
\begin{equation}\label{Gmunuup}
G^{\mu \nu} \pi_\mu \pi_\nu \equiv
\sum_{i=1}^d \pi_i^2 - \frac{1}{d-1} \left(\sum_{i=1}^d
\pi_i\right)^2 + \pi_\phi^2
\end{equation}
where $ \pi_\mu \equiv (\pi_i, \pi_\phi)$ are the
momenta conjugate to $\beta^i$ and $\phi$, respectively, i.e.
\be
\pi_\m = 2 \tilde{N}^{-1} G_{\m \n} \dot{\beta}^\n =
2 G_{\m \n} \frac { d {\beta}^\n}{d\tau} \equiv  2 G_{\m \n} v^\n
\ee
Here, the $\tau$-parameter velocities have been designated by
$v^\mu\equiv d {\beta}^\m/d\tau$.

\subsection{Diagonal Kasner solution}

The Kasner solution (with or without dilaton) is now easily obtained
by solving (\ref{EoM1}) and (\ref{EoM2}) in the diagonal case.  Indeed,
the equations of motion reduce to
\be
\frac{d^2 \b^{\m}}{ d \tau^2} =0.
\ee
They are solved by
\be
\beta^\mu = v^\mu \tau + \beta^\mu_0
\label{freemotion}
\ee
where $v^\m$ and $\beta^\mu_0$ are constants of the motion.
The ``zero-mass constraint'' becomes
\be
 G_{\mu \nu} v^\mu v^\nu = 0.
\label{constraint}
\ee
One can transform the simple affine parameter solution  (\ref{freemotion})
into the usual Kasner solution expressed in terms of the proper time by
integrating the relation $d t = - \sqrt{g} d\tau$, with
$\sqrt{g} = \exp (- \Sigma_i \b^i), $
whence $t \propto \exp\big(- (\Sigma_i v^i ) \tau \big)$, or
\be
\tau  = - \frac{1}{\sum_i  v^i} \ln t + const. \label{taut}
\ee
We need to require $\Sigma_i v^i > 0$ to remain consistent with our
convention that $\tau \to + \infty$ at $ t \to 0^+$ near the
singularity. This yields
\begin{eqnarray}
ds^2 &=& -dt^2 + \sum_{i=1}^d A_i^2(t) (dx^i)^2 \; , \quad
A_i(t) = b_i t^{p_i}  \label{Kasnersol} \\
\phi &=& - p_\phi \ln t + C_\phi
\label{dilatonsol}
\end{eqnarray}
where  $b_i \equiv \exp(-\b^i_0)$ and $C_\phi \equiv \b_0^{d+1}$
are integration constants and the minus sign in front of $p_\phi$
in (\ref{dilatonsol}) is included for the sake of uniformity in
the formulas below (if there is no dilaton one simply sets $p_\phi
= C_\phi = 0$). By rescaling the spatial coordinates, one can set $b_i
=1$ and obtain the standard (proper time) form of the Kasner
metric. The Kasner exponents $p_{\m} = (p_i,p_\phi)$ are given in
terms of the affine velocities $v^{\m} \equiv d {\beta}^\m/d\tau$
by \be p_{\m} =\frac{v^\mu}{\sum_i v^i}. \ee Note that the sum in
the denominator does not include the dilaton.

They are subject to the quadratic constraint\footnote{Contrary to
the variables $\b^\mu$ and the velocities $v^\mu$, we do not assign any
covariance properties to the standard Kasner exponents, but regard them
simply as parameters, leaving their labels always in the lower position.}
\begin{equation}
\sum_{i=1}^d p_i^2 - \left(\sum_{i=1}^d p_i\right)^2 + p_\phi^2 = 0.
\end{equation}
coming from the ``zero-mass condition'', and to the linear constraint
\begin{equation}\label{sumpi}
\sum_{i=1}^d p_i = 1
\end{equation}
coming from their definition above.

If there are no dilatons, it follows from the above equations
that there is at least one Kasner exponent which is negative, so at
least one of the scale factors $A_i(t)$ blows up as $t \rightarrow 0$.
The scale factors associated with positive Kasner exponents
contract to zero monotonically. By contrast, in the presence of a
dilaton, all the Kasner exponents can be positive simultaneously.
In both cases there is an overall contraction of the spatial volume since
the determinant $g$ of the spatial metric tends to zero.  Indeed, a consequence
of the linear constraint above on the Kasner exponents is
\begin{equation}
g  \propto t^2 .
\label{gt}
\end{equation}
Note that the relations (\ref{taut}) and (\ref{gt}) have been derived
only for the exact (homogeneous) Kasner solution in the vacuum.

\subsection{Hyperbolic space}

Still as a preparation for dealing with the general inhomogeneous case,
let us present an alternative way of solving the dynamics defined by
the kinetic terms of diagonal metrics, i.e. the action (\ref{betaaction}),
or its Hamiltonian form. This alternative way  will turn out to
be very useful for describing the asymptotic dynamics of general
inhomogeneous metrics. It consists in decomposing the motion of the
variables $\b^{\m}$ into two pieces, namely a radial part
$\rho$, and an angular one $\g^{\m}$. Here ``radial'' and
``angular'' refer to polar coordinates in the Minkowski space
of the $\b^{\m}$. More precisely, shifting, if necessary, the origin
in $\b$-space to arrange that $v_\mu \beta^\mu_0 <0$, the
$\b^{\m}$ trajectories (\ref{freemotion}) will, for large enough
values of $\tau$, get inside the future light
cone of the origin, i.e.
\be\label{rhotau1}
\beta^\mu \beta_\mu = 2 v_\mu \beta^\mu_0 \tau+
\beta_{0 \; \mu} \beta^\mu_0 < 0.
\ee
Let us then decompose $\b^{\m}$ in hyperbolic polar coordinates
$(\rho, \g^{\m})$, i.e
\begin{equation}
\label{hyperbolic}
\beta^\mu = \rho \gamma^\mu,
\end{equation}
where $\gamma^\mu$ are coordinates on the future sheet of the unit
hyperboloid, which are constrained by
\begin{equation}
\gamma^\mu \gamma_\mu = -1
\end{equation}
and $\rho$ is the timelike variable
\be\label{rhosquare}
\rho^2 \equiv - \b^\mu \b_\mu > 0
\ee
This decomposition naturally introduces the unit hyperboloid (``$\g$-space''),
which is a realization of the $m$-dimensional hyperbolic (Lobachevskii)
space $H_m$, with $m=d-1+n\;$ if there are $n\geq0$ dilatons.

In terms of the ``polar'' coordinates
$\rho$ and $\gamma^\mu$, the metric in $\b$-space becomes
\begin{equation}
d\s^2 = - d \rho^2 + \rho^2 d\S^2
\end{equation}
where $d\S^2$ is the metric on the $\g$-space $H_m$. In these
variables the Hamiltonian reads
\begin{equation}
H _0= \frac{\tilde{N}}{4} \left[- \pi_\rho^2 + \frac{1}{\rho^2}
\pi_{\gamma}^2\right]
\end{equation}
where $\pi_{\gamma}$ are the momenta conjugate to the constrained
hyperbolic coordinates $\gamma^\mu$. (It is straightforward to introduce
angular coordinates on $H_m$ to derive more explicit formulas for
$d\S^2$ and $\pi_{\gamma}^2$.) The extra index $0$ on $H$ is to underline
the fact that this Hamiltonian refers only to a small part of the total
Hamiltonian which we shall study below, because it describes only
the kinetic terms associated to the $(d+n)$ variables $\b^{\m}$.

An equivalent expression is
\begin{equation}
H_0 = \frac{\tilde{N}}{4\rho^2} \left[- \pi_\lambda^2 + \pi_{\bf
\gamma}^2\right]
\label{Hamiltonianrho}
\end{equation}
where we have introduced  the new configuration variable
\be
\label{lambda}
\lambda  \equiv \ln \rho \equiv \ft12 \ln \big( -G_{\m \n} \b^\m \b^\n\big)
\ee
with conjugate momentum $\pi_\lambda$.

The form (\ref {Hamiltonianrho}) of the Hamiltonian shows that, when
using a radial projection on $H_m$, the motion becomes simplest in the gauge
\be\label{gaugerho}
\tilde{N} = \rho^2 ,
\ee
in terms of which (\ref{Hamiltonianrho}) reduces to a free Hamiltonian
on the pseudo-Riemann\-ian space with metric $- d\l^2 + d\S^2$.
In the gauge (\ref{gaugerho}) the logarithmic radial momentum
 $\pi_\lambda = \rho \pi_\rho$ is a constant of the motion.
In this gauge, we see that the free motion of the $\b$'s is
projected onto a geodesic  motion on $H_m$.

The coordinate time $T$ associated to this gauge, see Eq.(\ref{timeT}),
is linked to the affine parameter $\tau$ via
\be
dT = \frac{d \tau}{\rho^2}.
\ee
{}From \Ref{rhotau1} we get that $\rho^2$ varies linearly
with $\tau$:
\be\label{rhotau}
\rho^2 = - \b_\m \b^\m = - 2 v_\mu \b^\mu_0 \tau - \b^\mu_0 \b_{0\mu}
\ee
which implies
\be
T = - \frac{1}{2 v_\mu \b^\mu_0 } \ln \tau + const.
\ee
Recalling that $\tau$ varies logarithmically with the proper time $t$,
we see that $ T \propto \ln \vert \ln t \vert$.

One can also use the configuration variable  (\ref{lambda}) as an
intrinsic time variable to describe the dynamics. In view of
(\ref{rhotau}) we have \be \lambda=\ft12 \ln \tau  + const. =
\ft12 \ln \vert \ln t \vert + const. \ee Note that the various
links above between the different time scales are derived only for
an exact Kasner solution. Similar relations will hold asymptotically
in the general inhomogeneous case (see discussion in chapter 7).

\section{Iwasawa decomposition and dynamics of \\non-diagonal Kasner metrics}
\setcounter{equation}{0} \setcounter{theorem}{0}
\setcounter{lemma}{0} \label{nondiagonal}

\subsection{Iwasawa decomposition} \label{ndiagIwa}

For homogeneous solutions {\it in vacuo}, the metric remains diagonal
if the initial data are so. This is in general not true when matter
(such as $p$-forms) or inhomogeneities are included, in which case
off-diagonal components generically appear even if there are none
initially. For this reason, it is important to understand the r\^ole of
off-diagonal terms already in this simpler homogeneous context, by
examining the evolution of initial data that are not diagonal.
And, as for the simple diagonal metrics above, it is instructive
to study the dynamics of non-diagonal metrics in several complementary ways.

A first way of dealing with the evolution of initially non-diagonal
metrics is to perform a suitable linear transformation on the
diagonal Kasner solution. If $L$ is the linear transformation
needed to diagonalize the initial data, it is easy to see from (\ref{EoM2})
that the solution is given, in terms of the parameter $\tau$, by
\begin{equation}
\rg(\tau) =  L^T \, \rg_K (\tau) \, L
\label{offdiagonal}
\end{equation}
where $^T$ denotes transposition and $\rg_K(\tau)$ is the diagonal
Kasner solution \Ref{diagonalKasner}, \Ref{freemotion}. The dilaton,
being a scalar, is still given by the same expression as before, i.e.
$\phi = \b^{d+1} = v^ {d+1} \tau + \b_0^{d+1}$. Note the relation
\be
\det \rg(\tau) = (\det L)^2 \det \rg_K(\tau),
\ee
so that the relation \Ref{taut} between $\tau$ and $t$ still holds.
Therefore $\det \rg$ still goes to zero like $t^2$.

A second way of describing the evolution of non-diagonal
metrics is to perform an Iwasawa decomposition of the metric,
\begin{equation}
\rg = \cn^T \, \ca^2 \, \cn
\label{Iwasawa}
\end{equation}
where $\cn$ is an upper triangular matrix with $1$'s on the diagonal and
$\ca$ is a diagonal matrix with positive entries, which we parametrize as
\begin{equation}
\ca = \exp( -\beta ) \; ,\quad \beta = {\rm diag}\, (\beta^1, \beta^2, \cdots,
\beta^d).
\end{equation}
The explicit form of \Ref{Iwasawa}  reads
\be
\label{Iwasawaex}
g_{ij} = \sum_{a=1}^d e^{- 2 \b^a}  {\cn^a}_i  \, {\cn^a}_j
\ee
The evolution of the new configuration variables
$\b^a = f( g_{ij})$ defined by the Iwasawa decomposition
differs in general from that  of the quantities $\b^i$ entering
the diagonal Kasner solution (\ref{freemotion}), i.e.
$\b^a \ne \b^i$ except for the special case ${\cn^a}_i = \d^a_i$.
Henceforth in this paper, the notation $\b$ will always refer
to the logarithmic scale factors $\b^a $ w.r.t. to the Iwasawa frame
\Ref{Iwasawaex} (to be augmented, if needed, by the dilaton as a $(d+1)$-th
coordinate $\b^{d+1} \equiv \phi$). Furthermore, we adopt the convention
here and in the remainder that summation over the spatial coordinate
indices $i,j,\dots$ is always understood, whereas sums over the
Iwasawa frame indices $a,b,\dots$ will always be  written out.

One can view the Iwasawa decomposition\footnote{The Iwasawa
decomposition applies to general symmetric spaces (see e.g.
\cite{Helgason}). In our case the relevant symmetric space is the
coset space $SL(d, \Rn)/SO(d)$ since the space of positive
definite symmetric matrices can be identified with
$GL(d,\Rn)/O(d)$, which is isomorphic to $SL(d,\Rn)/SO(d) \times
\Rn^+$.} as the Gram-Schmidt orthogonalisation of the initial
coordinate coframe  $dx^i$, which is indeed a triangular process,
\begin{equation}\label{Iwasawa0}
g_{ij} dx^i dx^j = \sum_{a=1}^d e^{ -2\beta^a} \theta^a \!\otimes\! \theta^a
\end{equation}
Starting with $\theta^d = dx^d$, one successively constructs the next
$\theta$'s by adding linear combinations of the $dx^k$ (for $k>j$)
in such a way that $\theta^{d-1}$ is orthogonal to $\theta^d$,
$\theta^{d-2}$ is orthogonal to both $\theta^d$ and $\theta^{d-1}$,
etc.  Explicitly,
\begin{eqnarray}
\theta^d &=& dx^d \nn\\
\theta^{d-1} &=& dx^{d-1} + {\cn^{d-1}}_d \, dx^d , \nn\\
\theta^{d-2} &=& dx^{d-2} + {\cn^{d-2}}_{d-1} \, dx^{d-1} +
{\cn^{d-2}}_d \, dx^d, \nn\\
&\cdots&
\label{changeofframe}
\end{eqnarray}
Equivalently,
\be\label{Iwasawa1}
\theta^a = {\cn^a}_i \, dx^i.
\ee
Thus ${\cn^a}_i$ vanishes  for $a>i$, is equal to one for
$a=i$, and has a non-trivial coordinate dependence only
for $a<i$. It is easily seen that the determinant of the matrix $\cn$
is equal to $1$. Therefore the sum of the $\b$'s is directly
related to the metric determinant (as in the diagonal case):
$g = \det \rg  = \det \ca^2 = \exp \big(- 2 \Sigma_a \b^a\big)$.

In the following we shall also need the vectorial frame $\{ e_a \}$
dual to the  coframe ${\theta^a}$:
\be\label{Iwasawa2}
e_a = {\cn^i}_a \frac{\partial}{\partial x^i}.
\ee
where the matrix ${\cn^i}_a$ is the inverse of  ${\cn^a}_i$, i.e.,
${\cn^a}_i {\cn^i}_b = \delta^a_b$.  It is again an upper
triangular matrix with 1's on the diagonal.

\subsection{Asymptotics of non-diagonal Kasner metrics} \label{3dKasner}

Later in this paper we will deal with the generic inhomogeneous, non-diagonal
metric $g_{ij}(x^0,x^i)$ and show how to study its Hamiltonian dynamics
directly in terms of the Iwasawa variables $(\b^a, {\cn^a}_i)$ and of their
conjugate momenta. But let us first take advantage of the known explicit
solution  \Ref{offdiagonal} of the nondiagonal evolution in order to
understand the qualitative behaviour of the Iwasawa variables
$(\b^a, {\cn^a}_i)$ as $ t \to 0$, i.e. $\tau \to  +\infty$.
To be concrete let us  explicitly consider the case of three spatial
dimensions ($d=3$, $D=4$) using the results of \cite{HPT} for the Iwasawa
decomposition of 3-dimensional metrics.

Setting
\begin{equation}
\cn = \left( \matrix{1 &n_1 & n_2 \cr 0 & 1 & n_3 \cr 0 & 0 & 1  \cr
} \right)
\end{equation}
together with
\begin{equation}
\ca = \left( \matrix{\exp(-\beta^1) &0 & 0 \cr 0 & \exp(-\beta^2) &
0 \cr 0 & 0 & \exp(-\beta^3)  \cr } \right),
\end{equation}
one finds
\begin{eqnarray}
g_{11} &=& e^{-2 \beta^1}, \; \; \; g_{12} = n_1 e^{-2 \beta^1},
\; \; \;
g_{13} = n_2 e^{-2 \beta^1}, \non
g_{22} &=& n_1^2 e^{-2 \beta^1} + e^{-2 \beta^2}, \; \; \;
g_{23} = n_1 n_2 e^{-2 \beta^1} + n_3 e^{-2 \beta^2}, \non
g_{33}  &=& n_2^2 e^{-2 \beta^1} + n_3^2 e^{-2 \beta^2} +
e^{-2 \beta^3}
\end{eqnarray}
from which one gets
\begin{eqnarray}
\beta^1 &=& - \frac {1}{2} \ln g_{11}, \; \;  \; \beta^2 = - \frac
{1}{2} \ln \left[ \frac{g_{11} g_{22} -g_{12}^2} {g_{11}} \right] ,  \non
\beta^3 &=& - \frac {1}{2} \ln \left[ \frac {g} {g_{11} g_{22}
-g_{12}^2}\right], \; \;  \; n_1 = \frac {g_{12}}{g_{11}}, \non
n_2 &=& \frac {g_{13}}{g_{11}}, \; \;  \; n_3 = \frac {g_{23}
g_{11} -g_{12} g_{13}}{g_{11} g_{22} -g_{12}^2}.
\end{eqnarray}
On the other hand, (\ref{offdiagonal}) and (\ref{Kasnersol}) yield
\begin{equation}
g_{ij} (t) = t^{2p_1} l_i l_j + t^{2p_2} m_i m_j + t^{2p_3} r_i r_j
\label{A9}
\end{equation}
with the constant matrix
\begin{equation}
L = \left( \matrix{l_1 &l_2 & l_3 \cr m_1 & m_2 & m_3 \cr r_1 &
r_2 & r_3  \cr } \right),
\end{equation}
where we have absorbed the $b_i$'s in $l_i$, $m_i$ and $r_i$.

Combining these relations, we can deduce the explicit time dependence
of the Iwasawa variables
\begin{eqnarray}
\beta^1 (t)&=& - \frac {1}{2} \ln X , \; \;
\beta^2 (t) = - \frac {1}{2} \ln \left[ \frac{Y}{X} \right], \label{A11}\non
\beta^3 (t) &=& - \frac {1}{2} \ln \left[ \frac{t^{2(p^1 + p^2 + p^3)}
(\det L)^2}{Y} \right], \non
n_1 (t) &=& \frac{t^{2p_1} l_1 l_2 + t^{2p_2} m_1 m_2
+ t^{2p_3} r_1 r_2}{X}, \non
n_2 (t) &=& \frac{t^{2p_1} l_1 l_3 + t^{2p_2} m_1 m_3 + t^{2p_3} r_1
r_3}{X}, \; \; n_3 (t) = \frac {Z}{Y}
\end{eqnarray}
with
\begin{eqnarray}
X (t) &=& t^{2p_1} l_1^2 + t^{2p_2} m_1^2  + t^{2p_3} r_1^2, \non
Y (t) &=& t^{2p_1 + 2 p_2} (l_1 m_2 -l_2 m_1)^2 +
t^{2p_1 + 2 p_3} (l_1 r_2 -l_2 r_1)^2 \nonumber \\
&& \hspace{2cm} + t^{2p_2 + 2 p_3} (m_1 r_2 -m_2 r_1)^2, \non
Z (t) &=& t^{2p_1+ 2 p_2 } (l_1 m_2 - l_2 m_1)(l_1 m_3 - l_3 m_1)
\nonumber \\
&&  + t^{2p_1 + 2p_3} (l_1 r_2  - l_2 r_1)(l_1 r_3 - l_3 r_1)
\nonumber \\
&&  + t^{2 p_2 + 2p_3} (m_1 r_2 - m_2 r_1)(m_1 r_3 - m_3 r_1).
\label{A17}
\end{eqnarray}

These explicit formulas show that the evolution of the Iwasawa
variables $(\b, \cn)$ in the generic non-diagonal case is rather
complicated.  However, what will be important in the following is
that they drastically simplify in the asymptotic limit  $t \to 0$,
i.e. $\tau \to + \infty$. Without loss of generality, one can
assume $p_1 \leq p_2 \leq p_3$. If necessary, this can be achieved
by multiplying $L$ by an appropriate permutation matrix.  We shall
in fact only consider $p_1<p_2 < p_3$, leaving the discussion of
the limiting cases to the reader. For generic $L$, i.e.
$l_1\not=0$, $r_1 \not= 0$, $l_1 m_2 - l_2 m_1 \not=0$, and $m_1
r_2 -m_2 r_1 \not= 0$, one then finds the following asymptotic
behaviour as $t \to 0$, i.e. $\tau \to + \infty$ :

\begin{eqnarray}
\tau\rightarrow +\infty &:& \beta^1 \sim v^1
\tau, \; \beta^2 \sim v_2 \tau, \; \beta^3 \sim v_3 \tau, \nonumber \\
&&n_1 \rightarrow \frac{l_2}{l_1}, \; n_2 \rightarrow
\frac{l_3}{l_1}, \; n_3 \rightarrow \frac{l_1 m_3 - l_3 m_1}{l_1
m_2 - l_2 m_1}. \label{future}
\end{eqnarray}
Here we used the links derived above between $(p_\m, t)$
and $(v^\m, \tau)$.

The remarkable feature of these results is that, in the limit
$\tau \to + \infty$, the evolution of the Iwasawa variables
$(\b, \cn)$ is essentially as simple as the evolution we obtained
in the diagonal case. Namely, the diagonal degrees of freedom $\b$
in the Iwasawa decomposition behave linearly with $\tau$,
while the elements of the upper triangular matrix $\cn$
tend to constants. That the $n_i$'s should asymptotically tend to
constants should be clear because they are homogeneous functions of degree
zero in the metric coefficients --- in fact, ratios of polynomials of degree
one or two in the $t^{2p_i}$. It is more subtle that the scale factors
$\exp(-2 \beta^i)$, which are homogeneous of degree one in the $g_{ij}$'s,
are not all driven by the fastest growing (or least decreasing) term
($t^{2p_1}$ for $t \rightarrow 0^+$). This is what happens for the
first scale factor $\exp(-2 \beta^1)$. However, the second scale factor
$\exp(-2 \beta^2)$ feels the subleading term $t^{2p_2}$ because the
leading term drops from its numerator, equal to the minor
$g_{11}g_{22} - (g_{12})^2$. Similar cancellations occur for the
last scale factor $\exp(-2 \beta^3)$, which  feels only the
smallest term  $t^{2p_3}$ as $t \rightarrow 0^+$.

The results \Ref{future} admit a simple generalization to $d$
dimensions. By repeating the above explicit calculation, one can
prove that, in any dimension $d$, the $\beta$'s become
asymptotically linear functions of $\tau$, as in the diagonal
case, with coefficients that are given by a permutation of the
underlying diagonal Kasner exponents. Furthermore, the ${\cn^a}_i$
tend to constants, a phenomenon which we will refer to as the
``asymptotic freezing'' of the off-diagonal metric variables. More
precisely, let $v^1 \leq v^2 \leq \cdots \leq v^d$ be the ordered
(unnormalized) underlying Kasner exponents; then, for generic $L$,
we have \be \beta^a \sim v^a \, \tau \quad {\rm \, and} \quad
{\cn^a}_i \rightarrow \hbox{const}. \ee as $\tau \rightarrow
+\infty$ or, equivalently, $t \rightarrow 0^+$. An alternative
derivation of these results will follow, as a particular case,
from the general result we shall derive below concerning the
qualitative evolution of generic, inhomogeneous, matter-driven
solutions in Iwasawa variables (after projecting out the radial
motion of the Iwasawa $\b$ variables).

Note also that, in the simple case of homogeneous nondiagonal metrics
{\em in vacuo} (without curvature and/or matter) one can discuss not only
the  limit  $ t \to 0$, but also the limit $ t \to + \infty$
(i.e $\tau \to - \infty$).
In this second limit, one finds again that the $\beta$'s become
asymptotically linear functions of $\tau$,
with coefficients that are given by a permutation of the underlying
Kasner exponents, and that the ${\cn^a}_i$ tend to constants.
More precisely, for any $d$, $\beta^a \sim v^{(d-a)} \, \tau$ ,
while the explicit results in $d=3$ read
\begin{eqnarray}
\tau \rightarrow - \infty &:&
\beta^1 \sim p_3 \tau\; , \;\;\; \beta^2 \sim p_2 \tau \; ,
\;\;\; \beta^3 \sim p_1 \tau, \nonumber \\
&& n_1 \rightarrow
\frac{r_2}{r_1}\; , \;\;\; n_2 \rightarrow \frac{r_3}{r_1}\; , \;\;\;
n_3 \rightarrow \frac{m_1 r_3 - m_3 r_1}{m_1 r_2 - m_2 r_1}
\label{past}
\end{eqnarray}
Note that in {\it both } limits, one has
$\beta^1 \leq \beta^2 \leq \cdots \leq \beta^d$.
The fact that the $\beta$'s are ordered in this way is due to
our choice of an {\it upper triangular} $\cn$; had we taken $\cn$ to be
{\it lower} triangular instead, these inequalities would have been reversed.

\section{Asymptotic dynamics in the general case}
\label{general}
\setcounter{equation}{0}
\setcounter{theorem}{0}
\setcounter{lemma}{0}

Having warmed up with the simple diagonal and homogeneous Kasner
solution in Section 3, and having introduced the Iwasawa
decomposition of the metric (Section 4), we are now ready to apply
these techniques to the description of the asymptotic dynamics in
the limit $t \to 0$ to the general inhomogeneous, matter-driven
case. The main ingredients (already introduced above for the
homogeneous solutions) of our study are:
\begin{itemize}
\item  use of the Hamiltonian formalism, \item Iwasawa
decomposition of the metric, i.e. $ g_{ij}\to (\b^a, {\cn^a}_i)$,
\item decomposition of $\b^\m = (\b^a, \phi)$ into radial ($\rho$) and angular
($\g^\m$) parts, and \item use of the pseudo-Gaussian gauge
\Ref{tildeN=rho2}, i.e. of the time coordinate $T$ as the
evolution parameter.
\end{itemize}

More explicitly, with the conventions already described before, we
assume that in some spacetime patch, the metric is given by
(\ref{metric1}) (pseudo-Gaussian gauge), such that the local
volume $g$ collapses at each spatial point as $x^0 \rightarrow
+\infty$, in such a way that the proper time $t$ tends to $0^+$.
We work in the Hamiltonian formalism, i.e. with first order
evolution equations in the phase-space of the system. For
instance, the gravitational degrees of freedom are initially
described by the metric $g_{ij}$ and its conjugate momentum
$\pi^{ij}$. We systematically use the Iwasawa decomposition
\Ref{Iwasawaex} of the metric to replace the $d(d+1)/2$ variables
$g_{ij}$ by the $ d + d(d-1)/2$ variables $(\b^a, {\cn^a}_i)$.
Note that $(\b^a, {\cn^a}_i)$ are {\it ultralocal} functions of
$g_{ij}$, that is they depend, at each spacetime point, only on
the value of $g_{ij}$ at that point, not on its derivatives. This
would not have been the case if we had used a ``Kasner frame''
(as defined below) instead
of an Iwasawa one. The transformation $g \to (\b, \cn)$ then
defines a corresponding transformation of the conjugate momenta,
as we will explain below. We then augment the definition of the
$\b$'s by adding the dilaton field, i.e. $ \b^\m  \equiv (\b^a,
\phi)$, and define the hyperbolic radial coordinate $\rho$ as in
(\ref{rhosquare}). Note that $\rho$ is also an ultralocal function
of the configuration variables $(g_{ij}, \phi)$. We assume that
the hyperbolic coordinate $\rho$ can be used everywhere in a given
region of space near the singularity as a well-defined (real)
quantity which tends to $+ \infty$ as we approach the singularity.
We then define the slicing of spacetime by imposing the gauge
condition \Ref{tildeN=rho2}. The time coordinate corresponding to
this gauge is called $T$ as above (see \Ref{timeT}). Our aim is to
study the asymptotic behaviour of all the dynamical variables
$\b(T), \cn(T),.....$ as $T \to + \infty$ (recall that this limit
also corresponds to $ t \to 0$, $\sqrt{g} \to 0$, $\rho \to +
\infty$, with $\b^\m$ going to infinity inside the future light
cone). Of course, we must also ascertain the self-consistency of
this limit, which we shall refer to as the ``BKL limit''.

\subsection{Hamiltonian action}

To focus on the features relevant to the billiard picture, we
assume first that there are no Chern-Simons terms or couplings of
the exterior form gauge fields through a modification of the
curvatures $F^{(p)}$, which are thus taken to be Abelian, $F^{(p)}
= d A^{(p)}$. We verify in subsection \ref{CSCM} below that these
interaction terms do not change the analysis. The Hamiltonian
action in any pseudo-Gaussian gauge, and in the temporal gauge
\Ref{temporal}, reads
\beq
&& S\left[ g_{ij}, \pi^{ij}, \phi, \pi_\phi, A^{(p)}_{j_1 \cdots j_p},
\pi_{(p)}^{j_1 \cdots j_p}\right] = \nonumber \\
&& \hspace{1cm}
\int dx^0 \int d^d x \left( \pi^{ij} \dot{g_{ij}}
+ \pi_\phi \dot{\phi} + \frac{1}{p!}\sum_p \pi_{(p)}^{j_1 \cdots j_p}
\dot{A}^{(p)}_{j_1 \cdots j_p} - H \right)
\label{GaussAction}
\eeq
where the Hamiltonian density $H$ is
\beq\label{Ham}
H &\equiv&  \tilde{N} \ch \\
\ch &=& \ck + \cm \\
\ck &=& \pi^{ij}\pi_{ij} - \frac{1}{d-1} \pi^i_{\;i} \pi^j_{\;j}
+ \frac{1}{4} \pi_\phi^2 + \nonumber \\
&& \hspace{2cm} + \sum_p \frac{e^{- \lambda_p \phi}} {2 \, p!}
\, \pi_{(p)}^{j_1 \cdots j_p}
\pi_{(p) \, j_1 \cdots j_p} \\
\cm &=& - g R + g g^{ij} \partial_i \phi \partial_j \phi + \sum_p
\frac{e^{ \lambda_p \phi}}{2 \; (p+1)!} \, g \, F^{(p)}_{j_1
\cdots j_{p+1}} F^{(p) \, j_1 \cdots j_{p+1}}
\eeq
where $R$ is the spatial curvature scalar. The dynamical
equations of motion are obtained by varying the above action
w.r.t. the spatial metric components, the dilaton, the spatial
$p$-form components and their conjugate momenta. In addition,
there are constraints on the dynamical variables,
\beq
\ch &\approx& 0  \; \; \; \; \hbox{(``Hamiltonian constraint")}, \\
\ch_i &\approx& 0  \; \; \; \; \hbox{(``momentum constraint")}, \\
\varphi_{(p)}^{j_1 \cdots j_{p-1}} &\approx& 0 \; \; \; \;
\hbox{(``Gauss law" for each $p$-form) } \label{Gauss}
\eeq
with
\beq
\ch_i &=& -2 {\pi^j}_{i|j} + \pi_\phi
\partial_i \phi + \sum_p \frac1{p!} \
\pi_{(p)}^{j_1 \cdots j_p} F^{(p)}_{i j_1 \cdots j_{p}} \\
\varphi_{(p)}^{j_1 \cdots j_{p-1}} &=&
{\pi_{(p)}^{j_1 \cdots j_{p-1} j_p}}_{\vert j_p}
\eeq
where the subscript $|j$ stands for the spatially covariant derivative.

Let us now see how the Hamiltonian action gets transformed when one
performs, at each spatial point, the Iwasawa decomposition
(\ref{Iwasawa}),\Ref{Iwasawaex} of the spatial metric. The
``supermetric'' (\ref{DeWitt}) giving the kinetic terms
of the metric and of the dilaton then becomes
\begin{eqnarray}\label{dsigma1}
d\s^2 &=& \tr d \beta^2 - (\tr d\beta)^2 + d \phi^2 \non
&&     + \frac{1}{2} \tr \left[ \ca^2  \big( d\cn \cn^{-1} \big)
\ca^{-2}  \big(d\cn \cn^{-1}\big)^T\right]
\end{eqnarray}
i.e.,
\beq\label{dsigma2}
d\s^2 &=& \sum_{a=1}^d (d \beta^a)^2 -
\left(\sum_{a=1}^d d \beta^a\right)^2  + d\phi^2  \non
&& + \frac{1}{2} \sum_{a<b} e^{2(\beta^b - \beta^a)}
\left( {d\cn^a}_i \, {\cn^i}_b \right)^2
\eeq
where we recall that ${\cn^i}_a$  denotes, as in  \Ref{Iwasawa2},
the inverse of the triangular matrix ${\cn^a}_i$ appearing in the
Iwasawa decomposition  \Ref{Iwasawaex} of the spatial metric $g_{ij}$.
For $d =3$, this expression reduces to the one of \cite{HPT}.
This change of variables corresponds to a point canonical transformation,
which can be extended to the momenta in the standard way via
\be\label{cantra}
\p^{ij}\dot{g}_{ij} \equiv \sum_a \pi_a \dot{\b}^a
+ \sum_{a} {P^i}_a \dot{{\cal N}^a}_{i}
\ee
Note that the momenta
\be\label{Nmomenta}
{P^i}_a = \frac{\partial\cal L}{\partial \dot{{\cal N}^a}_i}
= \sum_{b} e^{2(\beta^b - \beta^a)} {\dot\cn^a}_{\;\;j} {\cn^j}_b {\cn^i}_b
\ee
conjugate to the non-constant off-diagonal Iwasawa components
${\cn^a}_i$ are only defined for $a<i$; hence the second sum in
(\ref{cantra}) receives only contributions from $a<i$.

We next split the Hamiltonian  density\footnote{We use the term
``Hamiltonian density'' to denote both $H$ and $\ch$. Note that
$H$ is a usual spatial density (of weight 1, i.e. the same weight 
as $\sqrt{g}$), while $\ch \equiv \sqrt{g} H/N$ is  a density 
of weight 2 (like $ g = (\sqrt{g})^2$). Note also that 
$\p^{ij}$ is of weight 1, while $ \tilde{N} \equiv N/\sqrt{g}$
is of weight $-1$.} $\ch$  (\ref{Ham}) in two parts, one
denoted by ${\cal H}_0$, which is the kinetic term for the local
scale factors $\beta^\mu$ (including dilatons) already encountered
in section 3, and a ``potential density'' (of weight 2) denoted by
$\cv$, which contains everything else.
Our analysis below will show why it makes sense to group the kinetic
terms of both the off-diagonal metric components and the $p$-forms
with the usual potential terms, i.e. the term $\cal M$ in (\ref{Ham}).
[Remembering that, in a Kaluza-Klein reduction, the off-diagonal
components of the metric in one dimension higher become a
one-form, it is not surprising that it might be useful to group
together the off-diagonal components and the $p$-forms.]
Thus, we write
\be\label{HplusV}
\ch =  {\cal H}_0 + \cv
\label{calH}
\ee
with the kinetic term of the $\b$ variables
\be
{\cal H}_0 = \frac{1}{4} G^{\mu\nu} \pi_\mu \pi_\nu
\ee
where the r.h.s. is that already defined in (\ref{Gmunuup}), with the
replacement of the coordinate index $i$ by the frame index $a$.
The total (weight 2) potential density,
\be
\cv = \cv_S + \cv_G + \sum_p \cv_{p}  + \cv_\phi ,
\ee
is naturally split into a centrifugal part linked to the kinetic
energy of the off-diagonal components (the index ``$S$'' referring
to ``symmetry'', as discussed below)
\be
\label{centrifugal}
\cv_S = \frac{1}{2}
\sum_{a<b} e^{-2(\beta^b - \beta^a)}
\left( {P^j}_b {{\cal N}^a}_j\right)^2,
\ee
a ``gravitational'' (or ``curvature'') potential
\be
\label{gravitational}
\cv_G =  - g R ,
\ee
and a term from the $p$-forms,
\be
 \cv_{(p)} = \cv_{(p)}^{el} + \cv_{(p)}^{magn}
\ee
which is a sum of an ``electric'' and a ``magnetic'' contribution
\beq
\cv_{(p)}^{el} &=&  \frac{e^{- \lambda_p \phi}} {2 \, p!} \,
\pi_{(p)}^{j_1 \cdots j_p}
\pi_{(p) \, j_1 \cdots j_p} \\
\cv_{(p)}^{magn} &=&  \frac{e^{ \lambda_p \phi}}{2 \; (p+1)!} \,
g \, F^{(p)}_{j_1 \cdots j_{p+1}} F^{(p) \, j_1 \cdots j_{p+1}}
\eeq
Finally, there is a  contribution to the potential linked to the
spatial gradients of the dilaton:
\be
\cv_\phi  = g g^{ij} \partial_i \phi \partial_j \phi.
\ee
We will analyze in detail these contributions to the potential, term by
term,  in section~6.

\subsection{Appearance of sharp walls in the BKL limit}
\subsubsection{Derivation of central result}

In the decomposition of the Hamiltonian given above, we have split
off the kinetic terms of the  scale factors $\b^a$ and of the dilaton
$\b^{d+1} \equiv \phi$ from the other variables, and assigned
the off-diagonal metric components and the $p$-form fields to
various potentials, each of which is a complicated function of
$\b^\m, {\cn^a}_i, {P^i}_a,  A^{(p)}_{j_1 \cdots j_{p}},
\pi_{(p)}^{j_1 \cdots j_p}$ and of some of their spatial gradients.
The reason why this separation is useful is that, as we are going to show,
in the BKL limit, and in the special Iwasawa  decomposition which
we have adopted, the asymptotic dynamics is governed by the scale
factors $\b^\m$, whereas all other variables ``freeze'', just
like for the general Kasner solution discussed in section 3.
Thus, in the asymptotic limit, we have schematically
\be
\cv\Big( \b^\m, {\cn^a}_i, {P^i}_a,  A^{(p)}_{j_1 \cdots j_{p}},
\pi_{(p)}^{j_1 \cdots j_p},\dots\Big) \longrightarrow \cv_{\infty}(\g^\m)
\ee
where $\cv_{\infty}(\g^\m)$ stands for a sum of certain
``sharp wall potentials'' which depend only on the angular hyperbolic
coordinates $\g^\m \equiv \b^\m/ \r$. As a consequence, the asymptotic
dynamics can be described as a ``billiard'' in the hyperbolic space
of the $\g^\m$'s, whose walls (or ``cushions'') are determined by the
energy of the fields that are asymptotically frozen.

This reduction of the complicated potential to a much simpler
``effective potential'' $\cv_{\infty}(\g^\m)$ follows essentially
from the exponential dependence of $\cv$ on the diagonal Iwasawa
variables $\b^\m$, from its independence from the conjugate
momenta of the $\b$'s, and from the fact that the radial magnitude
$\rho$ of the $\b$'s becomes infinitely large in the BKL limit.

To see the essence of this reduction, with a minimum of technical
complications, let us consider a potential density (of weight 2) 
of the general form
\be\label{V1}
\cv(\beta, Q, P) = \sum_A c_A(Q, P) \exp\big(- 2 w_A (\beta) \big)
\ee
where $(Q,P)$ denote the remaining phase space variables (that is,
other than $(\b, \p_\b)$). Here $w_A (\beta) = w_{A \m} \b^\m$
are certain linear forms which depend only on the (extended) scale factors,
and whose precise form will be derived in the following section. Similarly
we shall discuss below the explicit form of the pre-factors $c_A$, which
will be some complicated polynomial functions of the remaining
fields, i.e. the off-diagonal components of the metric, the $p$-form
fields and their respective conjugate momenta, and of some of their
spatial gradients.  The fact that the $w_A (\beta)$ depend
{\em linearly} on the scale factors $\beta^\m$ is an important property
of the models under consideration. A second non-trivial fact is that,
{\em for the leading contributions}, the pre-factors are always
non-negative, i.e. $c_A^{\rm leading}\geq 0$. Since the values of the
fields for which $c_A =0$ constitute a set of measure zero, we will
usually make the ``genericity assumption'' $c_A >0$ for the leading terms
in the potential $\cv$\footnote{Understanding the effects of the possible failure of
this assumption is one of the subtle issues in establishing
a rigorous proof of the BKL picture.}. The third fact following from
the detailed analysis of the walls that we shall exploit is that all
the leading walls are {\it timelike}, i.e. their normal vectors
(in the Minkowski $\b$-space) are spacelike.

As shown in section~3.3, the part of the Hamiltonian describing the
kinetic energy of the $\b$'s, $H_0 = \tilde{N} \ch_0 $, takes the form
\Ref{Hamiltonianrho} when parametrizing $\b^\m$ in terms of
$\r$ and $\g^\m$, or equivalently, $\lambda \equiv \ln \r$ and $\g^\m$
(cf. Eq. (\ref{lambda})). Choosing the gauge \Ref{tildeN=rho2} to simplify
the kinetic terms $H_0$ we end up with an Hamiltonian of the form
\beq\label{V2}
H\big(\lambda, \pi_{\lambda}, \g, \pi_{\g},Q,P\big) &=& \tilde{N} \ch \\
&& \!\!\!\!\!\!\!\!\!\!\!\!\!\!\!\!\!\!\!\!\!\!\!\!
= \frac{1}{4} \left[- \pi_\lambda^2 + \pi_{\bf \gamma}^2\right] +
\rho^2 \sum_A c_A(Q,P) \exp\big(-\rho w_A (\gamma) \big) \nonumber
\eeq
where $\pi_{\bf \gamma}^2$ is the kinetic energy of a particle
moving on $H_m$. In \Ref{V2} and below we shall regard $\lambda$
 as a primary dynamical variable
(so that $\r \equiv e^\l$).

The essential point is now that, in the BKL limit, $\l \to + \infty$ i.e. $\r \to + \infty$,
each term $\rho^2 \exp\big(- 2 \rho w_A (\gamma) \big)$ becomes a sharp
wall potential, i.e. a function of  $w_A (\gamma)$ which is zero
when $w_A (\gamma) >0$, and  $+\infty$ when $w_A (\gamma) < 0$.
To formalize this behaviour we define the sharp wall
$\Theta$-function\footnote{One should more properly write $\Theta_\infty(x)$,
but since this is the only step function encountered in this article,
we use the simpler notation $\Theta(x)$.} as
\be
\Theta (x) := \left\{ \begin{array}{ll}
                      0  & \mbox{if $x<0$} \\
                      +\infty & \mbox{if $x>0$}
                      \end{array}
                      \right.
\ee A basic formal property of this $\Theta$-function is its
invariance under multiplication by a positive quantity. With the
above assumption checked below that all the relevant prefactors
$c_A(Q,P)$ are {\it positive} near each leading wall, we can
formally write \be \lim_{\rho\rightarrow\infty} \Big[ c_A(Q,P)
\rho^2 \exp\big(-\rho w_A (\gamma) \Big] = c_A(Q,P)\Theta\big(-2
w_A (\gamma) \big) \equiv \Theta\big(- 2 w_A (\gamma) \big). \ee
Of course, $\Theta (-2 w_A (\gamma)) = \Theta (- w_A (\gamma))$,
but we shall keep the extra factor of $2$ to recall that the
arguments of the exponentials, from which the $\Theta$-functions
originate, come with a well-defined normalization. Therefore, the
limiting Hamiltonian density reads \be \label{V3}
H_{\infty}(\lambda, \p_{\lambda}, \g, \pi_{\g},Q,P) =
 \frac{1}{4} \left[- \pi_\lambda^2 + \pi_{\bf
\gamma}^2\right] +
\sum_{A'} \Theta\big(-2 w_{A'} (\gamma) \big),
\ee
where $A'$ runs over the {\it dominant walls}.  The set of dominant walls
is defined as the {\it minimal set} of wall forms which suffice to define the billiard table,
i.e. such that the restricted set of inequalities $  \lbrace w_{A'} (\gamma) \geq 0 \rbrace$
imply the full set $   \lbrace w_{A} (\gamma) \geq 0 \rbrace$.  The concept of dominant
wall will be illustrated below. [Note that the concept of ``dominant'' wall is a refinement
of the distinction, which  will also enter our discussion, between a leading wall and a subleading  one.]

The crucial point is that the limiting Hamiltonian \Ref{V3} no longer depends
on $\lambda, Q$ and $P$. Therefore the Hamiltonian equations of motion
for $\lambda, Q$ and $P$ tell us that  the corresponding conjugate momenta,
i.e. $\p_{\lambda}, P $ and $Q$, respectively, all become
{\it constants of the motion} in the limit $\lambda \to +\infty$.
The total Hamiltonian density $H_{\infty}$ is also a constant
of the motion (which must be set to zero). The variable $\lambda$
evolves according to $d \lambda/ dT = - \ft12 \p_{\lambda}$. Hence,
in the limit, $\lambda$ is a linear function of $T$. The only non-trivial
dynamics resides in the evolution of $(\g, \p_{\g})$ which
reduces to the sum of a free (non relativistic)
kinetic term  $\p_{\g}^2/4$ and a sum of sharp wall potentials,
such that the resulting motion of the $\gamma$'s
indeed constitutes a billiard, with geodesic motion on the unit
hyperboloid $H_m$ interrupted by reflections on the walls defined by
$w_A(\gamma)=0$. These walls are hyperplanes (in the sense of
hyperbolic geometry) because they are geometrically given by
the intersection of the unit hyperboloid $\b^\mu \b_\mu =-1$
with the usual Minkowskian hyperplanes $w_A(\b)=0$.

We note in passing that an alternative route for reducing the
dynamics to a billiard in $\g$-space would be to eliminate
the variable $\lambda$ by solving the Hamiltonian constraint
for $\p_{\lambda}$. This allows one to use $\lambda$ as a time variable,
and is similar to going from the quadratic form of the action of
a relativistic particle to its ``square-root form''
\beq
H_{\lambda} &=& \left( \pi_{\bf \gamma}^2/4 +
\sum_A c_A(Q,P) \rho^2 \exp(- 2 \rho w_A (\gamma))\right)^{1/2}
\longrightarrow \nonumber\\
&& \!\!\!\!\!\!\!\!\!\!\!\!\!\!\!\!\!\!\!\!\!\!\!\!
\left(\pi_{\bf \gamma}^2/4 +
\sum_{A'} \Theta \big(-2  w_{A'} (\gamma)\big) \right)^{1/2} \equiv
\ft12 \big| \pi_\gamma \big| +  \sum_{A'} \Theta \Big(-2  w_{A'} (\gamma)\Big)
\eeq

As the above derivation of the asymptotic constancy of all the
phase-space variables $(Q,P)$ may seem a bit formal, we study,  in Appendix  A,
a simplified model explicitly showing how this asymptotic constancy
arises. This toy model also exemplifies the residual,
asymptotically decaying variations of $(Q,P)$ as $\r \to \infty$.

\subsubsection{Finite volume vs. infinite volume}

Geodesic motion in a billiard in hyperbolic space has been much
studied. It is known that this motion is chaotic or non-chaotic
depending on whether the billiard has finite or infinite volume
\cite{Ma69,HM79,Z84,Esk}.  In the finite volume case, the generic
evolution exhibits an infinite number of collisions with the walls
with strong chaotic features (``oscillating behavior").

By contrast, if the billiard has infinite volume, the evolution is
non-chaotic.  For a generic evolution, there are only finitely
many collisions with the walls.  The system generically settles
after a finite time in a Kasner-like motion that lasts all the way
to the singularity.

\subsubsection{$\b$-space description}

The above derivation relied on the use of hyperbolic polar coordinates
$(\rho,\gamma)$. This use is technically useful in that it represents the
walls as being located at an asymptotically  fixed position in hyperbolic
space, namely $w_A(\gamma)=0$. However, once one has derived the final result
\Ref{V3}, one can reexpress it in terms of the original variables $\b^\m$,
which run over a linear (Minkowski) space. Owing to the linearity of
$w_A (\beta) = \rho w_A (\gamma)$ in this Minkowskian picture, the
asymptotic motion takes place in a ``polywedge'', bounded by the hyperplanes
$w_A(\b)=0$. The billiard motion then consists of free motions of  $\b^\m$
on straight lightlike lines within this polywedge, which are interrupted
by specular reflections off the walls. [See formula \Ref{collision} 
below for the explicit effect of these reflections on the components 
of the velocity vector of the $\b$-particle.]
Indeed, when going back to $\b$-space
(i.e. before taking the BKL limit), the dynamics of the scale factors at
each point of space is given by the Hamiltonian
\be
 \label{V4}
H(\b^\m, \p_{\m}) =
\tilde{N} \ch = {\tilde{N}}   \left[\ft14 G^{\mu \nu} \pi_\mu \pi_\nu  +
\sum_A c_A   \exp\big(- 2 w_A (\b)  \big)    \right]
\ee
The $\b$-space dynamics simplifies in the gauge $\tilde{N} =1$, corresponding
to the time coordinate $\tau$. In the BKL limit, the Hamiltonian \Ref{V4}
takes the limiting form (in the gauge $ \tilde{N} = 1$)
\be\label{V5}
H_{\infty}(\b^\m, \p_{\m}) = \ft14 G^{\mu \nu} \pi_\mu \pi_\nu  +
\sum_{A'}    \Theta \big(- 2 w_{A'} (\b)  \big)
\ee
where the sum is again only over the dominant walls.
When taking equal time slices of this polywedge (e.g., slices on
which $\Sigma_i \b^i$ is constant), it is clear that with increasing
time (i.e increasing $\Sigma_i \b^i $, or increasing $\tau$ or $\rho$)
the walls recede  from the observer. The $\b$-space  picture
is useful for simplifying the mathematical representation of the dynamics
of the scale factors which takes place in a linear space. However, it is
inconvenient both for proving that the exponential walls of \Ref{V4}
do reduce, in the large $\Sigma_i \b^i $ limit to sharp walls, and
for dealing with the dynamics of the other phase-space variables
$(Q,P)$, whose appearance in the coefficients  $ c_A $ has been suppressed
in \Ref{V4} above. Let us only mention that, in order to prove,
in this picture, the freezing of the  phase space variables $(Q,P)$
one must consider in detail the accumulation of the  ``redshifts''
of the energy-momentum $\p_\m$ of the $\b$-particle when it undergoes
reflections on the receding walls, and the effect of the resulting
decrease of the magnitudes of the components of $\p_\m$ on the
evolution of $(Q,P)$. (In the notation of Appendix A, it is important
to take into account the fact that $p_0$ decreases with time.)
By contrast, the $\g$-space picture that we used above allows for a more
streamlined treatment of the effect of the limit $\r \to + \infty$
on the sharpening of the walls and on the freezing of $(Q,P)$.

In summary, the dynamics simplifies enormously in the asymptotic
limit. It becomes ultralocal in that it reduces to a continuous
superposition of  evolution systems (depending only on a time
parameter) for the scale factors and the dilatons, at each spatial
point, with asymptotic freezing of the off-diagonal and $p$-form
variables. This ultralocal description of the dynamics is valid
only asymptotically. It would make no sense to speak of a billiard
motion prior to this limit, because one cannot replace the
exponentials by $\Theta$-functions. Prior to this limit, the
evolution system for the scale factors involves the coefficients
in front of the exponential terms, and the evolution of these
coefficients depends on various spatial gradients of the other
degrees of freedom. However, one may contemplate setting up an
expansion in which the sharp wall model is replaced by a model
with exponential (``Toda-like") potentials, and where the
evolution of the quantities entering the coefficients of these
``Toda walls'' is treated as a next to leading effect. See
\cite{DHN2} for the definition of the first steps of such an
expansion scheme for maximal supergravity in eleven dimensions.

\subsection{Kasner frames vs. Iwasawa frames}\label{KversusI}

At this point, we want to clarify an issue that might at first
sight seem paradoxical to the reader. In most of the BKL
literature (notably in the original analysis of \cite{BKL,BKL2}),
one tries to construct the metric in frames equal (or close)
to  ``Kasner frames''\footnote{We know, however,  no detailed
development of the Hamiltonian formalism within such a frame whose
definition involves {\it both} the metric variables and their
conjugate momenta.}.  These are defined as frames with respect to which
both the spatial metric and the  extrinsic curvature are diagonal.  Once the time
slicing has been fixed, this geometric frame is unique, up to {\it
time-dependent rescalings} of each basis vector, when the
eigenvalues of the extrinsic curvature are distinct and some
definite ordering of the (time-independent) eigenvalues $p_a$ has
been adopted.  It is not clear how to fix in a rigourous (and useful)
manner  the arbitrary time-dependent rescalings of each basis vector.
However, this can be done (following \cite{BKL72})
 in an approximate manner by considering
the evolution of the geometry at each spatial point as a
succession of free Kasner flights interrupted by collisions
against the walls. More precisely, one can uniquely fix the
normalization of the Kasner frame by imposing two requirements:
(i) that   the Kasner frame be {\it time-independent} during each
Kasner epoch, i.e. that the Lie derivative of the frame vectors
along $\partial/ \partial t$ be zero, and (ii)  that the linear
transformation between two successive Kasner frames has the
special form given in \Ref{framechange} below (and not
\Ref{framechange} up to some time-independent rescaling).

 More explicitly, if we consider the
three-dimensional case for definiteness and denote the covariant
components of the frame by $\{l_i,m_i,r_i\}$, one  has (suppressing
 the $x$-dependence) \be g_{ij} (t) = A_1^2(t) l_i l_j
+A_2^2(t) m_i m_j +A_3^2(t) r_i r_j \label{K} \ee
during a certain Kasner epoch, and   \be g_{ij} (t) = A_1^{2}(t) \, l'_i l'_j
+A_2^{2}(t) \, m'_i m'_j +A_3^{2}(t) \, r'_i r'_j \label{K'}\ee
during the subsequent Kasner epoch. Here the two successive Kasner frames
$\{l_i,m_i,r_i\}$,  $\{l'_i, m'_i,r'_i\}$ are (in this approximation) 
independent of time, while the scale factors $A_a(t)$ vary like a power law
during each Kasner free flight (say $A_a(t) \approx b_a t^{p_a} $ 
during the first epoch, and $A_a(t) \approx b'_a t^{p'_a} $ during the next,
with some interpolating behaviour during the collision).
Ref.~\cite{BKL72}  has argued (by studying the effect of one collision
in the ``incoming'' frame $\{l_i,m_i,r_i\}$, and by rediagonalizing 
the ``outgoing'' metric) that the transformation between the two 
successive Kasner frames could be written as:
\beq 
l'_i = l_i, \; \; m'_i = m_i + \s_m
l_i, \; \; r'_i = r_i + \s_r l_i \label{framechange} 
\eeq 
Here, $\s_m$ and $\s_r$ are quantities which can a priori be of order
unity.

The seeming paradox is that according to \cite{BKL72} the $\s_m$
and $\s_r$ do not get smaller for collisions closer to the
singularity, so that the transformation (\ref{framechange}) from
the old Kasner axes $l$ to the new Kasner axes $l'$ is of order
one, no matter how close one gets to the singularity; therefore,
the Kasner axes generically never ``come to rest'' if there is an
infinite number of collisions. On the other hand, as we just saw,
the Iwasawa frames become approximately time-independent for
asymptotic values of $\tau$, and the extrinsic curvature
approximately diagonal.

We now show that there is no contradiction between the oscillatory
behavior of the Kasner axes (in the generic inhomogeneous case)
and the asymptotic freezing of the off-diagonal components ${\cal
N}$ in the Iwasawa frame. The key point is the very restricted
form of the change (\ref{framechange}) of the covariant components
during a collision. Recall first that the covariant components of
the Iwasawa frame are given   the three
covectors appearing as  the three lines of the matrix $\cn$, i.e.
$\theta^1 = dx^1 + n_1 dx^2 + n_2dx^3, \theta^2 = dx^2 + n_3 dx^3,
\theta^3 = dx^3$. [Note also in passing that, contrary to the
Kasner frames, the Iwasawa frames are not geometrically
uniquely defined since one can redefine the coordinates $x^i$.]

Let us now see what \Ref{framechange} implies for the
change in the Iwasawa variables. For doing this we need to relate these to the
$l_i$, $m_i$, $n_i$.  This was done in section \ref{3dKasner} for
the simple exact Kasner solution and is easy to generalize to the
case of an oscillatory metric. Indeed, when one is not in the
``collision region" (and this can be applied both {\it before} and
{\it after} the specific collision under consideration in
\Ref{framechange}), the coordinate components of the metric take
the form  \Ref{K} or \Ref{K'} with $A_2^2 \ll A_1^2$ and $A_3^2
\ll A_2^2$. As the definition of the Iwasawa components $n_a$ is
purely algebraic, we can apply the formulas
(\ref{A11})-(\ref{A17}) of section \ref{3dKasner} to the present
case. It suffices to replace $t^{p_a}$ by $A_a$
everywhere. Then, it is easy to see that $n_1$, $n_2$ and $n_3$
are still given by the same final formula as  above, i.e \beq
\label{IK} n_1 = \frac{l_2}{l_1}, \; n_2 = \frac{l_3}{l_1} , \;
n_3 = \frac{l_1 m_3 - l_3 m_1}{l_1 m_2 -l_2 m_1}, \eeq  (before
the collision) and \beq \label{IK'} n'_1 = \frac{l'_2}{l'_1}, \;
n'_2 = \frac{l'_3}{l'_1} , \; n'_3 = \frac{l'_1 m'_3 - l'_3
m'_1}{l'_1 m'_2 -l'_2 m'_1}, \eeq (after the collision). If we
substitute in this second formula $l_i'$, $m_i'$ and $r_i'$ in
terms of $l_i$, $m_i$ and $r_i$ according to (\ref{framechange}),
we get the {\it same} values for the Iwasawa off-diagonal
variables $n_1$, $n_2$ and $n_3$, before and after the collision,
namely, $n'_1 = n_1$, $n'_2 = n_2$, $n'_3 = n_3$. There is thus no
contradiction between the change of Kasner axes
(\ref{framechange}) and the freezing of the off-diagonal Iwasawa
variables. The same conclusion holds for collisions against the
other types of walls, where the Kasner axes ``rotate" as in
(\ref{framechange}).

\subsection{Constraints}
\label{constraints}

We have just seen that in the BKL limit, the evolution equations
become ordinary differential equations with respect to time.
Although the spatial points are decoupled in the evolution
equations, they are, however, still coupled via the constraints.  These
constraints just restrict the initial data and need only be
imposed at one time, since they are preserved by the dynamical
equations of motion. Indeed, one easily finds that, in the BKL limit,
 \be
\dot{\cal H } = 0
\ee
since $[{\cal H}(x), {\cal H}(x')] = 0$ in the
ultralocal limit.  This corresponds simply to the fact that the
collisions preserve the lightlike character of the velocity
vector. Furthermore, the gauge constraints (\ref{Gauss}) are also
preserved in time since the Hamiltonian constraint is gauge-invariant.
In the BKL limit, the momentum constraint fulfills
\be
\dot{{\cal H}}_k(x) =
\partial_k {\cal H} \approx 0 \ee
It is important that the restrictions on the initial data do not bring
dangerous constraints on the coefficients of the walls in the sense
that these may all take non-zero values. For instance, it is well
known that it is consistent with the Gauss law to take non-vanishing
electric and magnetic energy densities; thus the coefficients of
the electric and magnetic walls are indeed generically non-vanishing.
In fact, the constraints are essentially conditions on the
spatial gradients of the variables entering the wall coefficients,
not on these variables themselves.  In some non-generic contexts,
however, the constraints could force some of the wall coefficients
to be zero; the corresponding walls would thus be absent. [E.g.,
for vacuum gravity in four dimensions, the momentum constraints
for some Bianchi homogeneous models force some symmetry wall
coefficients to vanish.  But this is peculiar to the homogeneous
case.]

It is easy to see that the number of arbitrary physical functions
involved in the solution of the asymptotic  BKL equations of motion is the
same as in the general solution of the complete Einstein-matter
equations.  Indeed, the number of constraints on the initial data
and the residual gauge freedom are the same in both cases. Further
discussion of the constraints in the BKL context may be found in
\cite{AR,DHRW}.

\subsection{Consistency of BKL behaviour in spite of the increase
of spatial gradients}

The essential assumption in the BKL analysis, also made in the present
paper, is the asymptotic dominance of time derivatives with respect
to space derivatives near a spacelike singularity.
This assumption has been mathematically justified, in a rigorous manner,
in the  cases where  the billiard is of infinite volume, i.e. in the
(simple) cases where the asymptotic behaviour is not chaotic,
but is  monotonically Kasner-like \cite{AR,DHRW}.

On the other hand, one might a priori worry that this assumption
is self-contradictory in those cases where the billiard is of
finite volume, when the asymptotic behaviour is chaotic, with an
infinite number of oscillations.  Indeed, it has been pointed out
\cite{turb1,turb2} that the independence of the billiard evolution
at each spatial point will have the effect of {\it infinitely
increasing} the spatial gradients of various quantities, notably
of the local values of the Kasner exponents $p_\m(x)$. This
increase of spatial gradients towards the singularity has been
described as a kind of turbulent behaviour of the gravitational
field, in which energy is pumped into shorter and shorter length
scales  \cite{turb1,turb2}, and, if it were too violent, it would
certainly work against the validity of the BKL assumption of
asymptotic dominance of time derivatives. For instance, in our
analysis of gravitational walls in the following section, we will
encounter subleading walls, whose prefactors depend on spatial
gradients of the logarithmic scale factors $\b$.

To address the question of consistency of the BKL assumption
we need to know how fast the spatial gradients of $\b$, and of similar
quantities entering  the prefactors, grow near the singularity.
Let us consider the spatial gradient of $\b \equiv \r \g$, which is
\be\label{gradbeta}
\partial_i \b = \r \partial_i \g + \r \g \partial_i \lambda
\ee
As we are working here in the gauge \Ref{tildeN=rho2}, the spatial
derivatives  must be taken with fixed $T$. We know that, asymptotically,
$\lambda$ is a linear function of $T$, i.e. $\l = a(x) T + b(x)$
where $a(x) = - \ft12 \p_{\l}(x)$ is linked to the (spatially dependent)
conjugate momentum of $\l$.  Therefore the spatial gradient
$\partial_i \l = T \partial_i a(x) +  \partial_i  b(x) $ behaves linearly
in $T$, so that   $\partial_i \l  \propto \l \equiv \ln \r$.
The second term in  \Ref{gradbeta} consequently behaves as $\g \r \ln \r$
when $\r \to \infty$.  Let us now estimate the first term (which will
turn out to dominate the sum).

To estimate $\partial_i \g $ we can use the standard results on billiards
on hyperbolic space. Indeed, $dx^i  \partial_i \g$ can be thought of as
the infinitesimal deviation between two billiard trajectories.
The chaotic behaviour of the billiard implies that this deviation
will grow exponentially with $T$. In fact, because we work on $H_m$
with curvature $-1$ the Liapunov exponent for billiard trajectories
is equal to one, when using geodesic length as time parameter.
Remembering the Hamiltonian constraint $\p_{\l}^2 = \p_{\g}^2$,
the geodesic length is simply equal to $\l$. This yields the
growth estimate
\be
\partial_i \g  =   {\cal O}(1) \exp \l =   {\cal O}(1) \r
\label{result} \ee where the coefficient ${\cal O}(1)$ is a
chaotically oscillating quantity. This estimate is not affected by
the collisions against the walls because these preserve the angles
made by neighboring trajectories (the walls are hyperplanes).
Inserting \Ref{result} in \Ref{gradbeta}, we see that the first
term indeed dominates the second. We conclude that the chaotic
character of the billiard indeed implies an unlimited growth of
the spatial gradients of $\b$, but that this growth is only of {\it
polynomial} order in $\r$ \be
\partial_i \b = {\cal O}(1)  \r^2  . \ee
This polynomial growth of $\partial_i \b$ (and of its second-order
spatial derivatives) entails a polynomial growth of the prefactors
of the sub-dominant walls in section 6.2. Because it is
polynomial (in $\r$), this growth  is, however, negligible compared
to the {\it exponential} (in $\r$) behaviour of the various
potential terms. It does not jeopardize our reasoning based on
keeping track of the various exponential behaviours. As we will
see the potentially dangerously growing terms that we have
controlled here appear only in subdominant walls. The reasoning of
the Appendix shows that the prefactors of the dominant walls are
self-consistently predicted to evolve very little near the
singularity.

We conclude that the unlimited growth of some of the spatial gradients
does not affect the consistency of the BKL analysis done here.
This does not mean, however, that it will be easy to promote our
analysis to a rigorous mathematical proof. The main obstacle to such
a proof appears to be the existence of exceptional points, where a prefactor
of a dominant wall happens to vanish, or points
where a subdominant wall happens to be comparable to a dominant one.
Though the set of such exceptional  points is (generically) of measure
zero, their density might increase near the singularity
because of the increasing spatial gradients. This situation
might be compared to the KAM (Kolmogorov-Arnold-Moser) one, where the
``bad'' tori have a small measure, but are interspersed densely among
the ``good'' ones.

\subsection{BKL limit vs. other limits}
\label{strong}

The last issue which we wish to address in this chapter concerns the
relation of the BKL limit to other limits considered in the literature,
such as the strong coupling limit, or the ``small tension limit'',
as well as the relation with asymptotically velocity dominated solutions.

It is sometimes useful to separate the time derivatives (conjugate momenta)
in the Hamiltonian from the space derivatives, viz.
\be
{\cal H} = \ck' + \e \cv'
\ee
where $\e = \pm 1$ according to whether the spacetime signature
is Lorentzian ($\e = 1$) or Euclidean ($\e = -1$). Here,
\be
\ck'= \ch_0 + \cv_S + \cv^{el}_{(p)}
\ee
contains all the kinetic terms, and
\be
\cv'= \cv_G + \cv^{magn}_{(p)} + \cv_\phi
\ee
the terms with spatial derivatives. We stress that this split is
different from the one introduced in subsection 5.1, where only the
kinetic terms of the sclae factors and the dilatons were kept
as such. The above split is useful because for
some models the asymptotic dynamics is entirely controlled by $\ck'$,
i.e., by the limit $\e = 0$. This happens whenever the billiard that
emerges in the BKL limit is defined by the symmetry and electric walls,
as for instance for $D=11$ supergravity \cite{DH1}, or the pure
Einstein-Maxwell system in spacetime dimensions $D \geq 5$ \cite{DH2,KiSe}.
Curvature and magnetic walls are then subdominant, i.e., spatial gradients
become negligible as one approaches the singularity.

If the curvature and magnetic walls can be neglected, the evolution
equations are exactly the same as the equations of motion obtained
by performing a direct torus reduction to $1+0$ dimensions. We stress,
however, that no homogeneity assumption has to be made here.
{\em The effective torus dimensional reduction follows from
the dynamics and is not imposed by hand.}

The limit $\e =0$ is known as the ``zero signature limit"
\cite{CT} and lies half-way between spacetimes of Minkowskian or
Lorentzian signature. It corresponds to a vanishing velocity of
light (or vanishing ``medium tension"); the underlying geometry is
built on the Carroll contraction of the Lorentz group \cite{MH0}.
The terminology ``strong coupling" is also used \cite{I76} and
stems from the fact that, with appropriate rescalings, $\ch$
can be rewritten as \be \ch = G_N \ck' + G_N^{-1} \cv' \ee such that
the limit in question corresponds to large values of Newton's
constant $G_N$. The interest in this ultrarelativistic
(``Carrollian") limit has recently been revived in
\cite{Dau,LiSv,An} and \cite{Gibb2}.

If there are only finitely many collisions with the walls
(corresponding to a billiard of infinite volume) the dynamics in
the vicinity of the singularity becomes even simpler. After the
last collision, the asymptotic dynamics is  controlled solely by
the kinetic energy $\ch_0$ of the scale factors. This case where
both spatial gradients and matter (here $p$-form) terms can be
neglected, has been called ``asymptotically velocity-dominated"
(or ``AVD'') in \cite{Eardley} and allows a rigorous analysis of
its asymptotic dynamics by means of Fuchsian techniques
\cite{AR,DHRW}. By contrast, rigorous results are rare for the
case of infinitely many collisions (see, however, the recent
analytic advances in \cite{XXX}). Besides the existing rigorous
results, there also exists a wealth of numerical support for the
BKL ideas \cite{BeGaIs,Berger}.

\section{Walls}
\setcounter{equation}{0}
\setcounter{theorem}{0}
\setcounter{lemma}{0}

The decomposition (\ref{HplusV}) of the Hamiltonian gives rise to
different types of walls, which we now discuss in turn.
Specifically, we will derive explicit formulas for the linear
forms $w_A(\b)$ and the field dependence of the pre-factors
$c_A$ entering the various  potentials.

\subsection{Centrifugal (or symmetry) walls}
\label{Geometryoff} We start by analyzing the effects of the
off-diagonal metric components which will give rise to the
so-called ``symmetry walls''. As they originate from the
gravitational action they are always present. The relevant
contributions to the potential is the centrifugal potential
\Ref{centrifugal}. When comparing  \Ref{centrifugal} to the
general form \Ref{V1} analyzed above, we see that firstly the
summation index $A$ must be interpreted as a double index $(a,b)$,
with the restriction $ a<b$, secondly that the corresponding
prefactor is $ c_{ab} =   ({P^j}_a {{\cal N}^b}_j )^2$ is
automatically non-negative (in accordance with our genericity
assumptions, we shall assume $ c_{ab} >  0$). The centrifugal wall
forms read: \be w_{(ab)}^S(\b) \equiv w_{(ab)\mu}^S \beta^\mu
\equiv \beta^b - \beta^a  \; \; (a<b). \label{symwall} \ee We
refer to these wall forms as the ``symmetry walls'' for the
following reason.  When applying the general collision law
\Ref{collision} derived below to the case of the collision on the
wall \Ref{symwall} one easily finds that its effect on the
components of the velocity vector $v^\m$ is simply to permute the
components $v^a$ and $v^b$, while leaving unchanged the other
components $\m \ne a,b$.

The hyperplanes  $w_{(ab)}^S(\beta) = 0$ (i.e. the symmetry walls)
are timelike since \be G^{\mu \nu} w_{(ab) \mu}^S w_{(ab) \nu}^S
=+ 2 \label{normofgrad} \ee This ensures that the symmetry walls
intersect the hyperboloid $G_{\m \n} \b^\m \b^\n = -1$, $\sum_a
\beta^a \geq 0$. The symmetry billiard (in $\b$-space)  is defined
to be the region of  Minkowski space determined by  the
inequalities \be w_{(ab)}^S(\beta) \geq 0, \label{fullsymmetry}
\ee with $\sum_a \beta^a\geq 0$ (i.e. by the region of $\b$-space
where the $\Theta$ functions are zero). Its projection on the
hyperbolic space $H_m$ is defined by the inequalities $
w_{(ab)}^S(\g) \geq 0$.

The explicit expressions above of the symmetry wall forms also allow us to
illustrate the
notion of a ``dominant wall'' defined in subsection 5.2.1 above.
Indeed, the  $d(d-1)/2$ inequalities
\Ref{fullsymmetry} already follow from the following  {\it minimal} set of
$d-1$ inequalities
\be
\b^2 -\b^1 \geq 0, \; \b^3 - \b^2 \geq 0,
\; \cdots , \b^d - \b^{d-1} \geq 0
\label{symmetry}
\ee
More precisely, each linear form which must be positive
in \Ref{fullsymmetry} can be written as a linear combination,
with positive (in fact, integer) coefficients of the linear
forms entering the subset \Ref{symmetry}. For instance,
$ \b^3 - \b^1 = (\b^3 - \b^2) + (\b^2 -\b^1) $, etc. In the last
section, we will reinterpret this result by identifying the
dominant linear forms entering \Ref{symmetry} with the simple
roots of $ SL(n,\RR)$.  Note that, in principle, the final set
of dominant walls can only be decided when one starts from
the complete list of all the dynamically relevant walls. In all
the models we examine the set of dominant symmetry walls
\Ref{symmetry} will, however, be part of the final minimal
set of dominant walls defining the complete billiard table.

If only the symmetry  walls  were present in the Hamiltonian, one
would easily deduce from the above the following picture of the
dynamics: When the trajectory hits a wall $\b^{a+1} = \b^a$ from
within the interior of the billiard $\b^{a+1} - \b^a > 0$ it
undergoes a reflection which reorders $ v ^{a+1} $ and $v^a$ from
the incident state $ v^{a+1}< v^a$ (in which $\b^{a+1} - \b^a $
{\it decreases} towards zero) into an outgoing state $  v^{a+1} >
v^a$ (in which $\b^{a+1} - \b^a $ {\it increases} away from zero).
Each collision reorders a pair of velocity components so that,
after a finite number of collisions, they will get reordered in a
stable configuration where $v^1 \leq v^2 \cdots \leq v^d$ and
$\b^1 \leq \b^2 \cdots \leq \b^d$. The motion would then continue
freely, i.e. without collisions, after the last reordering reflection. The
same conclusion was already reached in Section 4 by a direct
calculation of the dynamics of the Iwasawa scale factors of
non-diagonal Kasner metrics (which is, indeed, a case where only
symmetry walls are present).

If we still consider for a moment the simple case of homogeneous
but non-diagonal metrics and recall that the metric can
be diagonalized at all times by a time-independent coordinate
transformation $x^i \rightarrow x'^i = L^i_{\; \; j} x^j$,
it might appear that the symmetry walls, which are related to the
off-diagonal components, are only a gauge
artifact with no true physical content. This conclusion, however,
would be incorrect.  First, the transformation needed to
diagonalize the metric may not be a globally well-defined
coordinate transformation if the spatial sections have non-trivial
topology, e.g., are tori, since it would conflict in general with
periodicity conditions. Second, even if the spatial sections are
homeomorphic to $\Rn^d$, the transformation $x^i \rightarrow x'^i =
L^i_{\; \; j} x^j$, although a diffeomorphism, is not a proper
gauge transformation in the sense that it is generated by a
non-vanishing charge, and therefore two solutions that differ by
such a transformation should be regarded as physically distinct (although
related by a symmetry). Initial conditions, for which the metric
is diagonal and hence the symmetry walls are absent, form a set
of measure zero.

Let us finally  mention the possibility of alternative treatments
of the dynamics of off-diagonal metric components. We have just
shown that the Iwasawa decomposition of the spatial metric leads
to a projected description of the $GL(d,\Rn)/SO(d)$-geodesics as
motions in the space of the scale factors with exponential
(``Toda-like") potentials. An alternative description can be based
on the decomposition $G = R^T \, A \, R$ of the spatial metric,
where $R \in SO(d)$ and $A$ is diagonal \cite{BKRyan,Khve}. One
then gets Calogero-like potentials $\propto\sinh^{-2}(\b^a -
\b^b)$. In the BKL limit, these potentials can be replaced by
sharp wall potentials but whether the system lies to the left or
to the right of the wall $\b^a - \b^b = 0$ depends on the initial
conditions in this alternative description.

\subsection{Curvature (gravitational) walls}

Next we analyze the gravitational potential, which requires a
computation of curvature. To that end, one must explicitly express
the spatial curvature in terms of the scale factors and the
off-diagonal variables ${\cn^a}_i$. Again, the calculation is most
easily done in the Iwasawa frame (\ref{Iwasawa1}), in which the
metric assumes the form (\ref{Iwasawa0}).
 We use the short-hand notation $A_a \equiv e^{-\b^a}$
 for the (Iwasawa) scale factors.
 Let $C^a_{\; \; bc}(x)$
be the structure functions of the Iwasawa basis $\{\theta^a\}$, viz.
\be
d \theta^a = -
\frac{1}{2}C^a_{\; \; bc} \theta^b\wedge \theta^c
\ee
where $d$ is the spatial exterior differential. The structure
functions obviously depend only on the off-diagonal components
${{\cal N}^a}_i$, but not on the scale factors.
Using the Cartan formulas for the connection one-form
$\o^a_{\; \; b}$,
\beq
&& d\theta^a + \sum_b \o^a_{\; \; b} \wedge \theta^b = 0 \\
&& d \gamma_{ab} = \o_{ab} + \o_{ba}
\eeq
where $ \o_{ab}\equiv \gamma_{ac} {\o^c}_b$, and
\be
\gamma_{ab} = \delta_{ab} A_a^2\equiv \exp(-2\b^a) \delta_{ab}
\ee
is the metric in the frame $\{\theta^a\}$, one finds
\beq
\o^c_{\; \; d} &=& \sum_b \frac{1}{2} \left( C^b_{\; \; cd}
\frac{A_b^2}{A_c^2} + C^d_{\; \; cb} \frac{A_d^2}{A_c^2}
 - C^c_{\; \; db}  \frac{A_c^2}{A_d^2} \right) \theta^b \nonumber \\
&& + \sum_b \frac{1}{2 A^2_c} \Big[ \delta_{cd} (A^2_c)_{,b}
+ \d_{cb} (A^2_c)_{,d} - \d_{db} (A^2_d)_{,c} \Big] \theta^b
\eeq
In the last bracket above, the commas denote the frame
derivatives   $\partial_a \equiv {\cn^i}_a \partial_i$.

The Riemann tensor $R^c_{\; \; def}$, the Ricci tensor $R_{de}$
and the scalar curvature $R$ are obtained through
\beq
\Omega^a_{\; \; b} &=& d \o^a_{\; \; b}
+ \sum_c \o^a_{\; \; c} \wedge \o^c_{\; \; b} \\
&=& \frac{1}{2} \sum_{e,f} R^a_{\; \; bef} \theta^e \wedge \theta^f
\eeq
where $\Omega^a_{\; \; b}$ is the curvature $2$-form and
\be
R_{ab} =  \sum_c R^c_{\; \; acb}, \; \; \;
R = \sum_a \frac{1}{A^2_a} R_{aa}.
\ee
Direct, but somewhat cumbersome, computations yield
\be
R = - \frac{1}{4} \sum_{a,b,c} \frac{A_a^2}{A^2_b \, A^2_c}
(C^a_{\; \;bc})^2 + \sum_a \frac{1}{A_a^2} F_a\big(\partial^2 \beta,
\partial \beta, \partial C, C\big)
\label{formulaforR}
\ee
where $F_a$ is some complicated function of its arguments
whose explicit form will not be needed here.  The only property
of $F_a$ that will be of importance is that it is a polynomial
of degree two in the derivatives $\partial \beta$ and of degree
one in $\partial^2 \beta$.  Thus, the exponential dependence
on the $\beta$'s which determines the asymptotic behaviour
in the BKL limit, occurs only through the $A^2_a$-terms written
explicitly in (\ref{formulaforR}).

In \Ref{formulaforR} one obviously has $b\neq c$ because the
structure functions $C^a_{\; \; bc}$ are antisymmetric in the
pair $[bc]$. In addition to this restriction, we can assume,
without loss of generality, that $a \not=b,c$ in the first
sum on the right-hand side of (\ref{formulaforR}). Indeed, the
terms with either $a=b$ or $a=c$ can be absorbed into a
redefinition of $F_a$. We can thus write the gravitational
potential density (of weight 2) as
\be\label{VsubG}
\cv_G \equiv -g R =
\frac{1}{4} {\sum_{a,b,c}}' e^{-2\a_{abc}(\beta)}
(C^a_{\; \; bc})^2 - \sum_a e^{-2 \m_a(\beta)} F_a
\ee
where the prime on $\sum$ indicates that the sum is to be performed
only over unequal indices, i.e. $a\neq b, b\neq c , c\neq a$, and
where the linear forms $\a_{abc}(\beta)$ and $\m_a(\beta)$
are given by
\be
\a_{abc}(\beta) = 2 \beta^a + \sum_{e\neq a,b,c} \beta^e \quad\quad
(a\neq b \, , \; b\neq c \, , \; c\neq a)
\ee
and
\be
\m_a(\beta) = \sum_{c \not= a} \beta^c.
\ee
respectively. Note that $ \a_{abc}$ is symmetric under the exchange
of $b$ with $c$, but that the index $a$ plays a special role.

Comparing the result \Ref{VsubG} to the general form    \Ref{V1}
we see that there are, a priori, two types of gravitational walls:
the $\a$-type and the $\m$-type.  The $\a$-type   walls
clearly come with positive prefactors, proportional to
the square of a structure function $ C^a_{\; \; bc}$ .
The $\m$-type terms  seem to pose a problem because they do not
have a definite sign. It would therefore seem that, in the
BKL limit, the gravitational potential would tend to
\be
\lim_{\rho\rightarrow\infty} \cv_G = {\sum_{a,b,c}}'
\Theta[-2\a_{abc}(\b)] + \sum_a \Big(\pm \Theta[-2 \m_a(\b)]\Big)
\ee
However, the indefinite $\m$-type terms can generically be
neglected in the BKL limit. This is most simply seen by noting that
the inequalities $ \a_{abc}(\b) \geq 0$ imply $\m_a(\b) \geq 0$
because $\m_a$ is a linear combination with positive coefficients of the
$ \a_{abc}$'s. Indeed, we can write $\m_c = ( \a_{abc} + \a_{bca})/2$.
Therefore the $\a$-walls dominate the $\m$-ones.

In fact, one can establish a stronger result, namely $\m_a(\b)
\geq 0$ within the entire future light cone of the $\b$'s. For
this purpose, we note first that each linear form $\m_a(\b)$ is
{\em lightlike}, i.e. $G^{\mu\nu}({\mu_a})_\mu ({\mu_a})_\nu =0$. Therefore,
each hyperplane $\m_a(\b)=0$ is tangent to the light cone along
some null generator. This means that the future light cone is entirely
on one side of the hyperplane $\m_a(\b)=0$ (i.e., either
$\m_a(\b)> 0$ for all points inside the future light cone or
$\m_a(\b)< 0$). Now, the point $\b^1= \b^2 = \cdots = \b^d = 1$ is
inside the future light cone and makes all the $\m_a$'s positive.
Hence $\m_a(\b)>0$ inside the future light cone for each $a$ and
$\Theta[-2 \m_a(\b)] = 0$, and we really have \be
\lim_{\rho\rightarrow\infty} \cv_G = {\sum_{a,b,c}}'
\Theta[-2\a_{abc}(\beta)]. \ee Note, however, that the $\m$-type
walls may make their existence felt in the exceptional case when
$\b$ is close to the lightlike direction defined by $\m$. This is
the case of ``small oscillations" considered by BKL in
\cite{BKL,BKL2}, for which they verify (for $d=3$) that the
evolution is indeed controlled by the $\a_{abc}$-terms even in
that region.

{}From these considerations we deduce the additional constraints
\be
\a_{abc}(\b) \geq 0 \; \; \;  \; (D>3)
\ee
besides the symmetry inequalities (\ref{symmetry}). The hyperplanes
$\a_{abc}(\b) = 0$ are called the ``curvature" or ``gravitational"
walls. Like the symmetry walls, they are timelike since
\be
G^{\mu\nu} (\a_{abc})_\mu (\a_{abc})_\nu = +2
\ee
The restriction $D>3$ is due to the fact that in $D=3$ spacetime
dimensions, the gravitational walls $\a_{abc}(\b) = 0$ are absent,
simply because one cannot find three distinct spatial indices. In
this case all gravitational walls are of subdominant type $\m_a$
and thus, in the BKL limit,
\be
\cv_G \simeq \sum_a \big(\pm \Theta[-2 \m_a(\b)]\big) \simeq 0 \quad (D=3).
\ee
This is, of course, in agreement with expectations, because gravity
in three spacetime dimensions has no propagating degrees of
freedom (gravitational waves).

The fact that the gravitational potential becomes, in the BKL
limit, a positive sum of sharp wall potentials, is remarkable
for several reasons. First, the final form of the potential is
quite simple, even though the curvature is a rather complicated
function of the metric and its derivatives. Secondly, the limiting
expression of the potential is positive, even though there are
subdominant terms in $\cv_G$ with indefinite sign. Thirdly, it
is ultralocal in the scale factors, i.e. involves only the scale
factors but not their derivatives. It is this fact that accounts
for the decoupling of the various spatial points.

The coefficients of the dominant exponentials involve only the
undifferentiated structure functions ${C^a}_{bc}$. Consequently,
one can model the gravitational potential in leading order by {\em
spatially homogeneous} cosmologies, which have constant structure
functions, and by considering homogeneity groups that are
``sufficiently non-abelian'' so that none of the coefficients of
the relevant exponentials vanish (Bianchi types VIII and IX for
$d=3$, other homogeneity groups for $d>3$ - see \cite{DdRH0}). By
contrast, inspection of \Ref{formulaforR} and \Ref{VsubG} reveals
that the subleading terms not only lack manifest positivity, but
also do depend on spatial inhomogeneities via the spatial gradient
of the structure functions, and the first and second spatial
derivatives of the scale factors. It is quite intriguing that the
associated walls are {\em lightlike}, unlike the walls associated
with the leading terms all of which are timelike. Terms involving
lightlike walls will thus have to be taken into account in higher
orders of the BKL expansion. A Kac-Moody theoretic interpretation
of this fact was recently proposed in \cite{DHN2}.

The computations in this subsection involve only the Cartan formulas.
They remain valid if in \Ref{Iwasawa1} we replace the one-forms
$dx^i$ by some anholonomic frame $f^i = f^i_{\;j}(x) dx^j$.
This modifies the Iwasawa frame $\{\theta^a\}$, which has no
intrinsic geometrical meaning. The structure functions 
${C^a}_{bc}(x)$ of the new frame get extra contributions from 
the spatial derivatives acting on $f^i$. In fact, for $f^i = dx^i$ 
not all gravitational walls $\a_{ijk}$ appear because from 
(\ref{changeofframe}) we then have ${C^d}_{bc} = 0$ for the 
top component (since $\theta^d = dx^d$), and ${C^{d-1}}_{bc}= 0$
$b,c \neq d$.  Hence, the corresponding gravitational walls
are absent. To get all the gravitational walls, one therefore needs
an anholonomic frame $f^i$. However, the dominant gravitational wall
$\a_{1 \, d-1 \,d}$ is always present, and this is the one relevant
for the billiard, when gravitational walls are relevant at all.

\subsection{$p$-form walls}
While none of the wall forms considered so far involved the dilatons,
the electric and magnetic ones do as we shall now show. To make the
notation less cumbersome we will omit the super-(or sub-)script $(p)$
on the $p$-form fields in this subsection.

\subsubsection{Electric walls}
The electric potential density can be written as
\be
\cv^{el}_{(p)} =
\frac{1}{2 \, p!} \sum_{a_1, a_2, \cdots, a_p} e^{-2 e_{a_1  \cdots
a_p}(\b)} (\ce^{a_1  \cdots a_p})^2
\label{Velec}
\ee
where $\ce^{a_1  \cdots a_p}$ are the components of the electric field
$\pi^{i_1 \cdots i_p}$ in the basis $\{\theta^a\}$
\be
\ce^{a_1 \cdots a_p} \equiv {\cn^{a_1}}_{j_1}
{\cn^{a_2}}_{j_2} \cdots {\cn^{a_p}}_{j_p} \pi^{j_1 \cdots j_p}
\ee
(recall our summation conventions for spatial cordinate indices)
and where $e_{a_1  \cdots a_p}(\b)$ are the electric wall forms
\be
e_{a_1 \cdots a_p}(\b) = \b^{a_1} + \cdots + \b^{a_p} -
\frac{\l_p}{2} \phi  
\ee
Here the indices $a_j$'s are all distinct because
$\ce^{a_1 \cdots a_p}$ is completely antisymmetric. The variables
$\ce^{a_1\cdots a_p}$ do not depend on the $\beta^\mu$. It is thus rather
easy to take the BKL limit. The exponentials in (\ref{Velec}) are
multiplied by positive factors which generically are different
from zero. Thus, in the BKL limit, $\cv^{el}_{(p)}$ becomes
\be
\cv^{el}_{(p)} \simeq \sum_{a_1< a_2< \cdots< a_p} \Theta[-2
e_{a_1  \cdots a_p}(\b)] .
\ee
The transformation from the variables $\big({\cn^a}_i,
{P^i}_a, A_{j_1 \cdots j_p}, \pi^{j_1 \cdots j_p}\big)$ to
the variables $\big({\cn^a}_i,{{\cal P}^i}_a, {\cal A}_{a_1 \cdots a_p},
\ce^{a_1  \cdots a_p}\big)$ is a point canonical
transformation whose explicit form is obtained from
\be
\sum_{a} {P^i}_a \dot{\cn^a}_{i} + \sum_p \frac1{p!}
\pi^{j_1 \cdots j_p} \dot{A}_{j_1 \cdots j_p} =
\sum_{a}{{\cal P}^i}_a \dot{\cn^a}_i +
\sum_p\sum_{a_1,\dots,a_p}\frac1{p!}
\ce^{a_1 \cdots a_p} \dot{{\cal A}}_{a_1 \cdots a_p}
\ee
The new momenta ${{\cal P}^i}_a$   conjugate to ${\cn^a}_{i}$ differ from the
old ones  ${P^i}_a$  by terms involving $\ce$,  $\cn$
and $\ca$ since the components ${\cal A}_{a_1 \cdots a_p}$ of
the $p$-forms in the basis $\{\theta^a\}$ depend on the $\cn$'s,
$$
{\cal A}_{a_1 \cdots a_p} =
{\cn^{j_1}}_{a_1} \cdots {\cn^{j_p}}_{a_p} A_{j_1\cdots j_{p}}.
$$
However, it is easy to see that these extra terms do not affect
the symmetry walls in the BKL limit.

\subsubsection{Magnetic walls}
The magnetic potential is dealt with similarly.  Expressing it
in the $\{\theta^a\}$-frame, one obtains
\be
\cv^{magn}_{(p)} =
\frac{1}{2 \, (p+1)!} \sum_{a_1, a_2, \cdots, a_{p+1}} e^{-2
m_{a_{1}  \cdots a_{p+1}}(\b)} (\cf_{a_1  \cdots a_{p+1}})^2
\label{Vmagn}
\ee
where $\cf_{a_1  \cdots a_{p+1}}$ are the components
of the magnetic field $F_{m_1 \cdots m_{p+1}}$ in the basis
$\{\theta^a\}$,
\be
\cf_{a_1 \cdots a_{p+1}} =
{\cn^{j_1}}_{a_1} \cdots  {\cn^{j_{p+1}}}_{a_{p+1}} F_{j_1 \cdots j_{p+1}}
\ee
The $m_{a_{1} \cdots a_{p+1}}(\b)$ are the magnetic linear forms
\be
m_{a_{1} \cdots a_{p+1}}(\b) = \sum_{b \notin \{a_1,a_2,\cdots a_{p+1}\}}
\b^b + \frac{\l_p}{2} \phi
\ee
where again all $a_j$'s are distinct. One sometimes rewrites
$m_{a_{1} \cdots a_{p+1}}(\b)$ as $\tilde{m}_{a_{p+2} \cdots a_d}$,
where $\{a_{p+2}, a_{p+3}, \cdots, a_d \}$ is the set complementary
to $\{a_1,a_2,\cdots a_{p+1} \}$; e.g.,
\be
\tilde{m}_{1 \, 2 \, \cdots \, d-p-1} =
\b^1 + \cdots + \b^{d-p-1} +\frac{\l_p}{2} \phi
= m_{d-p \, \cdots \, d}
\ee
Of course, the components of the exterior derivative $\cf$ of $\ca$ in the non-holonomic
frame $\{\theta^a\}$ involves the structure coefficients,
i.e. $\cf_{a_1  \cdots a_{p+1}} =
\partial_{[a_1} {\cal A}_{a_2 \cdots a_{p+1}]} + C {\cal A}\hbox{-terms}$
where $\partial_a \equiv {\cn^i}_a \partial_i$ is the frame derivative.

Again, the BKL limit is quite simple and yields (assuming generic
magnetic fields)
\be
\cv^{magn}_{(p)} \simeq \sum_{a_1< \cdots < a_{d-p-1}}
 \Theta[-2 b_{a_1 \cdots a_{d-p-1}}(\b)].
\ee
Just as the off-diagonal variables, the electric and magnetic
fields freeze in the BKL limit since the Hamiltonian no longer
depends on the $p$-form variables. These drop out because one can
rescale the coefficient of any $\Theta$-function to be one (when
it is not zero), thereby absorbing the dependence on the $p$-form
variables.

The scale factors are therefore constrained by the further 
``billiard'' conditions
\be
e_{a_1  \cdots a_p}(\b) \geq 0, \; \; \;
\tilde{m}_{a_1 \cdots  a_{d-p-1}}(\b) \geq 0.
\ee
The hyperplanes $e_{a_1  \cdots a_p}(\b) = 0$ and
$\tilde{m}_{a_1 \cdots  a_{d-p-1}}(\b) =0$ are called ``electric"
and ``magnetic" walls, respectively. Both walls are timelike
because their gradients are spacelike, with squared norm
\be
\frac{p(d-p-1)}{d-1} + \Big(\frac{\lambda_p}{2}\Big)^2 > 0
\label{elecnorm}
\ee
(For $D=11$ supergravity, we have $d=10, p=3$ and $\lambda_p = 0$
and thus the norm is equal to +2). This equality explicitly shows
the invariance of the norms of the $p$-form walls under
electric-magnetic duality.

\subsection{Walls due to dilatons or to a cosmological constant}

The fact that the leading walls originating from the centrifugal,
gravitational and $p$-form or are all timelike, is an important
ingredient of the overall  BKL picture. As we saw above, however,
there exist subleading contributions in the gravitational potential
whose associated wall forms are lightlike. Such is the case also for
the dilaton contribution
\be
\cv_\phi = g g^{ij}\partial_i \phi\partial_j \phi,
\ee
which has the same form as the subleading gravitational walls, since
the exponentials that control its asymptotic behaviour are easily seen to
be $\exp[-2 \m_a(\b)]$. Consequently, at least to leading order,
we can neglect $\cv_\phi$ in the BKL limit.

The only example of a {\it spacelike} wall that we know of is the
cosmological constant term  (in its weight-2 form) \be
\label{Lambda} \cv_\Lambda = \Lambda g  =  \Lambda \exp
\Big[-2\sum_a \b^a \Big] \ee When $\Lambda$ is positive (de Sitter
sign) the spacelike wall   \Ref{Lambda} is repulsive. Depending on
the initial conditions -- which  set the scale -- this wall either
prevents the system from reaching the BKL small volume regime or
does not prevent the collapse. In the first case, the spacelike
wall acts as a ``barrier'' to the motion of the billiard ball, and
the reflection against it forces the billiard ball to run
``backwards in $\b$-time''  in the direction of {\it increasing}
spatial volume (this is analogous to the bounce in the (global) de
Sitter solution, which is a hyperboloid --- a sphere that first
contracts and then expands). In the second case, the billiard ball
is  already ``beyond the barrier'', and the presence of the
spacelike wall has only a subdominant effect  on its motion. In
the BKL picture it quickly becomes negligible: the cosmological
potential $\Lambda g$ tends to zero as $g$ goes to zero. When
$\Lambda$ is negative (Anti-de Sitter sign) the spacelike wall
\Ref{lambda} is attractive and tends to favour collapse. It,
however, quickly becomes negligible in the BKL limit.

\subsection{Chern-Simons and Yang-Mills couplings}
\label{CSCM}
The addition of Chern-Simons terms, Yang-Mills or Chapline-Manton
couplings does not bring in new (asymptotically relevant) walls. 
The only change in the
asymptotic dynamics is a modification of the constraints.

For Yang-Mills couplings, the contribution to the energy density
from the Yang-Mills field takes the same form as for a collection
of abelian $1$-forms, with the replacement of the momenta ${\p}^i$
by the Yang-Mills momenta ${\p}^i_a$ (where $a = 1, \cdots, N$ and $N$
is the dimension of the internal Lie algebra) and of the magnetic
fields by the corresponding non-abelian field strengths.
As their abelian counterparts, these do not involve the scale
factors $\b^\m$. Because of this key property, the same analysis
goes through. Each electric and magnetic $1$-form wall is simply
repeated a number of times equal to the dimension of the Lie algebra.
Gauss law is, however, modified and reads
\be
{\cal D}_i {\p}^i_a \equiv
\nabla_i {\p}^i_a + f^b_{\; \; ac}{\p}^i_b A^c_i = 0.
\ee
Here, $\nabla_i$ is the standard metric covariant derivative.
Similarly, the momentum constraints are modified and involves the
non-abelian field strengths.

The discussion of Chapline-Manton couplings or Chern-Simons terms
proceeds in the same way.  The energy-density of the $p$-forms has
the same dependence on the scale factors as in the absence of
couplings, i.e., provides the same exponentials. The only
difference is that the wall coefficients are different functions
of the $p$-form canonical variables; but this difference is again
washed out in the sharp wall limit, where the coefficients can be
replaced by one (provided they are different from zero). The momentum
and Gauss constraints are genuinely different and impose different
conditions on the initial data.

Nevertheless, the Chern-Simons terms may play a more
significant r\^ole in peculiar contexts when only specialized
field configurations are considered. This occurs for instance in
\cite{Ivashchuk:1999rm}, where it is shown that the $D=11$ supergravity
Chern-Simons term for spatially homogeneous metrics and magnetic
fields may constrain some electromagnetic walls to disappear
``accidentally".  This changes the finite volume billiard to one
of infinite volume.

\section{Cosmological billiards}
\setcounter{equation}{0}
\setcounter{theorem}{0}
\setcounter{lemma}{0}

Let us summarize our findings. The dynamics in the vicinity of
a spacelike singularity is governed by the scale factors,
while the other variables (off-diagonal metric components,
$p$-form fields) tend to become mere ``spectators'' which get asymptotically
frozen. This simple result is most easily derived in terms
of the hyperbolic polar coordinates ($\r,\g$), and in the
gauge \Ref{tildeN=rho2}. In this picture, the essential dynamics
is carried by the angular variables $\g$ which move on a fixed
billiard table, with cushions defined by the dominant walls
$w_{A'}(\g)$. However, it is often geometrically more illuminating
to ``unproject'' this billiard motion in the full Minkowski space
of the extended scale factors $\b^\m$.  In that picture, the asymptotic
evolution of the scale factors at each spatial point reduces to a
zigzag of null straight lines w.r.t. the metric $G_{\mu \nu} d\b^\m d\b^\n$.
The straight segments of this motion are interrupted by collisions
against the sharp walls
\be\label{cushions}
w_A(\b) \equiv w_{A \m} \b^\m = 0
\ee
defined by the symmetry, gravitational and $p$-form potentials,
respectively. As we showed all these walls are {\it timelike},
i.e. they have spacelike gradients:
\be
G^{\m\n} w_{A \m}  w_{A \n} > 0
\ee
Indeed, the gradients of the symmetry and gravitational wall forms
have squared norm  equal to $+2$, independently of the dimension $d$.
By contrast, the norms of the electric and magnetic gradients, which
are likewise positive, depend on the model. As we saw, there also
exist subdominant walls, which can be neglected as they are
located ``behind'' the dominant walls.

In the $\b$-space picture, the free motion before a
collision is described by a null straight line of the type of
\Ref{freemotion}, with the constraint \Ref{constraint}. The effect
of a collision on a particular wall $w_A(\b)$ is easily obtained
by solving \Ref{V4}, or \Ref{V5}, with only one term in the sum.
This dynamics is exactly integrable: it suffices to decompose the
motion of the $\b$-particle into two (linear) components: $(i)$ the
component parallel to the (timelike) wall hyperplane, and $(ii)$ the
orthogonal component. One easily finds that the parallel motion
is left unperturbed by the presence of the wall, while the orthogonal motion
suffers a (one-dimensional) reflection, with a change of the sign of the
outgoing orthogonal velocity with respect to the ingoing one. The net effect
of the collision on a certain wall  $w(\b)$ then is to change the ingoing velocity vector
$v^\m$ entering the ingoing free motion \Ref{freemotion} into an outgoing
velocity vector $v'^\m$ given (in any linear frame) by the usual formula
for a geometric reflection in the  hyperplane $w(\b) =0$:
\begin{equation}
v'^{\mu} =  v^{\mu} - 2 \ \frac{(w \cdot v) \,
w^{\mu}}{(w \cdot w)}  \, .
\label{collision}
\end{equation}
Here, all scalar products, and index raisings, are done with the $\b$-space
metric $G_{\m\n}$. Note that the collision law \Ref{collision} leaves
invariant the (Minkowski) length of the vector $ v^{\mu}$. Because the
dominant walls are timelike, the geometric reflections that the velocities
undergo during a collision, are elements of the orthochronous Lorentz
group. Each reflection preserves the norm and the time-orientation;
hence, the velocity vector remains null and future-oriented.

{}From this perspective, we can also better understand the relevance of
walls which are {\em not} timelike. {\em Lightlike} walls (like some
of the subleading gravitational walls) can never cause reflections
because in order to hit them the billiard ball would have to move
at superluminal speeds in violation of the Hamiltonian constraint.
The effect of {\em spacelike} walls (like the cosmological constant
wall) is again different: they are either irrelevant (if they are
``behind the motion''), or otherwise they reverse the time-orientation
inducing a motion towards increasing spatial volume (``bounce'').

The {\it hyperbolic billiard} is obtained from the $\b$-space
picture by a radial projection onto the unit hyperboloid of the 
piecewise straight motion in the polywedge defined by the walls. 
The straight motion thereby becomes a
geodesic motion on hyperbolic space. The ``cushions'' of the
hyperbolic billiard table are the intersections of the hyperplanes
\Ref{cushions} with the unit hyperboloid, such that the billiard
motion is constrained to be in the region defined as the intersection
of the half-spaces $w_A(\b) \geq 0$ with the unit hyperboloid.
As we already emphasized, not all walls are relevant since some
of the inequalities $w_A(\b) \geq 0$ are implied by others
\cite{DH3}. Only the dominant wall forms, in terms of which all
the other wall forms can be expressed as linear combinations with
non-negative coefficients, are relevant for determining the
billiard. Usually, these are the minimal symmetry walls and some of the
$p$-form walls. The billiard region, as a subset of hyperbolic
space, is in general {\it non-compact} because the cushions meet at
infinity (i.e. at a cusp); in terms of the original scale factor
variables $\b$, this means that the corresponding hyperplanes
intersect on the lightcone. It is important that, even when the
billiard is non-compact, the hyperbolic region can have finite volume.

Given the action \Ref{keyaction} with definite spacetime
dimension, menu of fields and dilaton couplings, one can determine
the relevant wall forms and compute the billiards.  For generic
initial conditions, we have the following results, as to which of
the models (\ref{keyaction}) exhibit oscillatory behaviour (finite
volume billiard) or Kasner-like behaviour (infinite volume
billiard)
\begin{itemize}
\item Pure gravity billiards have finite volume for spacetime dimension $D
\leq 10$ and infinite volume for spacetime dimension $D \geq 11$
\cite{DHS}. This can be understood in terms of the underlying
Kac-Moody algebra \cite{DHJN}: as shown there, the system is chaotic
precisely if the underlying indefinite Kac-Moody algebra is hyperbolic.
\item The billiard of gravity coupled to a dilaton always has infinite
volume, hence exhibits Kasner-like behavior  \cite{BK1,AR,DHRW}.
\item If gravity is coupled to $p$-forms (with $p \not = 0$ and $p <D-2$)
{\em without a dilaton} the corresponding billiard has a finite
volume \cite{DH2}. The most prominent example in this class is
$D=11$ supergravity, whereas vacuum gravity in $11$ dimensions
is Kasner-like. The $3$-form is crucial for closing the billiard.
Similarly, the Einstein-Maxwell system in four (in fact any number of)
dimensions has a finite-volume billiard (see \cite{Jantzen0,Leblanc,Weaver}
for a discussion of four-dimensional homogeneous models with Maxwell
fields exhibiting oscillatory behaviour).
\item The volume of the mixed Einstein-dilaton-$p$-form system depends
on the dilaton couplings. For a given spacetime dimension $D$ and
a given menu of $p$-forms there exists a subcritical domain
$\cal D$ in the space of the dilaton couplings, i.e. an open
neighbourhood of the origin $\lambda_p = 0$ such that:
$(i)$ when the dilaton couplings $\lambda_p$ belong to $\cal D$
the general behaviour is Kasner-like, but $(ii)$ when the $\lambda_p$
do not belong to $\cal D$ the behaviour is oscillatory \cite{DH1,DHRW}.
For all the superstring models, the dilaton couplings do not belong to
the subcritical domain and the billiard has finite volume.  Note, however, that
the superstring dilaton couplings are precisely ``critical'', i.e. on the borderline
between the subcritical and the overcritical domain.
\end{itemize}

As a note of caution let us point out that some indicators of chaos
must be used with care in general relativity, because of reparametrization
invariance, and in particular redefinitions of the time coordinate;
see \cite{Cor,Imp} for a discussion of the original Bianchi IX model.

We next discuss the link between the various time coordinates used
in the analysis. When working in the gauge \Ref{tildeN=rho2}, the
basic time coordinate is $T$. Let us see how the other time
coordinates depend on $T$. First, we note that the dynamical
variable $\l$ is asymptotically a linear function of $T$. Indeed,
its conjugate momentum $\p_{\l} <0$ is asymptotically constant, so
that the integration of $d \l/ dT = - \ft12 \p_{\l}$ yields
\be
\l =  - \ft12 \p_{\l} T + const.
\ee
Hence $\r$ is an exponential function of $T$:
\be
\r = \exp \l \, \propto \, \exp (- \ft12 \p_{\l} T)
\ee
{}From this behavior we infer the time dependence of the
intermediate time coordinate $\tau$ by integrating its defining
relation $d \t = \r^2 dT = - (2/ \p_{\l}) \r^2 d\l= - (2/ \p_{\l})
\r d \r $. This yields \be \t = - (1/ \p_{\l}) \r^2 + const. \ee
At this stage, the asymptotic links between $T,\t,\l$ and $\r$ are
similar to the results derived above for a Kasner solution and
does not depend on whether the system is chaotic or non-chaotic.

The situation is more subtle for the proper time $t$ in the
chaotic case (in the non-chaotic case where one settles in a
Kasner regime after a finite number of collisions, nothing is
changed, of course). The proper time is obtained by integrating
\be
dt = \sqrt{g} d\t = \sqrt{g} \r^2 dT
    = - (2/  \p_{\l})    \sqrt{g} \r d\r.
\ee
The main term in the integrand is $\sqrt{g} =\exp( - \r \sigma)$,
where $\sigma \equiv \Sigma_i \g^i$. The quantity
$\sigma $ oscillates chaotically and is difficult to control.  As
argued in \cite{KhalaSinai}, however, one nevertheless gets
(in the four-dimensional case) the  usual Kasner
 relation $\sqrt{g} \sim t$ (and thus $\t \sim \ln t$ and $T \sim \ln \vert \ln t
\vert$) up to subdominant corrections. We refer the interested
reader to \cite{KhalaSinai} for the details.

Concerning the frequency of collisions, note that the billiard
picture makes it clear that the typical time interval between two
collisions is constant as a function of $T$. In other words, the
number of collisions goes like $\l \approx \frac{1}{2} \ln \t \sim \ln \vert \ln t
\vert$.

The hyperbolic billiard description of the (3+1)-dimensional homogeneous
Bianchi IX system was first worked out by Chitre \cite{Chitre}
and Misner \cite{Misnerb}. It was subsequently generalized to
inhomogeneous metrics in \cite{Kirillov1993,IvKiMe94}.
The extension to higher dimensions with perfect fluid sources was
considered in \cite{KiMe}, without symmetry walls. Exterior $p$-form
sources were investigated in \cite{IvMe,Ivashchuk:1999rm} for special
classes of metric and $p$-form configurations. As far as we know,
however, the uniform approach used in the present paper based on a
systematic use of the Iwasawa decomposition of the spatial metric is new.

\section{Kac-Moody theoretic formulation}
\label{KM0} \setcounter{equation}{0} \setcounter{theorem}{0}
\setcounter{lemma}{0}

Although the billiard description holds for all systems governed
by the action (\ref{keyaction}), the billiard in general has no notable
regularity property. In particular, the dihedral angles between
the faces, which can depend on the (continuous) dilaton couplings,
need not be integer submultiples of $\pi$. In some instances,
however, the billiard can be identified with the fundamental Weyl
chamber of a symmetrizable Kac-Moody (or KM) algebra of indefinite 
type\footnote{Throughout this chapter, we will use the abbreviations
``KM'' for ``Kac-Moody'', and ``CSA'' for Cartan subalgebra.}, with 
Lorentzian signature metric \cite{DH3,DHJN,DdBHS}. Such billiards are
called ``Kac-Moody billiards''. More specifically, in \cite{DH3},
superstring models were considered and the rank $10$ KM algebras
$E_{10}$ and $BE_{10}$ were shown to emerge, in line with earlier
conjectures made in \cite{Julia,Julia2}\footnote{Note that the Weyl
groups of the $E$-family have been discussed in a similar vein in
the context of $U$-duality \cite{LPS,OPR,BFM}.}. This result was further
extended to pure gravity in any number of spacetime dimensions, for
which the relevant KM algebra is $AE_d$, and it was understood that
chaos (finite volume of the billiard) is equivalent to hyperbolicity
of the underlying Kac-Moody algebra \cite{DHJN}. For pure gravity
in $D=4$ the relevant algebra is the hyperbolic algebra $AE_3$
first investigated in \cite{FF}. Further examples of emergence of
Lorentzian Kac-Moody algebras, based on the models of \cite{BGM,CJLP3},
are given in \cite{DdBHS}.

The main feature of the gravitational billiards that can be associated
with KM algebras is that there exists a group theoretical interpretation
of the billiard motion: the asymptotic BKL dynamics is equivalent (in a sense
to be made precise below), at each spatial point, to the asymptotic
dynamics of a one-dimensional nonlinear $\sigma$-model based on a certain
infinite dimensional coset space $G/K$, where the KM group $G$ and its 
maximal compact
subgroup $K$ depend on the specific model. As we have seen, the walls
that determine the billiards are the dominant walls. For
KM billiards, they correspond to the simple roots of the KM algebra.
Some of the subdominant walls also have an algebraic interpretation in
terms of higher-height positive roots. This enables
one to go beyond the BKL limit and to see the beginnings of a possible
identification of the dynamics of the scale factors {\em and} of all
the remaining variables with that of a non-linear $\sigma$-model
defined on the  cosets of the Kac-Moody group divided its maximal
compact subgroup \cite{DHN2}.

The KM theoretic reformulation will not only enable us to give a
unified group theoretical derivation of the different types of walls
discussed in the preceding section, but also shows that
{\em the $\beta$-space of logarithmic scale factors, in which
the billard motion takes place, can be identified
with the Cartan subalgebra of the underlying indefinite
Kac-Moody algebra.} The various types of walls can thus be
understood directly as arising from the large field limit
of the corresponding $\s$-models. It is the presence of gravity, which
comes with a metric in scale-factor space of Lorentzian signature,
which forces us to consider {\it infinite dimensional} groups
if we want to recover all the walls found in our previous analysis,
and this is the main reason we need the theory of KM algebras.
For finite dimensional Lie algebras we obtain only a subset of
the walls: one of the cushions of the associated billiard is missing, 
and one always ends up with a monotonic
Kasner-type behavior in the limit $t\rightarrow 0^+$ (for instance,
the non-diagonal Kasner solution discussed in chapter~4 has only
symmetry walls corresponding to the finite dimensional coset
space $GL(d,\Rn)/SO(d)$). The absence of chaotic oscillations
for models based on finite dimensional Lie groups is consistent 
with the classical
integrability of these models. While they remain formally integrable
for infinite dimensional KM groups, one can understand the
chaotic behavior as resulting from the projection of a motion in
an infinite dimensional space onto a finite dimensional subspace.

Before proceeding we should like to emphasize that the equivalence
between the models discussed in the foregoing sections and the KM
$\sigma$-models to be presented in this section has so far only
been established, in the case of general models, for their
asymptotic dynamics. A proposal relating the infinitely many
off-diagonal degrees of freedom arising in the KM $\sigma$-model
to spatial gradients of the metric and other fields, as well as
possibly other degrees of freedom has recently been made in
\cite{DHN2}. The first steps of this proposal have been explicitly
checked for the relation between $D=11$ supergravity and the
$\E$ $\s$-model. The relevance of non-linear $\s$-models
for uncovering the symmetries of $M$-theory has also been 
discussed from a different, spacetime-covariant point of view in
\cite{West,SWest,SWest2}, but there it is $E_{11}$ rather than $\E$ 
that has been proposed as a fundamental symmetry.

\subsection{Some basic facts about KM algebras}

We first summarize some basic results from the theory of KM
algebras, referring the reader to \cite{Kac,MP} for comprehensive
treatments. As explained there, every KM algebra $\frakg \equiv
\frakg (A)$ is defined by means of an integer-valued Cartan matrix
$A$ and a set of generators and relations \cite{Kac,MP}. We shall
assume that the Cartan matrix is symmetrizable since this is the
case encountered for cosmological billiards.  The Cartan matrix
can then be written as ($i, j =1,\dots r $, with $r$ denoting the
{\it rank}
 of   $\frakg (A)$)
 \be A_{ij} = \frac{2   \langle
\a_i | \a_j \rangle }{\langle \a_i | \a_i \rangle }
\ee
where $\{\a_i  \}$ is a set of $r$
simple roots, and  where the angular brackets denote the invariant
symmetric bilinear form of $\frakg (A)$ \cite{Kac}. Here the bilinear
form acts on the roots, which are linear forms on the Cartan subalgebra
(CSA) $\frakh\subset\frakg (A)$. The generators, which are also 
referred to as Chevalley-Serre generators, consist of triples 
$\{h_i, e_i, f_i\}$ with $i=1,\dots,r$, and for each $i$ form 
an $sl(2,\Rn)$ subalgebra. The CSA $\frakh$ is then spanned by the elements
$h_i$, so that 
\be 
[h_i, h_j] = 0 
\ee 
Furthermore, 
\be 
[e_i , f_j]
= \d_{ij} h_j \ee and \be [h_i, e_j] =  A_{ij} e_j \; , \quad
[h_i, f_j] = - A_{ij} f_j 
\ee
so that the value of the linear form $\a_j$, corresponding to 
the raising operator $e_j$, on the element $h_i$  of the 
preferred basis $\{h_i \}$ of  $\frakh$ is $\a_j(h_i) = A_{ij}$.  
More abstractly, and independently of the choice
of any basis in the CSA,  the roots appear as eigenvalues of the
adjoint action of any element $h$ of the CSA on the raising ($e_i$) or
lowering  ($f_i$) generators: $[h,e_i] = + \a_i(h) e_i$,
$[h,f_i] = - \a_i(h) f_i$. Last but not least we have the 
so-called Serre relations
 \be 
{\rm ad \,}(e_i)^{1-A_{ij}} \big(e_j\big) = 0 \; ,
\quad {\rm ad \,}(f_i)^{1-A_{ij}} \big(f_j\big) = 0 
\ee 

Every KM algebra possesses the triangular decomposition 
\be 
\frakg (A) = \frakn^- \oplus \frakh \oplus \frakn^+ 
\ee 
where $\frakn^+$ and
$\frakn^-$, respectively, are spanned by the multiple commutators
of the $e_i$ and $f_i$ which do not vanish on account of the Serre
relations and the Jacobi identity. To be completely precise,
$\frakn^+$ is the quotient of the free Lie algebra generated by
the $e_i$'s by the ideal generated by the Serre relations ({\it
idem} for $\frakn^-$ and $f_i$). In more mundane terms, when the
algebra is realized, in a suitable basis, by infinite dimensional
matrices, $\frakn^+$ and $\frakn^-$ simply consist of the
``nilpotent'' matrices with nonzero entries only above or below
the diagonal. Exponentiating them formally, one obtains infinite
dimensional matrices again with nonzero entries above or below the
diagonal, and 1's on the diagonal -- as exemplified in the finite
dimensional case by the matrices $\cn$ in chapter 4.

One of the main results of the theory is that, for positive definite
$A$, one just recovers from these relations Cartan's list of finite
dimensional Lie algebras (see e.g. \cite{Humphreys} for a clear
introduction). For non positive-definite $A$, on the other hand, the 
associated KM algebras are infinite dimensional. If $A$ has only one 
zero eigenvalue (with all other eigenvalues strictly positive) one
obtains the so-called affine algebras, whose structure and properties
are rather well understood \cite{Kac,GO}. For indefinite $A$
({\it i.e.} at least one negative eigenvalue of $A$), on the other
hand, very little is known, and it remains an outstanding problem
to find a manageable representation for them \cite{Kac,MP}
(see also \cite{GN} for a physicist's introduction). In particular,
there is not a single example of an indefinite KM algebra for which
the root multiplicities, {\it i.e.} the number of Lie algebra elements
associated with a given root, are known in closed form. The scarcity of
results is even more acute for the ``Kac-Moody groups'' obtained by
formal exponentiation of the associated Lie algebras.
In spite of these caveats, we will proceed formally, making sure
that our definitions reduce to the standard formulas in the truncation
to finite dimensional Lie groups, and more generally remain well defined
when we restrict the number of degrees of freedom to any finite subset.

In the remainder we will thus assume $A$ to be {\it Lorentzian}, i.e. 
non-degenerate and indefinite, with one negative eigenvalue. We shall 
see that this choice is physically motivated by the fact that 
the negative eigenvalue can be associated with the conformal 
factor which, as we saw, is the one degree of freedom making the 
reduced Einstein action unbounded from below. As a special, and 
important case, this class of  Lorentzian KM algebras includes
{\it hyperbolic} KM algebras whose Cartan matrices are such that the
deletion of any node from the Dynkin diagram leaves either a finite
or an affine subalgebra, or a disjoint union of them.

The ``maximal compact'' subalgebra $\frakk$ is defined as the
invariant subalgebra of $\frakg (A)$ under the standard Chevalley
involution, {\it i.e.} \be \theta (x) = x  \quad {\rm for} \; \,
{\rm all} \;\, x\in\frakk \ee with \be \theta (h_i) = - h_i \; ,
\quad \theta (e_i) = - f_i \; , \quad \theta (f_i) = - e_i \ee
More explicitly, it is the subalgebra generated by multiple
commutators of $(e_i - f_i)$. For finite dimensional $\frakg (A)$,
the inner product induced on the maximal compact subalgebra
$\frakk$ is negative-definite, and the orthogonal complement to
$\frakk$ has a positive definite inner product.  This is not so,
however, for indefinite $A$ (see the footnote on page 438 of
\cite{JN}).

It will be convenient in the following to introduce the operation of 
{\it transposition} acting on any Lie algebra element $E$ as 
\be
E^T := - \theta(E)
\ee
In this notation the relations above become
$ h_i^T = h_i,  e_i^T = f_i, f_i^T = e_i $,
and the subalgebra $\frakk$ is generated by the ``anti-symmetric'' elements
satisfying $E^T = - E$. After exponentiation, the elements of the
corresponding maximally compact subgroup $K$ formally appear as
``orthogonal matrices'' obeying $k^T = k^{-1}$.

Sometimes it is convenient to use a so-called Cartan-Weyl basis
for $\frakg(A)$. Using Greek indices $\mu,\nu,\dots$ to label the
root components corresponding to an arbitrary basis $H_{\m}$ in
the CSA, with the usual summation convention and a Lorentzian
metric $G_{\mu\nu}$ for an indefinite $\frakg$, we have $h_i :=
\a_i^\mu H_\mu$, where $ \a_i^\mu$ are the ``contravariant
components'', $G_{\mu\nu} \a_i^\n \equiv \a_{i \, \m}$, of the
simple roots $\a_i$ ($i=1,\dots r$), which are linear forms on the
CSA, with usual ``covariant components'' defined as $ \a_{i \, \m}  
\equiv \a_i(H_\m)$.

To an arbitrary root $\a$ corresponds a set of Lie-algebra generators
$E_{\a ,s}$, where $s=1, \dots, \mult (\a)$ labels the (in general) 
multiple Lie-algebra elements associated with $\a$. The root multiplicity
$\mult (\a)$ is always one for finite dimensional Lie algebras, and 
is also one for the positive norm roots 
$ \a^2 \equiv  \langle \a | \a \rangle > 0$ (i.e. the ``real
roots'') of general KM algebras (including, of course, the simple roots),
but generically exhibits exponential growth as a function of $ -\a^2$
for indefinite $A$. In this notation, the remaining Chevalley-Serre 
generators are given by $e_i := E_{\a_i}$ and $f_i := E_{-\a_i}$.
Then we have
\be\label{HwE}
[ H_\mu , E_{\a,s} ] = \a_\mu E_{\a,s}
\ee
and
\be
[E_{\a,s} , E_{\a',t}] = \sum_u c_{\a\a'}^{s,t,u} E_{\a + \a',u}
\ee
The elements of the Cartan-Weyl basis are normalized such that
\be\label{CWnormalization}
\langle H_\mu | H_\nu \rangle = G_{\mu\nu} \; , \quad
\langle E_{\a,s} | E_{\b,t} \rangle = \d_{st} \d_{\a + \b, 0}
\ee
where we have assumed that the basis satisfies $E^T_\a  = E_{-\a}$.
Let us finally recall that the Weyl group of a KM algebra is the discrete group
generated by reflections in the hyperplanes orthogonal to the simple roots.

\subsection{Decomposition of $AE_3$ into $ SL(3)$ representations}

As a special example, let us consider the hyperbolic KM algebra
$AE_3$, with Cartan matrix\footnote{This algebra is also called
$\cal F$ \cite{FF}, $H_3$ \cite{Kac},  $HA_1^{(1)}$ \cite{Kang} or
$A_1^{\wedge \wedge}$ \cite{DHJN}.} 
\be\label{AE3} 
\hspace{1.5cm}
A_{ij} = \left( \matrix{2 &-1 & 0 \cr
                                     -1 & 2 & -2 \cr
                                     0 & -2 & 2  \cr} \right)
\ee
and work out the first few terms of its decomposition into $SL(3,\Rn)$ 
representations. This decomposition refers to the {\it adjoint action} 
of the   $sl(3,\Rn)$  subalgebra defined below on the complete KM algebra.

{}From the mathematical point of view, the algebra (\ref{AE3}) was
first studied in \cite{FF} as it is the simplest hyperbolic KM algebra
containing a non-trivial affine subalgebra (see also \cite{Kang} and
references therein for more recent work). From the physical perspective
this algebra is of special interest for pure gravity in four dimensions both
because  its regular subalgebras and its Chevalley-Serre generators 
can be physically identified with known symmetry groups arising in 
dimensional reductions of general relativity \cite{Nic}, and because 
its Weyl chamber is related to the original BKL billiard \cite{DHJN} 
(see footnote \ref{foot11}).  First of all, the $SL(2,\Rn)$ subgroup
corresponding to the third diagonal entry of $A_{ij}$ can be identified 
with the Ehlers group \cite{Ehlers} which acts on solutions of Einstein's
equations with one Killing vector. The affine subgroup corresponding
to the submatrix
\be\label{Geroch}
\hspace{1.5cm} \left( \matrix{2 &-2\cr
                              -2 & 2  \cr} \right)
\ee
is the Geroch group acting on solutions of Einstein's equations
with two commuting Killing vectors \cite{Geroch} (axisymmetric
stationary, or colliding plane wave solutions, respectively), see e.g.
\cite{Julia,Julia1,BM,Nic1} and references therein. Both the Ehlers
group and the Geroch group act in part by {\it nonlocal} transformations
on the metric (or vierbein) components.

A crucial role, in our analysis, is played by the $SL(3,\Rn)$ subgroup
generated by $(e_1,f_1,h_1)$ and $(e_2,f_2,h_2)$, corresponding to the 
submatrix
\be\label{SL3}
\hspace{1.5cm} \left( \matrix{2 &-1\cr
                             -1 & 2  \cr} \right)
\ee
This group can be realized as acting on the spatial components of 
the metric (or vierbein), as an extension of the so-called 
Matzner-Misner $SL(2,\Rn)$ group for solutions with two
commuting Killing vectors.

The billiard picture for pure gravity in four dimensions can be
readily understood in terms of the Weyl group of $AE_3$
\cite{DHJN}. For $SL(3,\Rn)$, which has two simple roots, the Weyl
group is the permutation group on three objects. The two
hyperplanes orthogonal to the simple roots of $SL(3,\Rn)$ can be
identified with the symmetry walls encountered in section 6.1. The
third simple root extending (\ref{SL3}) to the full rank 3 algebra
(\ref{AE3}), which ``closes off'' the billiard, can then be
identified with curvature wall orthogonal to $\a_{123}$
\cite{DHJN}. Readers may indeed check that the scalar products
between the two symmetry wall forms and the curvature wall form
reproduce the above Cartan matrix\footnote{\label{foot11} Note
that in the original analysis of \cite{BKL,BKL2,Chitre,Misnerb},
the symmetry walls are not included; the KM algebra that arises
has the $3 \times 3$ Cartan matrix
$$
\hspace{1.5cm} A_{ij} = \left( \matrix{2 & -2 & -2 \cr
                                     -2 & 2 & -2 \cr
                                     -2 & -2 & 2  \cr} \right)
$$
and its fundamental Weyl chamber (radially projected on the hyperbolic
plane $H_2$) is the ideal equilateral triangle having its three
vertices at infinity.}.

Let us now expand the nilpotent subalgebra $\frakn^+$ in terms
of representations of the $A_2 \equiv sl(3,\Rn)$ subalgebra of $AE_3$
exhibited in \Ref{SL3}. For this purpose, given any root $\a$,
we define its $ sl(3,\Rn)$ level $\ell$ to be the number of times the
root $\a_3$ appears in it, to wit
\be\label{root}
\a = m \a_1 + n \a_2 + \ell \a_3
\ee
Note that this notion of level is different from the affine level ($\equiv m$)
which counts the number of appearances of the over-extended root
$\a_1$ \cite{FF}. The hyperbolic algebra is thus decomposed
into an infinite tower of irreducible representations of its
$sl(3,\Rn)$ subalgebra. Such a decomposition is simpler than one
in terms of the affine subalgebra, whose representation theory
is far more complicated (and, unlike that of $sl(3,\Rn)$, only
incompletely understood).

After inclusion of the third Cartan generator $h_3$, the
level $\ell =0$ sector is just a $gl(3,\Rn)$ subalgebra with
generators ${K^i}_j$ (where $i,j =1,2,3$) and commutation relations
\be
[{K^i}_j , {K^k}_l ] =  \d^k_j {K^i}_l -\d^i_l  {K^k}_j
\ee
These Lie algebra elements will be seen to generate the $GL(3,\Rn)$ 
group acting on the spatial components of the vierbein. The restriction 
of the $AE_3$-invariant bilinear form to the level-0 sector is
\be
\label{KK}
\langle {K^i}_j | {K^k}_l \rangle = \d^i_l \d^k_j - \d^i_j \d^k_l
\ee
The identification with the Chevalley-Serre generators is
\beq
e_1 &=& {K^1}_2 \;\; ,  \quad f_1 = {K^2}_1 \;\; ,  \quad
       h_1 = {K^1}_1 - {K^2}_2  \nn\\
e_2 &=& {K^2}_3 \;\; ,  \quad f_2 = {K^3}_2 \;\; ,  \quad
       h_2 = {K^2}_2 - {K^3}_3  \nn\\
h_3 &=& -  {K^1}_1 - {K^2}_2 + {K^3}_3
\eeq
manifestly showing how the over-extended CSA generator $h_3$ enlarges the
original  $sl(3,\Rn)$  generated by $(e_1,f_1,h_1)$ and  $(e_2,f_2,h_2)$  
to the Lie algebra $gl(3,\Rn)$. The CSA generators are related to
the ``central charge'' generator $c$ by
\be\label{centralcharge}
c= h_2 + h_3 = - {K^1}_1
\ee
This is indeed the expected result: in the reduction of gravity to
{\it two} dimensions (i.e. with two commuting Killing vectors), the
central charge does not act on the internal degrees of freedom, but
as a scaling on the conformal factor \cite{Julia1,BM,Nic1} (here
realized as the 1-1 component of the vierbein). The affine level
counting operator $d$ is given by \cite{FF}
\be\label{affinelevel}
d =  h_1 + h_2 + h_3 = - {K^2}_2
\ee
and in contradistinction to $c$ it does act on the internal volume
in the reduction to two dimensions (i.e. the ``dilaton'' $\rho$ in
the notation of \cite{BM,Nic1}). The operator $d$ may be viewed
as the $L_0$ operator of a full Witt-Virasoro algebra enlarging
the Geroch group via a semidirect product \cite{JN}.

The irreducible representations of $SL(3,\Rn)$ are most conveniently
characterized by their Dynkin labels.  Let us recall that these labels
are non-negative integers which characterize any highest-weight 
representation of any finite-dimensional Lie algebra. Let us consider 
a representation with maximal vector $v_\Lambda$, of (highest) weight 
$\Lambda$, i.e. such that for any element $h$ in the CSA,
$h (v_\Lambda) = \Lambda(h) v_\Lambda$, where the Lie algebra 
acts on the representation vector space (or ``module''), and where  
$\Lambda$ is a linear form on the CSA. The Dynkin labels of this 
representation are defined as
$ p_i(\Lambda) \equiv 2 \langle \a_i | \Lambda \rangle / \langle 
\a_i | \a_i \rangle $, where $\a_i$ are simple roots of the  
Lie algebra under consideration.Note that in the case of simply 
laced algebras such as  $sl(3,\Rn)$, and $AE_3$,
the simple  roots have all a squared length equal to 2, so that 
the definition of the labels reduces to $ p_i(\Lambda) = 
\langle \a_i | \Lambda \rangle $ (while the Cartan matrix similarly 
simplifies to $A_{ij} = \langle \a_i | \a_j \rangle $).
In the case of $sl(3,\Rn)$ we have two simple roots, $\a_1$ and $\a_2$,
and therefore two Dynkin labels:  $ p_1(\Lambda) = 
\langle \a_1 | \Lambda \rangle $, $ p_2(\Lambda) = 
\langle \a_2 | \Lambda \rangle $.  In terms of the Young
tableau description of  $sl(3,\Rn)$ representations, the first Dynkin label
$p_1$ counts the number of columns having  two boxes, while
$p_2$ counts the number of columns having only one box.
For instance, $(p_1,p_2) = (1,0)$ labels an antisymmetric two-index tensor,
while $(p_1,p_2) = (0,2)$ denotes a symmetric two-index tensor.
Note also that the dimension of the  $sl(3,\Rn)$  representation described by
the labels $(p_1,p_2)$  is $ ( p_1 +1) ( p_2 +1) ( p_1 +p_2 +2) / 2 $.

Let us determine the representations of $sl(3,\Rn)$ appearing at the
first level ($\ell =1$) of $AE_3$ \footnote{This
analysis is analogous to the one performed in \cite{FF} for the
affine subalgebra $A_1^{(1)}$.} At this level,  one  sees that,
under the adjoint action  of $sl(3,\Rn)$, i.e. of $(e_1,f_1,h_1)$ and    
$(e_2,f_2,h_2)$, the extra Chevalley-Serre generator $f_3$ is a 
maximal vector. Indeed,
\beq
e_1 (f_3) \equiv [e_1, f_3] &=& 0 \non
e_2 (f_3) \equiv [e_2, f_3] &=& 0
\eeq
  The weight of $f_3$
is $\Lambda = -\a_3$ because, by definition 
$h (f_3) \equiv [h,f_3] = -\a_3(h) f_3$.
Hence the Dynkin labels of the representation built on the maximal
vector $f_3$ are $p_i = -   \langle \a_i | \a_3 \rangle = - A_{i3} $ 
with $i=1,2$. Explicitly, this gives $(p_1,p_2) = (0,2)$. These labels 
also show up as the eigenvalues of the actions of the 
(specially normalized) Cartan generators
$(h_1,h_2)$ on the maximal vector  $f_3$, namely:
\beq
h_1 (f_3)\equiv [h_1, f_3] &=& 0 \non
h_2 (f_3)\equiv [h_2, f_3] &=& 2  f_3
\eeq
As we said above, the representation   $(p_1,p_2) = (0,2)$ corresponds to
a symmetric (two-index) tensor.

Thus, at the levels $\pm 1$ we have $AE_3$ generators which can be 
represented as  symmetric tensors $E^{ij}= E^{ji}$ and $F_{ij}= F_{ji}$. 
One verifies that all algebra relations are satisfied with
($a_{(ij)} \equiv (a_{ij} +a_{ji}) /2$)
\beq\label{AE3com}
[{K^i}_j, E^{kl}] &=& \d^k_j E^{il} + \d^l_j E^{ki} \non
{}[{K^i}_j, F_{kl}] &=&  - \d^i_k F_{jl} - \d^i_l F_{kj} \non
{}[E^{ij} , F_{kl} ] &=& 2 \d^{(i}_{(k} {K^{j)}}_{l)} -
  \d^{(i}_{k} \d^{j)}_{l} \big({K^1}_1 + {K^2}_2 + {K^3}_3 \big)\non
\langle F_{ij} | E^{kl}\rangle &=&  \d^{(k}_{i} \d^{l)}_{j}
\eeq
and the identifications
\be
e_3 = E^{33} \;\; , \quad f_3 = F_{33}
\ee

As one proceeds to higher levels, the classification of $SL(3,\Rn)$
representations becomes rapidly more complicated due to the exponential
increase in the number of representations with level $\ell$.
Generally, the representations that can occur at level $\ell + 1$
must be contained in the product of the level-$\ell$ representations
with the level-one representation $(0,2)$. Working out these products
is elementary, but cumbersome. Moreover, many of the representations
constructed in this way will drop out (i.e. not appear as elements
of the $AE_3$ Lie algebra). The complications are not yet visible
at low levels; for instance, the level-two generator
$ E^{ab | jk} \equiv \varepsilon^{abi} {E_i}^{jk}$,
with labels $(1,2)$,
is straightforwardly obtained by commuting two level-one elements
\be
[ E^{ij} , E^{kl} ] =  \varepsilon^{mk(i} {E_m}^{j)l} +
                       \varepsilon^{ml(i} {E_m}^{j)k}
\ee
A more economical way to sift the  relevant representations
is to work out the relation between Dynkin labels and the associated
highest weights, using the fact that the highest weights of the
adjoint representation are the roots.  More precisely, the maximal
vectors being (as exemplified above at level 1) of the ``lowering type'',
the corresponding highest weights are {\it negative} roots,
say $\Lambda =  -\a$ with $\a$ of
the form (\ref{root}) with {\it non-negative} integers $\ell, m,n$.
Working out the Dynkin labels of $\Lambda =  -\a$ by computing,
the scalar products with the two simple roots of $sl(3,\Rn)$ then yields
\be
p_1 \equiv p = n -2m \;\; , \quad p_2 \equiv q = 2\ell + m - 2n
\ee
As indicated, we shall henceforth use the notation $ (p_1,p_2) \equiv (p,q)$
for the Dynkin labels. This formula is restrictive because all the integers
entering it must be non-negative.

Inverting this relation we get
\beq\label{mn}
m &=& \ft23 \ell - \ft23 p - \ft13 q  \non
n &=& \ft43 \ell - \ft13 p - \ft23 q
\eeq
with $n \geq 2m \geq 0$. A further restriction derives from
the fact that the highest weight must be a root of $AE_3$, viz.
its square must be smaller or equal to 2:
\be\label{Lambda2}
\Lambda^2 = \ft23 \big( p^2 + q^2 + pq - \ell^2 \big) \leq 2
\ee
Consequently, the representations occurring at level $\ell$ must belong
to the list of all the solutions of \Ref{mn} which are
such that the labels $m,n,p,q$ are non-negative integers and
the highest weight $\Lambda$ is a root, i.e. $\Lambda^2 \leq 2$.
These simple diophantine equations/inequalities can be easily
evaluated by hand up to rather high levels.

However, the task is not finished. Although the above procedure
substantially reduces the number of possibilities, it does not
tell us how often (including zero times!) a given representation appears (i.e. its
{\it outer multiplicity}). For this purpose we have to make use
of more detailed information about $AE_3$, namely the root
multiplicities computed in \cite{FF,Kac}. Matching the combined
weight diagrams with the root multiplicities listed in table $H_3$
on page 215 of \cite{Kac}, we obtain the following representations
in the decomposition of $AE_3$ w.r.t. its $sl(3,\Rn)$ subalgebra up
to level $\ell\leq 5$:
\beq
\ell = 1 \quad \rightarrow \quad (p,q) &=& (0,2) \non
\ell = 2 \quad \rightarrow \quad (p,q) &=& (1,2) \non
\ell = 3 \quad \rightarrow \quad (p,q) &=& (2,2) \non
                                       &&  (1,1) \non
\ell = 4 \quad \rightarrow \quad (p,q) &=& (3,2) \non
                                       &&  (1,3) \non
                                       &&  (2,1)
        \quad ({\rm occurs} \;{\rm twice}) \non
                                       &&  (0,2) \non
                                       &&  (1,0) \non
\ell = 5 \quad \rightarrow \quad (p,q) &=& (4,2) \non
                                       &&  (2,3)
        \quad ({\rm occurs} \;{\rm twice}) \non
                                       &&  (0,4)
        \quad ({\rm occurs} \;{\rm twice}) \non
                                       &&  (3,1)
        \quad ({\rm occurs} \;{\rm three} \; {\rm times}) \non
                                       &&  (1,2)
        \quad ({\rm occurs} \;{\rm four} \; {\rm times}) \non
                                       &&  (2,0)
         \quad ({\rm occurs}\;{\rm three} \; {\rm times}) \non
                                       &&  (0,1)
        \quad ({\rm occurs} \;{\rm twice})
\eeq
(we have not listed the representations compatible with \Ref{mn} and 
\Ref{Lambda2}, which drop out, i.e. have outer multiplicity zero).
Going to yet higher levels will require knowledge of $AE_3$ root
multiplicities beyond those listed in \cite{FF,Kac,Kang}. There is, 
however, an infinite set of admissible Dynkin labels that can be readily
identified by searching for ``affine'' highest weights for which
$m=0$ in (\ref{root}). They are
\be\label{affine1}
(p,q) = (\ell -1,2) \quad \Longleftrightarrow \quad (m,n) = (0,\ell -1)
\ee
Because the associated highest weights all obey $\Lambda^2 =2$
all these representations (and their ``transposed'' representations)
appear with outer multiplicity one independently of $\ell$. A second
series of affine representations is 
\be
(p,q) = (\ell,0) \quad \Longleftrightarrow \quad (m,n) = (0,\ell)
\ee
Now we have $\Lambda^2 =0$, and the corresponding representations 
have outer multiplicity zero because the corresponding states are already
contained as lower weight states in the rperesentations \Ref{affine1}.

The Lie algebra elements corresponding to \Ref{affine1} are thus given by 
the two conjugate infinite towers of $sl(3,\Rn)$ representations 
(for $\ell= 1,2,3,\dots$)
\be\label{affine2}
{E_{i_1 \cdots i_{\ell-1}}}^{jk} \;\; , \quad
{F^{i_1 \cdots i_{\ell-1}}}_{jk}
\ee
The affine representations are distinguished because they contain
the affine subalgebra $A_1^{(1)}\subset AE_3$. This embedding
was already studied in \cite{FF}, but in view of future applications
and because the identification is a little subtle we here spell out some
more details. The affine subalgebra is identified by requiring
its elements to commute with the central charge $c$ (\ref{centralcharge}).
>From this requirement we infer that we must truncate the full
algebra to the subalgebra generated by the elements ${K^i}_j$ for
$i,j \in \{ 2,3\}$ and those generators in (\ref{affine2})
with $i_1 = \dots i_{\ell -1} =1$ and $j,k \in \{ 2,3\}$.
In physicists' notation, the affine subalgebra is spanned by the current
algebra generators $\{T^-_m, T^3_m, T^+_m\}$ and the central charge $c$.
The affine level $m\in\ZZ$ is the eigenvalue of the affine
level counting operator $d$: $[d, T_m] = m T_m$.
We therefore conclude that the affine level zero
sector\footnote{Recall that the affine level $m$ must not be
confused with the level $\ell$ used in our decomposition of
$AE_3$, cf. the definition (\ref{root}).} consists of the
generators of the Ehlers $SL(2,\RR)$ group, viz.
\be
T^-_0 = f_3 \;\; (\equiv F_{33}) \;\; , \quad
T^3_0 = h_3 \;\; , \quad T^+_0 = e_3 \;\; (\equiv E^{33})
\ee
At affine level $m=-1$ we have
\beq
T^-_{-1} &=& e_2 \;\; (\equiv {K^2}_3) \non
T^3_{-1} &=& [e_2,e_3] = 2 E^{23}  \non
T^+_{-1} &=& [[e_2,e_3],e_3] = 2 [E^{23}, E^{33}] = 2 {E_1}^{23}
\eeq
Similarly, at affine level $m=+1$ we get
\beq
T^-_1 &=& [[f_2,f_3],f_3] = 2 [F_{23}, F_{33}] = 2 {F^1}_{23} \non
T^3_1 &=& [f_2,f_3] = F_{23}  \non
T^+_1 &=& f_2 \;\; (\equiv {K^3}_2)
\eeq
and so on for higher affine levels. The precise identification
of the affine subalgebra is of crucial importance for understanding
the embedding of the Geroch group into the full hyperbolic algebra.

Other cases of interest, where similar decompositions can be
worked out include the indefinite Kac-Moody algebras $E_{10}$
and $E_{11}$, both of which have been conjectured to appear
in $D\!=\!11$ supergravity and M Theory, see \cite{Julia,Julia2,DHN2},
and \cite{West}, respectively. The first six rungs in the the $sl(10,\Rn)$ 
level decomposition of  $E_{10}$ have been worked out in \cite{DHN2},
and will be extended to higher levels, as well as to $E_{11}$, in \cite{FN}.

\subsection{Nonlinear $\sigma$-Models in one dimension}

Notwithstanding the fact that we know even less about the ``groups''
associated with indefinite KM algebras we will formulate nonlinear
$\s$-models in one time dimension and thereby provide an
effective and unified description of the asymptotic BKL dynamics
for several physically important models.
The basic object of interest is a one-parameter dependent KM group element
 $\cv =\cv(t)$, which is assumed to be an element
of the coset space $G/K$, where $G$ is the group obtained by formal
exponentiation of the KM algebra $\frakg$, and $K$ its maximal
compact subgroup, which is again obtained by formal exponentiation
of the associated maximal compact subalgebra $\frakk$ defined above.
[In the ``transpose'' notation defined above, the group $K$ is the
group of ``orthogonal elements'': $k^T = k^{-1}$.]
For finite dimensional $\frakg(A)$ our definitions reduce to the usual
ones, whereas for indefinite KM algebras they are formal constructs
to begin with. [Formal constructs similar to our objects $\cv$, 
$\dot\cv\cv^{-1}$, etc. have been used in somewhat 
different settings  in \cite{CJLP1,CJLP2,West}.]
However, to ensure that our definitions are meaningful
operationally, we will make sure at every step that any finite truncation
of the model is well defined and can be worked out explicitly in a
finite number of steps.

In physical terms, $\cv$ can be thought of as an extension of the
vielbein of general relativity, with $G$ and $K$ as generalizations
of the $GL(d,\Rn)$ and local Lorentz symmetries of general relativity.
For infinite dimensional $G$, the object $\cv$ thus becomes an
``$\infty$-bein''.  It is then natural to associate to this vielbein a
``metric'', viz.
\be
\cm := \cv^T \cv
\ee
which is invariant under the  left action ( $\cv \to  k \cv$)
of the ``Lorentz group'' $K$ (actually the truncation of  $\cm$ to the
relevant $GL(n,\Rn)$ subgroup of the KM algebras entering the models
we study turn out to correspond to the matrix defined by the
{\it contravariant} components, $g^{ij}$, of the spatial metric
used in section 4 above). Exploiting this invariance, 
we can formally bring $\cv$ into a ``triangular gauge''
\be\label{Iwasawa3}
\cv = \ca \cdot \cn \Longrightarrow \cm = \cn^T \ca^2 \cn
\ee
where the abelian part $\ca$ belongs to the exponentiation of the CSA,
and the nilpotent part $\cn$ to the exponentiation  of $\frakn^+$
thus recovering the formulas which we already used in sections 4.1 and 4.2.
This formal Iwasawa decomposition, which is an infinite-dimensional 
generalization of the one we used before, can be made fully explicit 
by decomposing $\ca$ and  $\cn$ in terms of bases of  
$\frakh$ and $\frakn^+$:
\beq\label{Iwasawa4}
\ca (t) &=& \exp \big( \b^\mu (t) \, H_\mu \big) \; , \non
\cn (t) &=& \exp \Big( \SD \Smult \nu_{\a ,s} (t) \, E_{\a , s}\Big)
\eeq
where $\Delta_+$ denotes the set of positive roots. The components    
$ \b^\mu $, parametrizing a generic element in the CSA  $\frakh$,
will turn out to be in direct correspondence with the logarithmic scale factors
 $ \b^\mu= (\b^a, \phi)$ introduced in section~3. We anticipated 
this correspondence by using the same notation. [As explained above, 
the apparently ``wrong sign'' of the exponents of $\ca$ is due to the 
fact that $\cm$ will correspond to the inverse of the spatial metric.]
The main technical difference with the kind of Iwasawa decompositions 
used in the foregoing sections is that now the matrix $\cv (t)$
is infinite dimensional for indefinite $\frakg (A)$. Observe that
for finite dimensional matrices, there was no need to worry
about root multiplicities, as these are always one. By contrast,
there are now infinitely many $\nu$'s, and consequently $\cn$ contains
an infinite tower of new degrees of freedom. Next we define
\be\label{Ndot}
\dot \cn  \cn^{-1} =  \SD\Smult j_{\a ,s} E_{\a,s} \quad \in \frakn^+
\ee
with
\be
j_{\a ,s} = \dot \nu_{\a ,s} + `` \n \dot \n + \n \n \dot \n  + \cdots ''
\ee
(we put quotation marks to avoid having to write out the indices).
To define a Lagrangian we proceed in the usual way. First we consider 
the quantity
\be
\dot\cv\cv^{-1} = \dot\b^\mu H_\mu +
    \SD\Smult \exp\big(\a (\b)\big)  j_{\a ,s} E_{\a,s}
\ee which has values in the Lie algebra $\frakg (A)$. Here we have
set \be \a (\b) \equiv \a_\mu \b^\mu \ee for the value of the root
$\a$ ( $\equiv$ linear form) on the CSA element  $\b = \b^\m
H_\m$. 
Next we define 
\beq 
P&:=& \frac12 \left( \dot\cv\cv^{-1} +
(\dot\cv\cv^{-1})^T \right) \nonumber \\ 
&=&   \dot\b^\mu H_\mu +
\frac12 \SD\Smult j_{\a ,s} \exp\big(\a (\b)\big) (E_{\a,s} + E_{-\a,s})
\quad  \in \frakg \ominus \frakk \nonumber
\eeq
where we arranged the basis so that $E_{\a,s}^T =  E_{-\a,s}$. 
Then we define a KM-invariant $\s$-model action as $\int dt \cl$ 
where the Lagrangian is defined by using the
KM-invariant bilinear form  $\langle.|.\rangle$ (cf. \Ref{CWnormalization})
\beq
\cl &=& \frac12 n^{-1} \langle P | P \rangle \non
&=& n^{-1} \Big(\frac12 G_{\m\n}  \dot\b^\mu \dot\b^\nu +
   \frac14 \SD\Smult \exp\big(2\a(\b)\big) j_{\a,s} j_{\a,s}\Big) \label{Lagrangian}
\eeq
Here the Lorentzian metric $ G_{\m\n}$ is the restriction of the invariant
bilinear form to the CSA. It can be identified with the metric in
the space of the scale factors, which is why we adopt the same
notation.  We have introduced in this Lagrangian a ``lapse function''
$n$ (not to be confused with the lapse function $N$ introduced before), 
which ensures that our formalism is invariant under  reparametrizations
of the time variable. Finally, we can simply describe our Lagrangian 
\Ref{Lagrangian} as that of a {\it null geodesic} in the coset space $G/K$.

Taking the algebra $AE_3$ as an example, this Lagrangian contains
the Kasner Lagrangian (\ref{KasnerAction}) (without dilaton)  as a 
special truncation. More specifically, retaining only the level zero 
fields (corresponding to a $GL(3,\Rn)/O(3)$ $\s$-model)
\be
\cv(t) \Big|_{\ell =0} =   \exp (  {h^a}_b(t)  {K^b}_a )
\ee
and defining from $ {h^a}_b$ a vielbein by matrix exponentiation
$ {e^a}_b \equiv {(\exp h)^a}_b$, and a corresponding contravariant metric
$g^{ab} = {e^a}_c {e^b}_c$,  one checks that  the bilinear form (\ref{KK})
reproduces (half) the Lagrangian (\ref{KasnerAction}). [This computation
works more generally for any $GL(n,\Rn)$,  and was already used in
\cite{DHN2} in the $GL(10,\Rn)$  decomposition of $\E$.] The level-0 
sector is thus associated with the Einsteinian dynamics of a  
spatial dreibein depending only on time.

Let us then consider the fields $\phi_{ij}$ associated with the 
level-one generators $E^{ij}$. This leads to a truncation of our 
KM-invariant $\s$-model to the levels $\ell = 0,1$, i.e.
 \be
\cv(t) \Big|_{\ell =0,1} = \exp ({h^a}_b(t) {K^b}_a ) \exp(\phi_{ab} E^{ab})
\ee
In the gauge $n=1$, the Lagrangian now has the form
$\cl \sim (g^{-1} \dot g  )^2 + g^{-1} g^{-1} \dot \phi \dot \phi$,
where $g$ denotes the {\it covariant} metric $g_{ij}$.
As the $\phi_{ij}$'s enter only through their time derivatives, their 
conjugate momenta $\Pi^{ij}$ are constants of the motion. Eliminating 
the $\phi$'s in favour of the constant momenta $\Pi$ by a partial
Legendre transformation yields the reduced Lagrangian (for the dynamics 
of $g_{ij}(t)$) of the form 
$\cl \sim (g^{-1} \dot g  )^2 - g^{+1} g^{+1} \Pi \Pi$. In other 
words, the elimination of the $\phi$'s has generated a potential 
$ V_{\phi}= V_{\phi}(g)$ (in the gauge $n=1$)
\be
V_{\phi}(g) \propto + g_{ij} g_{kl} \Pi^{ik} \Pi^{jl}
\ee
It is then easy to check that this potential can be identified
with the leading (weight-2) gravitational potential in \Ref{VsubG} 
(corresponding to the gauge $\tilde{N} =1$), namely
$ \cv_G^{\rm leading} \equiv (-g R)^{\rm leading} =
+ \frac{1}{4} g g_{ai}g^{bj}g^{ck}  C^a_{\; \; bc}  C^i_{\; \; jk} $ 
under the identification of the structure constants ${C^i}_{jk}$ 
with the momenta conjugate to $\phi_{ij}$:
\be
\Pi^{ij} \propto \varepsilon^{kl(i} {C^{j)}}_{kl}
\ee
Note that the trace ${C^j}_{jk}$ drops out of this relation, and 
that $\Pi^{ij}$ is of weight one, like $\varepsilon^{ijk}$.

Consequently, when neglecting the subleading gravitational walls  
$ \propto F_a$in \Ref{VsubG} and in four spacetime dimensions, 
the BKL dynamics at each given spatial point $x_0$ is equivalent 
to the $\ell=0, \pm 1$ truncation of the $AE_3$-invariant dynamics 
defined by \Ref{Lagrangian}. The  fields $\phi_{ij}(t)$ parametrizing 
the components of the $AE_3$ coset element along the $\ell =1$ generators 
are canonically conjugate to the structure constants ${C^i}_{jk}(t,x_0)$.

Similarly, one would like to associate the generators (\ref{affine2})
with higher order spatial gradients. However, the proper physical
interpretation of these fields as well as of the other higher level
components remains yet to be found.  In the case of the relation between
supergravity in $D=11$ and the $\E$ coset model one could pursue 
the correspondence between spacetime fields and coset coordinates 
up to the $gl(10,\Rn)$  level $ \ell =3$ included (corresponding 
to height $29$) \cite{DHN2}. The correspondence worked thanks
to several ``miraculous'' agreements between the numerical coefficients
appearing in both Lagrangians.

Varying (\ref{Lagrangian}) w.r.t. the lapse function $n$ gives rise to
the constraint that the coset Lagrangian vanish. Let us define the 
canonical momenta
\be
\pi_\mu := \frac{\d\cl}{\d\dot \b^\mu} = n^{-1} G_{\m\n} \dot\b^\nu
\ee
and  the (non-canonical) momentum-like variables
\be
\Pi_{\a,s} := \frac{\d\cl}{\d j_{\a,s}}
      = \frac12 n^{-1} \exp \big(2\a(\b)\big)j_{\a,s}
\ee
The latter variables are related to the momenta canonically conjugate 
to the $\n_{\a,s}$, i.e. $ p_{\a,s} := \d\cl/\d \dot \n_{\a,s}$, by 
expressions of the form
$\Pi_{\a,s} =  p_{\a,s} + ``\n p + \n \n p +\cdots ''$, 
see next subsection.

In terms of  these variables  the KM-invariant Hamiltonian corresponding
to the above Lagrangian reads
 \be
 \label{HKM}
\ch(\b,\p,\n,p) =  n  \Big(  \frac12  G^{\m\n} \pi_\mu \pi_\nu
      + \SD\Smult \exp\big(-2\a(\b)\big) \Pi_{\a ,s} \Pi_{\a ,s} \Big)
\ee
The constraint of a vanishing Lagrangian becomes that of a 
vanishing Hamiltonian:
\be\label{HC}
\ch \approx 0
\ee
The momentum-like variables $\Pi_{\a,s}$ can be thought of as 
infinite-dimensional generalizations
of the asymptotically frozen
combinations ${P^j}_a {\cn^b}_j$ used in section 5.1, cf.
Eq.(\ref{Nmomenta}). They do not Poisson-commute with one another
in general. Instead one has Poisson brackets of the form
$\{ \Pi_{\a,r} , \Pi_{\b,s} \} = \Omega^{\g,t}_{\a,r \, \, 
\b,s}(\n) \Pi_{\g,t} $.
Note that $\pi_\mu$ is timelike or null because the potential is
manifestly non-negative; the motion in the CSA is timelike --- in fact,
in the limit to be studied below, a broken future-oriented lightlike line.

Because the coefficients of the exponentials in \Ref{HKM} are
non-negative we can now apply exactly the same reasoning as in
chapter 5. The crucial point is that the Cartan variables $\b$ do 
not enter the commutation relations of the $\Pi$ as indicated above.
Therefore, as above, the off-diagonal components $\nu_{\a,s}$ and
the momentum-like variables $\Pi_{\a,s}$ get frozen asymptotically, 
provided all spacelike walls are ``behind the motion" so that they do not
conflict with the BKL limit. As before, we can introduce hyperbolic
polar coordinates for parametrizing the dynamics of the Cartan variables
$\b^\mu =\rho\gamma^\mu$. When taking the limit $\rho\rightarrow\infty$,  
the Hamiltonian \Ref{HC} behaves as
\be\label{HC1}
\ch \sim \ch_{\Delta_+} (\b, \pi) = \frac12   G^{\m\n} \pi_\mu \pi_\nu+
\SD K_\a \exp \big[-2\a(\b)\big]
\ee
with constants $K_\a \geq 0$.
This looks like a Toda Hamiltonian, except that the sum extends
over all roots, rather than only the simple ones, and that the
underlying KM algebra is indefinite, and not just finite or affine
as in standard Toda theory. However, the {\it dominant potential walls}    
in the limit $\r \to \infty$ are given by those terms 
$K_\a \exp \big[-2\a(\b)\big]$ for which $\a$ is a {\it simple root}.
Indeed, by definition, the non-simple roots $\a = n_1 \a_1 + n_2 \a_2 + \cdots$
give rise to potential terms 
$\propto [ \exp ( -2 \a_1(\b))]^{n_1}  [ \exp ( -2 \a_2(\b))]^{n_2} \cdots $
which are subdominant w.r.t. the set of potentials corresponding to the simple
roots.\footnote{ Perhaps a useful analogy is to think of a
mountainscape (defined by Toda exponential potentials
for {\it all} roots); when the mountaintops rise into the sky, only the
nearest mountains (corresponding to the simple roots) remain
visible to the observer in the valley (the Weyl chamber). } Therefore, 
the $\s$-model Hamiltonian is asymptotically equivalent to the truncation 
of \Ref{HC1} to the simple roots:
\be\label{HCsimple}
\ch \sim \ch_{\rm simple} (\b, \pi) = \frac12   G^{\m\n} \pi_\mu \pi_\nu+
\sum_{i=1}^r K_i \exp \big[-2\a_i(\b)\big]
\ee
Such hyperbolic Toda models, restricted to the simple roots,  were
first introduced and studied in \cite{GIM}.

Furthermore, as explained in subsection 5.2, the potentials become
sharp wall potentials in the BKL limit $\b^\mu\rightarrow\infty$.
Making again the ``genericity assumption'' $K_i >0$ for the dominant 
simple root contributions (which is subject to the same caveats 
as before), we finally obtain the asymptotic Hamiltonian 
\be\label{HC2} 
\ch_\infty (\b, \pi) :=
\lim_{\rho\rightarrow\infty} \ch (\b, \pi) = \frac12 \pi^\mu
\pi_\mu +
  \sum_{i=1}^r  \, \Theta \big(-2\a_i (\b)\big)
\ee
where the sum is over the simple roots only, and the
motion of the $\b^\mu$ is confined to the fundamental Weyl
chamber $\a_i(\b) \geq 0$. Note that, in the present KM setup,
all the walls enter on the same footing. There is nothing left of 
the distinctions that entered our foregoing BKL-type studies between 
different types of walls (symmetry walls, gravitational walls,
electric walls,...).  The only important characteristic of a wall is
its height ${\rm ht} \, \a \equiv n_1 + n_2 +\cdots$ for a root
decomposed along simple roots as $ \a = n_1 \a_1 + n_2 \a_2 + \cdots$.

The sum in (\ref{HC}) not only ranges over the real roots,
which give rise to infinitely many timelike walls, but also
over null and purely imaginary roots of the KM algebra giving
rise to lightlike and spacelike walls, respectively. In view
of our discussion in section~7, the significance of the latter
for the KM billiard remains to be fully elucidated. Here we only
remark that imaginary roots are of no relevance for the Weyl group
of the KM algebra, which by definition consists only of reflections
against the timelike hyperplanes orthogonal to real roots.

To conclude: {\em in the limit where one goes to infinity in the
Cartan directions, the dynamics of the Cartan degrees of freedom
of the coset model become equivalent to a billiard motion within the
Weyl chamber, subject to the zero-energy constraint  $\ch (\b, \pi) =0$.}
Therefore, in the cases where the cosmological billiards that
we discussed in the first part of this paper are of KM-type,
the gravitational models are asymptotically equivalent (modulo the 
imposition of the additional momentum and Gauss constraints)
to the product over the spatial points of independent $(1+0)$-dimensional 
KM coset models $G/K$.

\subsection{Integrability, chaos and consistent truncations}
In this final subsection, we show that the   one-dimensional KM
$\sigma$-models are formally integrable. Then we  address the issue of
why this formal integrability is not incompatible with the occurrence of
chaos in the billiard description. We also discuss various possible 
finite-dimensional truncations of our infinite-dimensional 
KM-invariant $\s$-model.

The equations of motion of the off-diagonal fields constitute an infinite
system of non-linear ordinary differential equations of second order.
As usual, they are equivalent to the conservation of the $\frakg(A)$-valued
Noether charge
\beq\label{current1}
J &=&  \cv^{-1} P \cv \non
 &\equiv&  J^\mu H_\mu +
  \SD\Smult \left( J_{\a,s} E_{\a,s} + J_{-\a,s} E_{-\a,s} \right)
\eeq
in the gauge $n=1$. It is easily checked from
the first line of this equation that
\be\label{currentM}
J = \cm^{-1} \dot\cm
\ee
This equation can be formally solved   as
\be\label{M(t)}
\cm (t) = \cm (0) \cdot \exp (tJ)
\ee
For finite dimensional matrices $\cm$, the system is thus completely
integrable. This fact was already used in section~4 where we wrote 
down the exact solution \Ref{offdiagonal} of a $GL(n,\Rn)/O(n)$ model.
More precisely, the $GL(n,\Rn)$ analog of the (Lie-algebra valued)  
conserved charge $J$ is $ {\p^i}_j = \frac{1}{2} g^{ik} \dot g_{kj}$, 
whose conservation is clear from Eq.~\Ref{EoM2} . [Note again that 
the KM  gauge $n=1$ corresponds to $\tilde{N}=1$ so that  the KM time 
variable $t$ corresponds to the gravitational time scale $\t$.] 
In the finite-dimensional case we could derive the general solution 
by {\it diagonalizing} the constant matrix ${\p^i}_j$, which led to the 
Kasner solution, written in terms of the eigenvalues $v^a$ of ${\p^i}_j$ 
and of the diagonalizing matrix $L$ as in \Ref{offdiagonal}.
Then, thanks to finite dimensionality of the diagonalizing matrix,
we showed there that, after a finite transition time (subsequently
interpreted as linked to the finite number of collisions on the
symmetry walls needed to reorder the eigenvalues $v^a$),
the resulting asymptotic motion assumed a very simple monotonic form.

By contrast, for infinite dimensional matrices the existence of the 
formal solution \Ref{M(t)} does not necessarily imply regular behavior 
for the CSA degrees of freedom in the asymptotic limit where all other 
degrees of freedom get frozen. First of all, and in marked contrast
to the finite dimensional case, it is not known for indefinite 
KM algebras whether a generic Lie-algebra element $J$ can always be 
``diagonalized'', i.e. conjugated into the CSA by an element of the 
KM group $G$, see however \cite{KP}. Indeed, if that were the case, 
it would mean that a generic solution of the KM coset dynamics could 
be conjugated to a simple monotonic and diagonal Kasner-like solution 
$\b^\m = v^\m t +  \b^\m_0$  in the CSA, a conclusion difficult to 
reconcile with the chaotic motion that obtains in the BKL limit if the
KM algebra is hyperbolic. Secondly, even if $J$ could be conjugated 
into the CSA (as might be the case for certain non-generic elements)
there remains the possibility that the required diagonalizing matrix
(an infinite dimensional analog of the matrix $L$ in \Ref{offdiagonal}) 
introduces an infinite number of effective collisions in the time
development of the analog of \Ref{offdiagonal}, because the number
of effective collisions was found to increase with the rank of the 
matrix. In the latter case, the chaotic motion for the CSA degrees of freedom
would be the result of the projection onto a finite-dimensional space
of a regular geodesic motion taking place in an infinite-dimensional
phase space. In the former case, it might be the geodesic motion  on the
infinite-dimensional coset-space $G/K$  which could be intrinsically chaotic.
We will leave further investigation of these delicate mathematical questions
to future work.

In order to better understand why the existence of an infinite 
number of conserved quantities $\{ J^\mu , J_{\a,s},  J_{-\a,s} \}$ 
does not rule out chaos, let us consider possible 
{\it consistent truncations } of our $\s$-model. By a  
``consistent truncation'' we here mean a sub-model whose solutions
are solutions of the full model. In order to formalize this notion, 
let us introduce a {\it gradation} $\cd$ of root space: the $\cd$-degree  
of a root $\a = n_1 \a_1 + n_2 \a_2 + \cdots$ is defined as
$\cd(\a):= n_1 \cd_1 + n_2 \cd_2 + \cdots $, where $  (\cd_1, \cd_2, \cdots)$
is a given set of non-negative integers. Examples are the 
$sl(3,\Rn)$-level $\ell$, the affine level $m$, or the height 
$\cd (\a) \equiv {\rm ht}\, (\a) = n_1 + n_2 + \cdots$. In terms 
of an expansion according to increasing gradation $\cd$, the full $\s$-model 
Lagrangian \Ref{Lagrangian} has the structure (with all coefficients 
suppressed and in the gauge $n=1$):
\beq
 \cl \sim  \dot\b^2  +    \exp\big(2\a_1(\b)\big)\, [\dot \n^1]^2
+    \exp\big(2\a_2(\b)\big)\, \big[\dot \n^2 + \n^1 \dot \n^1\big]^2  \non
+     \exp\big(2\a_3(\b)\big) \, \big[\dot \n^3 + \n^1 \dot \n^2 + \n^2
\dot \n^1 +  \n^1 \n^1 \dot \n^1\big]^2 + \cdots  \label{expandedLag}
\eeq 
The notation here is somewhat schematic: the lower indices on the
roots and the upper indices on the off-diagonal fields refer to 
the gradation $\cd$ (so $\a_1, \a_2, \dots$ are {\em not} simple 
roots here). Furthermore, the degree zero term coincides with the 
free CSA kinetic term $\dot\b^2$ only if $\cd_i>0$ for all $i$; 
for other gradations, the level zero sector is described by the $\s$-model 
Lagrangian for the respective level-0 subalgebra (which may be 
non-abelian as was the case for the level $ell$ used in section~4.2).
The various terms within parentheses correspond to the
gradation of the $ j_{\a,s}$ above: $j_1 \sim  \dot \n^1,  j_2
\sim  \dot \n^2 + \n^1 \dot \n^1, j_3 \sim  \dot \n^3 + \n^1 \dot
\n^2 + \n^2 \dot \n^1 +  \n^1 \n^1 \dot \n^1, \dots$. The momenta
canonically conjugate to the off-diagonal variables
$\n^1,\n^2,\cdots$ have a similar graded structure 
\beq 
p_1 &\sim&
\exp\big(2\a_1(\b)\big)\, j_1   +   \exp\big(2\a_2(\b)\big) \, j_2 \n^1
+    \exp\big(2\a_3(\b)\big) \, j_3 \big[\n^2 + \n^1 \n^1\big] +\cdots \non
&& \hspace{1cm}= \Pi_1 + \Pi_2 \n^1 + \Pi_3 (\n^2 + \n^1 \n^1) +\cdots, \non 
p_2 &\sim&   \exp\big(2\a_2(\b)\big) j_2  +
\exp\big(2\a_3(\b)\big) j_3   \n^1  + \cdots  \non 
&& \hspace{1cm} = \Pi_2 + \Pi_3 \, \n^1   + \cdots, \non 
p_3 &\sim& \exp\big(2\a_3(\b)\big)\, j_3 +\cdots = \Pi_3 +\cdots,
\label{momenta} 
\eeq 
where, consistently with the above definition, we have introduced 
the short-hand notation $\Pi_n \equiv \exp\big(2\a_n(\b)\big)\, j_n$. 
Formally, the triangular relations above can be inverted iteratively
to get infinite series of the form 
\beq 
\Pi_1 &\sim& p_1 + p_2 \, \n^1 + p_3 \,\big[\n^2 + \n^1 \n^1\big] +\cdots,\non 
\Pi_2 &\sim& p_2 + p_3 \, \n^1 + p_4 \, \big[\n^2 + \n^1 \n^1\big] +\cdot,\non
\Pi_3 &\sim&  p_3   + p_4 \, \n^1 +\cdots
\eeq
This yields a formal expression  for the Hamiltonian in terms of
the canonical variables $(\b,\p; \n^n, p_n)$, where $n$ again refers
to the gradation,
\beq
\ch(\b,\p; \n^n, p_n)  &\sim& \p^2 +   \exp\big(-2\a_1(\b)\big)\,
\Pi_1^2 +   \exp\big(-2\a_2(\b)\big) \, \Pi_2^2 + \non &&\hspace{2cm}
\exp\big(-2\a_3(\b)\big)\, \Pi_3^2  + \cdots \eeq
It is now easy to see that one can obtain a consistent 
{\it finite-dimensional} truncation of the dynamics by requiring that
{\em all canonical momenta  above a certain  $\cd$-degree $n_0$
vanish }.  For instance, if we require $p_n =0$ for $n\geq 3$
implying $\Pi_n  = 0$ for all $n\geq 3$ we get a consistent dynamics 
following from the finite Hamiltonian 
\beq
\ch^{(2)}(\b,\p; \n^1, p_1, \n^2, p_2)  &\sim& \p^2 +
\exp\big(-2\a_1(\b)\big)\, \big[ p_1 + p_2 \n^1\big]^2 + \non && \hspace{2cm}
\exp\big(-2\a_2(\b)\big) \, p_2^2 
\label{h2} 
\eeq 
The finite-dimensional Lagrangian corresponding to this Hamiltonian reads 
\be \label{l2}
\cl^{(2)} \sim \dot\b^2 +    \exp\big(2\a_1(\b)\big) \, [\dot \n^1]^2
+    \exp\big(2\a_2(\b)\big) \, \big[ \dot \n^2 + \n^1 \dot \n^1 \big]^2
\ee
and is obtained from the original Lagrangian \Ref{expandedLag} by 
setting to zero all $j_n$ for $n\geq 3$, where, for instance,
$j_3 = \dot \n^3 + \n^1 \dot \n^2 + \n^2 \dot \n^1 +  \n^1 \n^1 \dot \n^1$,
and so on. Note that the highest-degree variable $\n^2$ does not 
explicitly appear in the truncated Hamiltonian $\ch^{(2)}$. This 
implies that the corresponding  highest-degree canonical momentum $p_2$  
is a constant of the motion. However, non-trivial dynamics is obtained
for the remaining degrees of freedom  $\b,\p; \n^1, p_1 $. Note also 
that a {\it particular} set of solutions of the $\cd$-degree-2 
dynamics above is obtained when the constant of motion $p_2$ happens 
to vanish. This particular case of the $\cd$-degree-2 dynamics
is simply the $\cd$-degree-1 truncation of the original dynamics, 
obtained by setting to zero all momenta above level 2, i.e.   
$ p_2 = p_3  = \dots = 0$.  The corresponding Hamiltonian is simply
\be
\label{h1}
 \ch^{(1)}(\b,\p; \n^1,p_1)  \sim \p^2 +   
\exp\big(-2\a_1(\b)\big)\, p_1^2
 \ee
Again the highest-degree momentum, which is now $p_1$, is a constant 
of the motion, so that   $ \ch^{(1)}$ directly defines the 
reduced dynamics of the Cartan variables $\b, \p$.

For $\cd \equiv {\rm ht}$, the truncated dynamics \Ref{h1} at degree
$\cd =1$ defines the finite hyperbolic Toda model \Ref{HCsimple} 
involving only the simple roots. We have seen in the previous subsection 
that this height-1 Hamiltonian universally describes the asymptotic 
dynamics  of the full $\s$-model. [Its BKL-limit directly yields the 
universal Weyl-chamber billiard  \Ref{HC2}.] On the other hand,
when the chosen gradation is the $sl(3)$ level of $AE_3$, we get
the low-level truncations of subsection~8.2 above: the level-0 truncation
($p_1= \dots = 0$) yields the Kasner dynamics, while the level-1 
dynamics describes the leading effect of the gravitational walls.  
A similar hierarchy of truncations appears in the $E_{10}$ model
of \cite{DHN2}, where the analysis included levels $\ell=0,1,2,3$ 
w.r.t. to the $sl(10,\Rn)\equiv A_9$ (see also \cite{DH2}).

Let us now clarify the meaning of the {\it infinite} set of conserved 
charges $J_{\pm\a,s}$ within the context of the {\it finite-dimensional} 
truncations which we just discussed. To this aim, one expresses 
$J$ in terms of the (non-canonical) Hamiltonian variables $(\b,\p,\n,\Pi)$:
\beq\label{current2}
J   &=& \pi^\mu \cn^{-1} H_\mu \cn + \non
&& + \SD\Smult \exp \big[-2\a(\b)\big]\, \Pi_{\a ,s} \cn^{-1} E_{\a,s}\cn\non
   && + \SD\Smult \Pi_{\a ,s} \cn^{-1} E_{-\a,s} \cn
\eeq
Here only the third term  spreads over the full algebra 
$\frakg = \frakn^- \oplus \frakh \oplus \frakn^+$, while the first 
and second are entirely contained in the parabolic subalgebras
$\frakh \oplus \frakn^+$ and $\frakn^+$, respectively. From this fact, 
it is easy to see that any truncation where the $\Pi_{\a,s}$ vanish 
for all roots strictly above some degree, i.e. for  $\cd(\a) > n_0$, 
implies the vanishing
of the correspondingly low ``negative-root charges''  $J_{-\a,s}$. 
As for the non-zero highest-degree momenta $\Pi_{\a_0,s}$ with
$\cd(\a_0) =n_0$, Eq.~\Ref{current2} simply yields 
$ \Pi_{\a_0,s}= J_{-\a_0,s}$. We thus recover the result above 
concerning the constancy of the highest-degree momenta. We can 
then move up (always in $\cd$-degree) by adding simple roots 
$\a_i$ to $\a_0$. Eq.~\Ref{current2} then implies
$ \Pi_{\a_0 - \a_i,s} = J_{-\a_0 + \a_i,s} + c \nu_{\a_i} J_{-\a_0,s} $,
where the constant $c$ is determined by the commutations relations. 
Continuing in this manner until one reaches the degree zero, one obtains 
for all the negative-degree (non-vanishing) momenta  expressions
which are linear in the charges and polynomial of ascending order in
the off-diagonal fields $\nu_{\a,s}$. Note, however, that these relations 
do not suffice to  eliminate non-trivial degrees of freedom from 
the truncated Hamiltonian,i.e. to replace it by an (on-shell) 
equivalent  Hamiltonian depending  on fewer variables.
[The only degrees of freedom that one can straightforwardly eliminate 
are the trivial highest-degree ones, whose associated ``position 
variables'' do  not enter the Hamiltonian.]  One might think that 
one would get further relations, and eventually enough relations to 
eliminate the off-diagonal degrees of freedom, by considering  
the higher-degree components of Eq.~\Ref{current2}. However, it 
does not work this way. The degree-zero projection, namely
\be
\label{current3}
\pi^\mu H_\mu + \SD\Smult \Pi_{\a,s}
    \big( \cn^{-1} E_{-\a,s} \cn \big) \Big|_\frakh = c^\mu H_\mu
\ee
(where $c^\mu$ is a constant vector) gives an expression for 
the degree-zero momenta $\p^\m$ which is linear in the charges 
and polynomial in the $\nu_{\a,s}$'s. On the other hand, 
the positive-degree projection of   Eq.~\Ref{current2} yields
for the positive-degree components of the charges $J_{\a,s}$ 
expressions which (for any positive degree)  will involve ``hidden'' 
off-diagonal fields  $\nu_{\a,s}$ of degree $\cd(\a) > n_0$.
These new variables did not enter the truncated Hamiltonians \Ref{h2},
which shows explicitly that the infinite number of conserved 
charges $J_{\a,s}$ of the original $\s$-model do not provide 
sufficiently many {\it autonomous} constants of the motion for 
the truncated Hamiltonian to guarantee its ``integrability'' in the
trivial sense of allowing one to reduce the number of degrees 
of freedom to zero.

Let us also dispose of another paradox concerning the formal 
integrability  of KM-related models. Even if we consider the 
hyperbolic Toda model defined by the truncation of our general 
$\s$-model to height one, i.e. $\ch_{\rm simple} (\b, \pi)$
given by Eq.~\Ref{HCsimple} with (strictly)  positive constants $K_i$, 
one finds that it has formal integrability features.
Indeed, the model \Ref{HCsimple} admits a Lax pair. Setting
$s_i \equiv \sqrt{  \langle \a_i | \a_i\rangle /2}$,   
rescaling the Chevalley-Serre generators as $e'_i  = s_i e_i$,  
$f'_i  = s_i   f_i, h'_i  = s_i^2  f_i$, and defining 
$W_{ij} \equiv  \langle \a_i | \a_j\rangle$, we replace
the variables $(\b^\m, \p_\m)$ by $(l^i, p^i)$ by
$l^i \equiv \sqrt{K_i} \exp \big[-\a_i(\b)\big] $ and 
$\Sigma_j  W_{ij} p^j \equiv G^{\m\n} (\a_i)_\m \p_\n$.
One can then check that the Lie-algebra valued expressions
\be
M = \sum_i [ p^i h'_i + l^i (e'_i+f'_i) ] \,\, , \,\, K = \sum_i 
l^i (e'_i-f'_i),
\ee
satisfy the Lax-pair evolution equation $\dot M = [K,M]$. One would 
then expect to be able to write down (infinitely) many conserved 
quantities associated to the isospectrality of the evolution of $M(t)$.  
[In the finite-dimensional case all the traces of the matrix  $M(t)$ 
are constants of the motion, see e.g. \cite{OT} and references therein.] 
However, in the KM case this formal Lax-pair integrability does not 
yield any concrete constants of the motion because there is no useful 
analog of the matrix trace in the KM algebra. The integrability 
properties of hyperbolic Toda theories have also been 
discussed in \cite{GIM}.

Let us note also in passing that there is full compatibility between 
our general result of the asymptotic constancy of all the off-diagonal 
degrees of freedom $\nu_{\a,s}$, $\Pi_{\a,s}$ and the existence of 
an infinite number of conserved quantities $J_{\a,s}$.  Morally
speaking the two infinite sets of asymptotic constants of integration
$\{ \lim \nu_{\a,s}\}$, $\{\lim \Pi_{\a,s} \}$  correspond to the 
doubly infinite set of charges $\{ J_{\pm\a,s} \}$  (in non-zero degree), 
and one can check the compatibility of this correspondence by using  
the results above concerning truncated models. As for the zero-level
conservation law \Ref{current3} it is also compatible with the 
asymptotic limit because, in the gauge we use here (analogous to 
the $\b$-space picture of the billiard) all the components of the 
zero-degree momenta $\p^\m$ tend to zero asymptotically (because of 
the redhifts accumulated on the receding walls).

In summary, there is thus no contradiction between the formal 
integrability of our $\s$-model and the chaos that appears in its 
generic asymptotic dynamics. This is mainly due to the presence of
infinitely many degrees of freedom, and the dynamics of the
finite-dimensional truncations of the $\s$-model of strictly 
positive degree.

\section{Conclusions}
\label{KM} \setcounter{equation}{0} \setcounter{theorem}{0}
\setcounter{lemma}{0}

In this paper, we have shown that theories involving gravity
admit a remarkable asymptotic description in the vicinity of a
spacelike singularity in terms of billiards in hyperbolic space.
Depending on whether the actual billiard has finite or infinite
volume, the dynamical evolution of the local scale factors is
chaotic (oscillatory) or monotonic (Kasner-like).
The billiard, and in particular its volume, is a fundamental
characteristic of the theory, in the sense that it is determined
solely by the field content and the parameters in the Lagrangian,
and not by the initial conditions (in the generic case; i.e.,
there may be initial conditions for which some walls are absent --
and the billiard is changed --, but these are exceptional).
Although we have not investigated the physical implications of
this property for cosmological scenarios (in particular,
string-inspired cosmologies \cite{GaVe,BDV,Lids,Wa,GaspV}), nor
its quantum analog, we believe that this result is interesting in
its own right because it uncovers an intrinsic feature of
gravitational theories. As discussed in section \ref{KM0}, the
regularity properties of some billiards appear to give a powerful
handle on possible hidden Kac-Moody symmetries, which remain to 
be exploited for their full worth.

\section*{Acknowledgements}
The work of M.H. is supported in part by the ``Actions de Recherche
Concert{\'e}es" of the ``Direction de la Recherche Scientifique -
Communaut{\'e} Fran{\c c}aise de Belgique", by a ``P\^ole
d'Attraction Interuniversitaire" (Belgium), by IISN-Belgium
(convention 4.4505.86), by Proyectos FONDECYT 1970151 and 7960001
(Chile) and by the European Commission RTN programme
HPRN-CT-00131, in which he is associated to K. U. Leuven. H.N. is
partially supported by the EU contract HPRN-CT-2000-00122. Both
M.H. and H.N. would like to thank I.H.E.S. for hospitality while
this work was carried out.

\appendix

\section{Asymptotic freezing: a simple model}
\label{freezing}
\setcounter{equation}{0} \setcounter{theorem}{0}
\setcounter{lemma}{0}

We have seen in the text that the ``non-diagonal'' phase-space variables
$(Q,P)$ (i.e. all variables except the extended scale factors  $\b^\m$
parametrizing the diagonal-Iwasawa-part of the metric, and the dilaton),
 get frozen to constant values in the BKL limit.
We provide here a more detailed understanding of this property by
discussing a simpler model which captures the essential features
of the Hamiltonian \Ref{V1}.

Consider a system with two canonically conjugate pairs $(q,p)$,
$(Q,P)$ and time-dependent Hamiltonian
\be
H(q,p,Q,P;T) = \frac{1}{2}p^2 + \frac{1}{2} (f(Q,P))^2 \rho^k e^{-\rho q}
\; \; \; \; \; \hbox{($k$ being any real number)}
\label{B1}
\ee
where $\rho$ is $\rho \equiv
\exp(T)$, with $T$ the time variable. We have simplified here the
Hamiltonian  \Ref{V2} by eliminating the variable $\lambda$,
i.e. by replacing it by an explicit function of the coordinate time $T$,
in the approximation of a constant $\p_{\lambda}$, so that
$\lambda$ is a linear function of $T$ (we scaled things so that
$\lambda = T$).
One can think of $(q,p)$ as mimicking
the scale factors, while $(Q,P)$ mimicks the off-diagonal
components or the $p$-form variables.  In (\ref{B1}), there is
only one potential wall for $q$ (namely, the second term).  We
shall consider later the case with several walls.

Let us start by remarking the important fact that
the variable $f(Q,P)$ has zero Poisson bracket with $H$, therefore
it is a constant of the motion.
To simplify the following argument,
we assume that we perform a canonical transformation such that
$ P^{\rm new} = f(Q,P)$. Dropping henceforth the label ``new'',
we get $P=P_0$ where $P_0$ is a constant
which we assume to be different from zero. The basic aim of  this Appendix is
then to show that the (new)  conjugate variable $Q$ will also tend to a constant as
$ T \to +\infty$.

In the limit of large
times, the motion in $q$ is a free motion interrupted by a
collision against the potential wall,
\be
q = p_0 \vert T-T_0 \vert + q_0
\ee
where $T_0$ is the time of the collision, $q_0$ the
turning point, and $p_0$ the constant momentum of $q$ far from
the wall.  The location of
the turning point is determined by energy conservation (using the fact
that $p(T_0)$ vanishes before changing sign):
\be
P^2_0 \rho_0^k e^{-\rho_0 q_0} = p_0^2.
\label{turning}
\ee
The time scale $\Delta T$ of the
collision (during which one feels the influence of the exponential
potential)  is roughly of the order $1/(\rho_0 p_0)$: the later the
collision, the sharper the wall. Let us evaluate the change in $Q$
in the collision.  To that end, we need to integrate
 $\dot{Q} = P \rho^k \exp(-\rho q)$ over the collision, which yields
\be
\Delta Q = P_0 \int_{-\infty}^\infty d \,T \rho^k e^{-\rho
(p_0 \vert T-T_0 \vert + q_0)}
\ee
 The integrand
is maximum at $T=T_0$. We can approximate the integral by the
value at the maximum times the time scale of the collision. Using
(\ref{turning}), one gets
\be
\Delta Q \approx \frac{p_0}{P_0\rho_0} =  \frac{p_0}{P_0} e^{- T_0}
\ee
Hence, the variable $Q$ receives a kick during the
collision (which can be of order one at early times), but the
later the collision (i.e. the larger $T_0$), the smaller the kick.

Assume now that there is another wall with the same prefactor and
 the same time
dependence, say at $q=d$, so that $q$ bounces between these two
walls,
$$
V_{additional} = \frac{1}{2} P^2 \rho^k e^{-\rho(d-q)}.
$$

 At each collision $Q$ receives a kick of order $1/\rho_0 = e^{- T_0} $.
 Because the speed of $q$ remains constant (in the large $T$
limit), the collisions are equally spaced in $T$.  Therefore,
the time of the $n$th collision grows
linearly with $n$: roughly  $T_0^{(n)} \sim n d/p_0$
The total
change in $Q$ is then obtained by summing all the individual changes,
which yields
\be
(\Delta Q)_{Total} \sim \sum_n e^{- n \, d / p_0}
\ee
 This sum converges.  Therefore, after a while, one can
neglect the further change in $Q$, i.e., assume $\dot{Q} = 0$. The
Hamiltonian describing the large time limit is obtained by taking
the sharp wall limit in the above $H$, and reads therefore
\be
H = \frac{1}{2}p^2 + \Theta(- q) + \Theta (q-d).
\ee
The pair $(Q,P)$
drops out because it is asymptotically frozen.  Our analysis
justifies taking the sharp wall limit directly in $H$ for this
system, which is the procedure we followed in the text to get the
gravitational billiards.

Note that,  had we studied a more complicated model with different
prefactors for the different walls, the prefactors $f_A(Q,P)$ would
no longer have been exactly conserved. However, their non-conservation
would only have been driven by the ``far-away'' walls, so that their
time-variation would have been exponentially small.

\end{document}